\documentclass[a4paper,11pt]{article}

\makeatletter
\renewcommand\tableofcontents{%
    \@starttoc{toc}%
}

\def\bS{\boldsymbol{S}}

\usepackage{amsmath,bbm}
\usepackage[pdfborder={0 0 0}]{hyperref}
\usepackage[T1]{fontenc}
\usepackage{microtype}

\newtheorem{theorem}{Theorem}[section]
\newtheorem{proposition}[theorem]{Proposition}
\newtheorem{lemma}[theorem]{Lemma}

\newtheorem{definition}[theorem]{Definition}

\numberwithin{equation}{section}

\usepackage{float}
\restylefloat{table}

\usepackage{mathtools,picins,array}

\usepackage{oldgerm}
\usepackage[usenames,dvipsnames]{color}

\usepackage{amsmath}

\newcommand{\Cl}{\mathbbm{C}}
\newcommand{\Rl}{\mathbb{R}}
\newcommand{\Nl}{\mathbb{N}}

\definecolor{lightgray}{rgb}{0.8,0.8,0.8}

\newcommand{\Om}{\Omega}

\newcommand{\te}{\theta}

\newcommand{\la}{\lambda}

\newcommand{\La}{\Lambda}
\newcommand{\eps}{\varepsilon}

\newcommand{\Tu}{\mathcal{T}}

\newcommand{\A}{\mathcal{A}}

\newcommand{\B}{\mathcal{B}}
\newcommand{\C}{\mathcal{C}}

\newcommand{\M}{\mathcal{M}}

\newcommand{\K}{\mathcal{K}}

\newcommand{\NN}{\mathcal{N}}

\newcommand{\Hil}{\mathcal{H}}
\newcommand{\VV}{\mathcal{V}}

\newcommand{\Q}{\mathcal{Q}}

\newcommand{\U}{\mathcal{U}}
\newcommand{\DD}{\mathcal{D}}
\newcommand{\W}{\mathcal{W}}

\newcommand{\SF}{\mathcal{S}}

\newcommand{\Wti}{\tilde{W}}

\newcommand{\gti}{\tilde{g}}

\newcommand{\LGpo}{\mathcal{L}_+^\uparrow}

\newcommand{\PGp}{\mathcal{P}_+}   
\newcommand{\LG}{\mathcal{L}}

\newcommand{\PGpo}{\mathcal{P}_+^\uparrow}   

\newcommand{\PG}{\mathcal{P}}

\newcommand{\frS}{\textfrak{S}}

\newcommand{\frA}{\frak A}

\newcommand{\balpha}{{\boldsymbol{\alpha}}}
\newcommand{\bbeta}{{\boldsymbol{\beta}}}

\def\bte{{\boldsymbol{\theta}}}
\def\bla{{\mbox{\boldmath{$\lambda$}}}}

\newcommand{\OO}{O}

\newcommand{\dom}{\mathrm{dom}\,}

\newcommand{\Strip}{\mathrm{S}}

\newcommand{\ot}{\otimes}

\newcommand{\tp}[1]{^{\otimes #1}}    

\newcommand{\zd}{z^{\dagger}}

\newcommand{\ad}{a^{\dagger}}

\usepackage[charter]{mathdesign}

\usepackage[cal=boondoxo,bb=boondox]{mathalfa}

\renewcommand{\OO}{\mathcal O}

\renewcommand{\SF}{\mathcal S}
\renewcommand{\Cl}{\mathbb{C}}

\begin{document}

\title{Algebraic constructive quantum field theory:\\ Integrable models and deformation techniques}
\author{Gandalf Lechner\footnote{Institut f\"ur Theoretische Physik, Universit\"at Leipzig, gandalf.lechner@uni-leipzig.de.}}
\date{March 12, 2015}

\maketitle

\begin{abstract}
	Several related operator-algebraic constructions for quantum field theory models on Minkowski spacetime are reviewed. The common theme of these constructions is that of a {\em Borchers triple}, capturing the structure of observables localized in a Rindler wedge. After reviewing the abstract setting, we discuss in this framework {\em i)} the construction of free field theories from standard pairs, {\em ii)~}the inverse scattering construction of integrable QFT models on two-dimensional Minkowski space, and {\em iii)} the warped convolution deformation of QFT models in arbitrary dimension, inspired from non-commutative Minkowski space.
\end{abstract}

\tableofcontents

\section{Algebraic constructions of quantum field theories}\label{Section:Intro}

Models in quantum field theory (QFT) are usually constructed with the help of a classical analogue: One starts from a classical relativistic field theory, and then uses some quantization procedure (typically involving renormalization) to arrive at a corresponding QFT model. Such constructions have led to theoretical predictions that in some cases match experimental data to a remarkable degree of accuracy. 

This success must however be contrasted with the difficulties of rigorously defining any interacting QFT model. Although at the level of perturbation theory, many QFT models are nowadays well understood (see, for example, \cite{Hollands:2008, BrunettiDuetschFredenhagen:2009, FredenhagenRejzner:2015} for discussions of this subject), one typically has no control over the convergence of the perturbation series, and no control over the error made by truncating it. 

{\em Constructive QFT}, on the other hand, aims at non-perturbative constructions of models of interacting quantum fields. This program was very successful in two and three spacetime dimensions, where a large family of interacting QFTs with polynomial self-interaction (``$P(\phi)_2$ models'' and the ``$\phi^4_3$ model'') was constructed, chiefly by Glimm and Jaffe \cite{GlimmJaffe:1987}. We refer to the recent review \cite{Summers:2011} for a detailed account of these constructions and the relevant literature. In four dimensions, however, no comparable results are known.

The models obtained in constructive QFT were shown to fit \cite{GlimmJaffe:1987} into both, the operator-algebraic Haag-Kastler framework \cite{Haag:1996} as well as the field-theoretic Wightman setting \cite{StreaterWightman:1964}. The tools used in their construction, in particular the Euclidean formulation, are however more closely linked to the field-theoretic picture. 

This review article focuses on a different approach to the rigorous construction of interacting models, namely on those which are constructed by operator-algebraic methods. In comparison to ``constructive QFT'', this ``algebraic constructive QFT''\footnote{A term coined by S.~J.~Summers, see also his online article {\tt http:// people.clas.ufl.edu/sjs/constructive-quantum-field-theory/} for a review.} is a much more recent topic. Due to page constraints, this review is far from exhaustive. However, several of the topics not treated here are the subject of other chapters in this book: For example, we will not discuss the recent progress in perturbative algebraic QFT \cite{DtschFredenhagen:2001-1}, or conformal algebraic QFT \cite{KawahigashiLongo:2004,Rehren:2015}, or the recent algebraic construction of models on two-dimensional de Sitter space by J.~Barata, C.~J\"akel and J.~Mund \cite{BarataJakelMund:2013}.
\\
\\
To motivate the discussion of the topics that {\em are} treated in this review, we first mention that from a philosophical point of view, there is no good reason to construct quantum field theories on the basis of classical field theories (that is, by quantization). Since quantum theories are supposed to provide the more fundamental description of reality, classical theories should rather appear only as a limiting case. One would like to take the quantum theories as the fundamental data, and consider their classical limits only for comparison with classical notions.

Desirable as it is, a purely quantum description of relativistic physics poses several major challenges. The first question is what the relevant mathematical structures are that one tries to construct. As we will focus on the algebraic setting of QFT \cite{Haag:1996} here, the data comprising a model theory in our case are a family of operator algebras $\A(\OO)$, labeled by localization regions $\OO$ in spacetime, subject to a number of conditions. Such a net of local algebras possesses an enormous degree of complexity, which makes it both, suitable for the description of the complex dynamics of relativistic field theories, but also challenging to construct. 

In model-inde\-pen\-dent investigations of QFT, the algebras of observables localized in certain spacelike wedge-shaped regions of Minkowski space (wedges), play a prominent role, a point emphasized in particular by D.~Buchholz. We will recall their definition and relevant properties in Section~\ref{Section:Wedges}. By the results of Bisognano and Wichmann \cite{BisognanoWichmann:1975,BisognanoWichmann:1976}, and later Borchers \cite{Borchers:1992}, algebras localized in wedges provide a link between the geometric properties of Minkowski space, encoded in its Poincar\'e symmetry, and certain algebraic properties of the net, encoded in its modular data\footnote{For large parts of this review, we will rely on Tomita-Takesaki modular theory, see for example \cite{BratteliRobinson:1987} for an introduction and \cite{Borchers:2000} for an overview of applications to QFT.}. Via the idea of {\em modular localization} (see Section~\ref{Section:Modular-Free}), this link also connects Wigner's classification of 
elementary particles by positive energy representations 
of the Poincar\'e group to the modular structure of wedge algebras.

Moreover, because wedges are unbounded regions, observables localized in them can have much milder momentum space properties than point-like localized quantum fields, which typically fluctuate enormously in energy and momentum. As argued by B.~Schroer, one can in particular consider wedge-localized observables that are free of vacuum polarization, i.e. just create single particle states from the vacuum, also in interacting theories \cite{Schroer:1999}. Such {\em polarization-free generators} do not exist for smaller localization regions in general, but can be used to generate algebras localized in wedges, and are directly related to the two-particle S-matrix \cite{BorchersBuchholzSchroer:2001,Mund:2010}. 

Finally, the family of wedges on Minkowski space forms a causally separating set (see Section~\ref{Section:Wedges}), so that it is possible to construct a complete net of local 
algebras in terms of a {\em single} algebra and a suitable representation of the Poincar\'e group. Making use of this observation, the construction of a QFT model is reduced to the construction of a so-called Borchers triple \cite{BuchholzLechnerSummers:2011}, consisting of an algebra localized in a wedge, together with a suitable representation of the Poincar\'e group and a vacuum vector. This general construction scheme is reviewed in the Sections~\ref{Section:WedgeAlgebras}--\ref{Section:RelativeCommutants}.
\\\\
While model-independent investigations did lead to the idea of constructing local nets from wedge algebras, they did not (yet) shed much light onto the question {\em how} this single algebra, on which the whole construction rests, should be realized. This question is closely related to the question of how to model interaction without making use of classical concepts, and as of now, has found no general answer. 

Thinking of quantum descriptions of interactions, the S-matrix is an object of central importance. Unfortunately, in theories with particle production, the S-matrix is also of such a complicated form that it is not a manageable quantity for describing interactions. There is, however, an exception to this rule: For certain {\em integrable} models on two-dimensional Minkowski space, the S-matrix is of a simple factorizing form, and in particular does not allow for production processes. In that setting, it is therefore possible to use it as suitable description of the quantum dynamics, and generate wedge algebras based on such an S-matrix.

This approach was initiated by B.~Schroer \cite{Schroer:1997-1}, who introduced certain wedge-local fields in this context (see Section~\ref{Section:S>BT}). This idea was then thoroughly investigated and generalized, in particular with regard to the analysis of {\em local} observables, by several authors. We will in Section~\ref{Section:IntegrableModels} review the construction of integrable models on two-dimensional Minkowski space by these methods, which led to the solution of the corresponding inverse scattering problem \cite{Lechner:2008}.
\\\\
To complement the concrete construction of integrable models on the basis of a factorizing S-matrix, we will also review a different construction scheme. As in the case of integrable models, the central object is that of a Borchers triple. However, here the input does not consist of an S-matrix, but rather amounts to a {\em deformation procedure}: Starting from the Borchers triple of some arbitrary QFT (in arbitrary dimension), one modifies/deforms it to a new, inequivalent one. The method to be used here is inspired \cite{GrosseLechner:2007} from non-commutative Minkowski space \cite{DoplicherFredenhagenRoberts:1995, BahnsDoplicherMorsellaPiacitelli:2015}, and now goes under the name of {\em warped convolution} \cite{BuchholzSummers:2008, BuchholzLechnerSummers:2011}. We review this deformation procedure in Section~\ref{Section:Warping}, where it is also compared to the approach taken in Section~\ref{Section:IntegrableModels}.

\newpage
\section{Operator-algebraic constructions based on wedge algebras}\label{Section:Abstract}

Most operator-algebraic approaches to constructing quantum field theory models on Minkowski space split the construction problem into two steps: First one constructs a {\em single} von Neumann algebra $\M$ and a representation of the Poincar\'e group with specific properties, and then these data are used to generate a full local net. The algebra $\M$ considered in the first step contains all observables localized in a special wedge-shaped region of Minkowski space, {\em wedge} for short. Before going into the quantum field theoretic constructions, we define these regions and discuss their geometric properties.

\subsection{Wedges}
\label{Section:Wedges}

In this section we will be working in Minkowski spacetime $\Rl^d$ of general dimension $d\geq2$, equipped with proper coordinates $x=(x_0,x_1,...,x_{d-1})$, with $x_0$ being the time coordinate\footnote{Although we will mostly be working with Minkowski space here, it should be noted that similar families of regions can also be defined in other situations: On the one-dimensional line, the half lines $(a,\infty)$ and $(-\infty,a)$, $a\in\Rl$, have the same properties as the wedges in Minkowski space (see also the discussion in Section~\ref{Section:Mass0}). Also the family of all intervals on a circle, of prominent importance in chiral conformal field theory \cite{Rehren:2015}, shares many properties with the family of wedges on Minkowski space, see for example \cite{Longo:2008}.

Furthermore, on certain curved spacetimes, such as de Sitter space \cite{BorchersBuchholz:1999,BrunettiGuidoLongo:2002}, anti de Sitter space \cite{BuchholzSummers:2004-2,LauridsenRibeiro:2007}, and more general curved spacetimes \cite{BuchholzMundSummers:2001,DappiaggiLechnerMorfaMorales:2010}, families of regions with properties analogous to Minkowski space wedges exist.}. The following regions will play a special role.

\begin{definition}\label{Definition:Wedges}{\bf (Wedges)}
	The {\em right wedge} is the set
	\begin{align}\label{def:WR}
		W_R
		:=
		\{x\in\Rl^d\,:\,x_1>|x_0|\}
		\,,
	\end{align}
	Any set $W\subset\Rl^d$ which is a proper Poincar\'e transform of $W_R$, i.e. $W=\La W_R+x$ for some $\La\in\LG_+$, $x\in\Rl^d$, is called a {\em wedge}. The set of all wedges is denoted $\W$.
\end{definition}

Wedges can equivalently be defined as regions that are bounded by two non-parallel characteristic hyperplanes \cite{ThomasWichmann:1997}, thereby avoiding reference to the particular wedge $W_R$. However, for our purposes the above definition will be convenient.

One might wonder why wedge regions deserve particular attention, and as a first answer to this question, we note that wedges have the special property that their causal complements are of the same form. In fact, one directly checks that the causal complement of $W_R$ is $W_R'=-W_R$, that is, a proper Poincar\'e image of $W_R$. Thus $W_R'\in\W$; it is customary to call $W_R'$ the {\em left wedge} and denote it by~$W_L$. By covariance, one then finds that for any wedge $W$, also $W'$ is a wedge. In the later constructions, this symmetry between wedges and their causal complements will be parallel to that of von Neumann (wedge) algebras and their commutants.

\piccaptiontopside
\piccaption{\small The right and left wedge in two-dimensional Minkowski space. Both regions extend to (right, respectively left) spacelike infinity.\label{Figure:WRandWLinD=2}}
\parpic(7cm,40mm)[l]{\includegraphics[scale=.45]{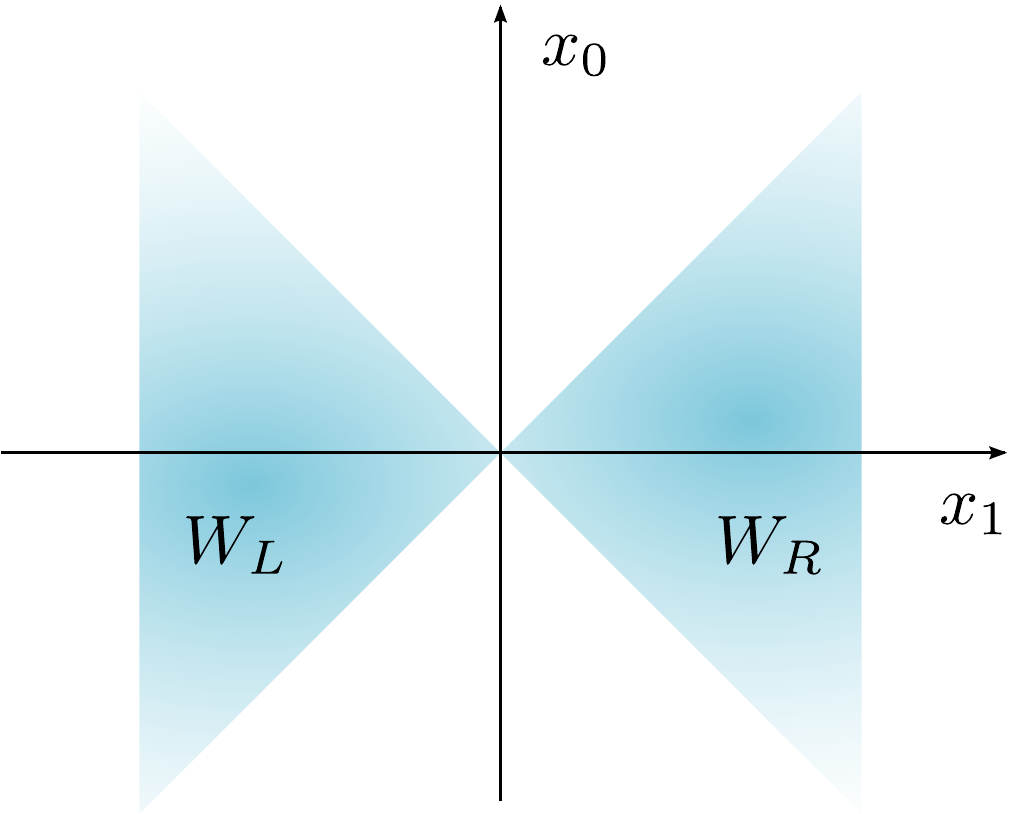}}
\vspace*{7mm}
\picskip{0}
As a consequence of $W_R'=-W_R$, we also have $W_R''=(-W_R)'=-W_R'=W_R$, so $W_R$ is {\em causally complete}\footnote{Hence $W_R$ is globally hyperbolic, and can be regarded as a spacetime in its own right.}. By covariance, this implies $W''=W$ for any wedge~$W$.

In the quantum field theory setting, we will be interested in mappings from $\W$ into the family of all von Neumann algebras on a fixed Hilbert space, complying with the usual assumptions of isotony, locality, and covariance \cite{Haag:1996}. As a preparation for this, we here consider the inclusion, causal separation, and covariance properties of wedges.

Beginning with inclusions, let $(x,\La)\in\PG_+$ denote a proper Poincar\'e transformation. It is clear that if $\La W_R=W_R$ and $x\in\overline{W_R}$, then $\La W_R+x\subset W_R$ (observe that wedges are in particular convex cones). In fact, also the converse is true, namely $\La W_R+x\subset W_R$ implies $\La W_R=W_R$ and $x\in\overline{W_R}$ \cite{ThomasWichmann:1997}. 

We thus see that there are only relatively few pairs of wedges $W_1,W_2$ that form inclusions. Namely, $W_1\subset W_2$ if and only if $W_1=W_2+x$ with $x\in\overline{W_2}$. Since causal complements of wedges are also wedges, the same applies to pairs of spacelike separated wedges: $W_1\subset W_2'$ if and only if $W_1=W_2'+x'$ with $x'\in\overline{W_2'}$. This simple structure of the family of causal configurations of wedges is used in constructions based on wedge algebras in a crucial manner.

Finally, we point out that the set $\W$ is {\em causally separating} in the following sense: Given any two bounded, convex, causally complete sets $\OO_1,\OO_2\subset\Rl^d$ (such as double cones), that are spacelike separated, $\OO_1\subset\OO_2'$, there exists $W\in\W$ such that $\OO_1\subset W\subset\OO_2'$ \cite[Prop.~3.7]{ThomasWichmann:1997} (see Fig.~\ref{Figure:WedgeInclusionsAndSeparations} below).

We collect these properties in a proposition.

\begin{proposition}\label{Proposition:WedgesInMinkowskiSpace}
	Any wedge $W\in\W$ is an open, convex, unbounded, causally complete set. The set $\W$ of all wedges in Minkowski space $\Rl^d$ is causally separating, and invariant under the action of the Poincar\'e group and causal complementation. Furthermore, given $W_1,W_2\in\W$ such that $W_1\subset W_2$, then $W_1=W_2+x$ for some $x\in\overline{W_2}$.	
\end{proposition}

By Definition~\ref{Definition:Wedges}, arbitrary Poincar\'e transformations leave $\W$ invariant as a set. For any given wedge $W\in\W$, there also exist specific Lorentz transformations which preserve $W$ or map it to its causal complement, respectively. Consider the proper Lorentz transformations, $t\in\Rl$,
\begin{align}\label{eq:jWR-LambdaWR}
	j_{W_R}(x)
	&:=(-x_0,-x_1,x_2,...,x_{d-1})
	\,,\\
	\La_{W_R}(t)x
	&:=
	\left(
	  \begin{array}{rrrrr}
	       \cosh(2\pi t) & \sinh(2\pi t) & 0 & \cdots & 0 \\
	       \sinh(2\pi t) & \cosh(2\pi t) & 0 & \cdots & 0 \\
	       0 & 0 & 1 & \cdots & 0\\
	       \vdots & \vdots &&\ddots &\vdots\\
	       0 & 0 &&&1
	  \end{array}
	\right)
	\left(\begin{array}{c}
	       x_0\\ x_1 \\x_2\\\vdots\\ x_{d-1}
	      \end{array}
	\right)
	\nonumber
	\,.
\end{align}
The map $j_{W_R}$ in the first line is the reflection about the $(d-2)$-dimensional {\em edge} $E(W_R):=\{x\,:\,x_0=x_1=0\}$ of $W_R$, and maps $W_R$ onto $-W_R=W_R'$. The second line defines the one parameter group of Lorentz boosts $\La_{W_R}(t)$ in the $x_1$-direction. By computing the eigenvectors and eigenvalues of $\La_{W_R}(t)$, $t\in\Rl$, one finds that these belong to the group of all Lorentz transformations that leave $W_R$ invariant as a set.

More generally, to any wedge $W=\La W_R+x$ we can assign its edge $E(W):=\La E(W_R)+x$, the reflection 
\begin{align}
 j_W:=(x,\La)j_{W_R}(x,\La)^{-1}
\end{align}
about $E(W)$, which satisfies $j_W(W)=W'$, and a one-parameter group of boosts, 
\begin{align}
 \La_W(t):=(x,\La)\La_{W_R}(t)(x,\La)^{-1}\,,
\end{align}
which preserve $W$. 
\\\\
\noindent{\bf Wedges in $d=2$ dimensions.}
While so far the spacetime dimension $d\geq2$ was arbitrary, we now specialize to the two-dimensional situation, where wedges have a number of additional properties. 

To begin with, the causal complement of any one point set $\{x\}$ in $\Rl^2$ consists precisely of the disjoint union of the two wedges $W_R+x$ and $W_L+x$ (cf. figure~\ref{Figure:WRandWLinD=2}). These are in fact all wedges in this setting: In $d=2$ dimensions, the proper Lorentz group is generated by the one parameter group $\{\La_{W_R}(t)\}_{t\in\Rl}$ and the spacetime reflection $j_{W_R}(x)=-x$ (which maps $W_R$ onto $W_L$, as just observed). Thus in this case,
\begin{align}
	\W=\{W_R+x,\;W_L+x\;:\;x\in\Rl^2\}
	\qquad\qquad (d=2)\,.
\end{align}
In two dimensions, wedges can also be most easily visualized.
\vspace*{6mm}
\piccaptiontopside
\piccaption{\small {\em left:} An inclusion of two right wedges $W_R+a\subset W_R$.\newline\newline
{\em right:} An illustration of the causal separation property of $\W$: The two spacelike regions $\OO_1,\OO_2$ are separated by the wedge $W$, i.e. $\OO_1\subset W\subset \OO_2'$.\label{Figure:WedgeInclusionsAndSeparations}}
\parpic(9cm,50mm)[l]{\includegraphics[scale=.5]{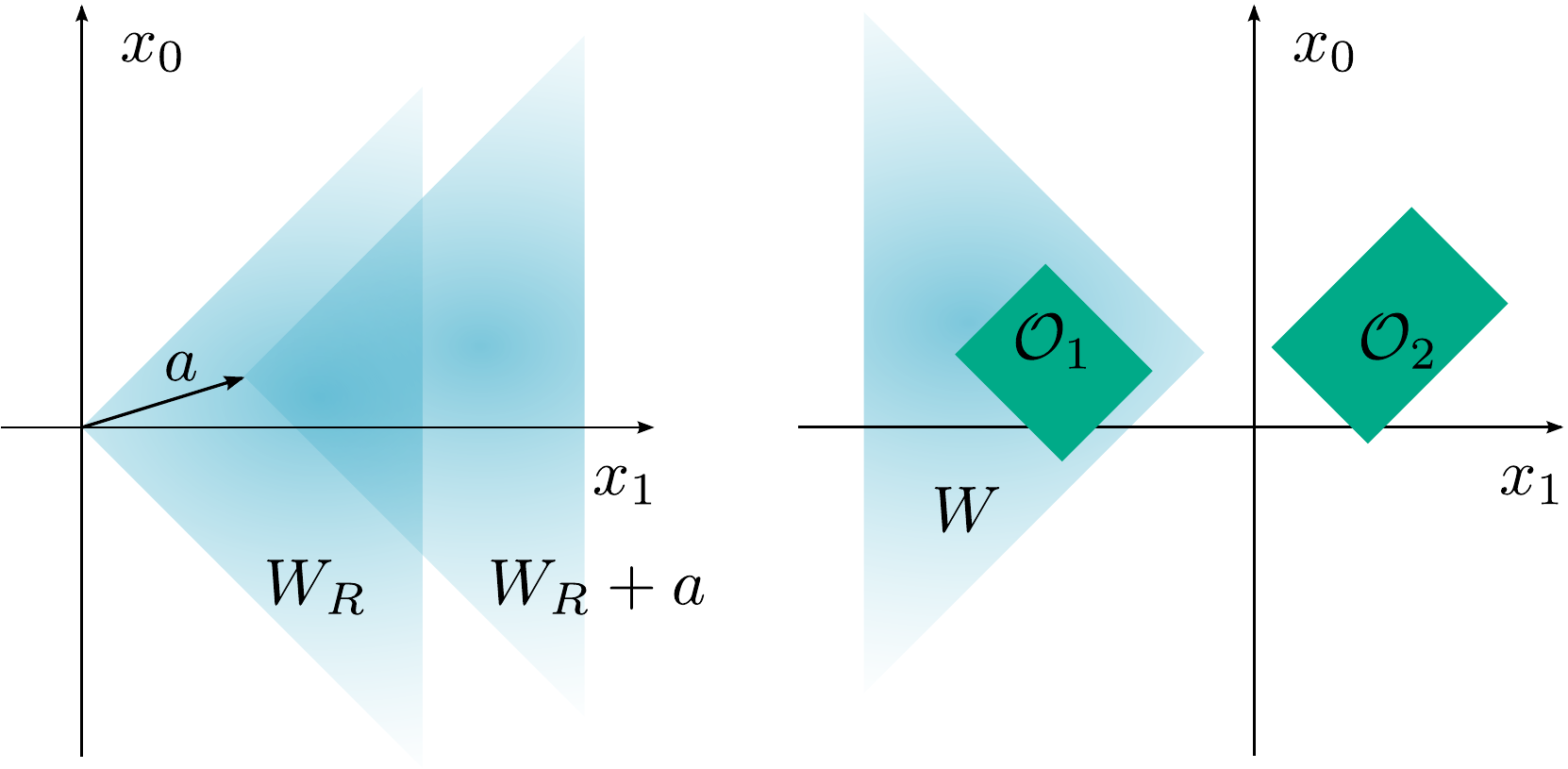}}
\vspace*{8mm}
\picskip{0}

{\em Double cones}, typically the most important localization regions in quantum field theory, are defined as intersections of forward and backward light cones. However, in two dimensions, this is the same as taking intersections of left and right wedges,
\begin{align}\label{eq:DoubleConesAsIntersectionsOfWedges}
	\OO_{x,y} 
	:=
	(W_R+x)\cap (W_L+y)
	=
	(W_R+x)\cap (W_R+y)'
	\,;
\end{align}
and this set is nonempty if and only if $(y-x)\in W_R$. Any double cone arises in this way, and from the second equality in \eqref{eq:DoubleConesAsIntersectionsOfWedges}, we see that the double cone $\OO_{x,y}$ is the {\em relative causal complement} of the inclusion $W_R+y\subset W_R+x$ (see figure \ref{Figure:RelativeCausalComplement}).

\vspace*{6mm}
\piccaptiontopside
\piccaption{\small {\em left:} The double cone $\OO_{x,y}$ as the wedge intersection $(W_R+x)\cap(W_L+y)$.\newline\newline
{\em right:} The double cone $\OO_{x,y}$ and the associated inclusion of right wedges $W_R+y\subset W_R+x$..\label{Figure:RelativeCausalComplement}}
\parpic(9cm,50mm)[l]{\includegraphics[width=85mm]{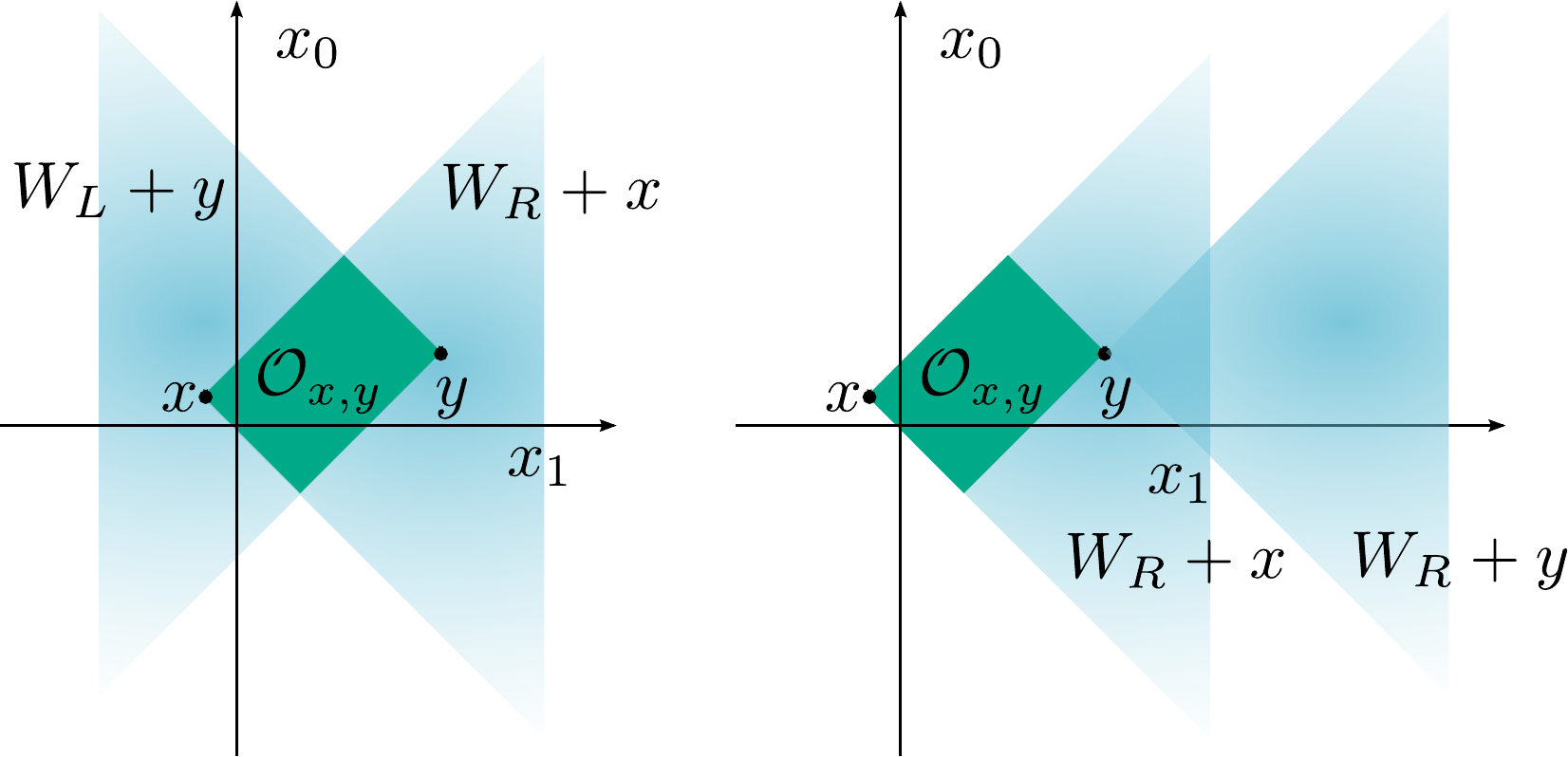}}
\vspace*{8mm}
\picskip{0}

In $d>2$ dimensions, nonempty intersections of two wedges are unbounded regions, here one needs an intersection of several wedges to arrive at a bounded region. In particular, double cones are not given as relative causal complements of inclusions of wedges in $d>2$. Quite generally, inclusions are easier to analyze than intersections. This is true on the geometric level of wedges\footnote{Note, for example, that for any $n\in\Nl$, there exists a family of $n$ wedges $W_1,...,W_n\subset\Rl^4$ such that $W_i\cap W_j=\emptyset$ for $i\neq j$ \cite{ThomasWichmann:1997}.}, but even more on the level of von Neumann algebras, and provides one of the many reasons why certain construction procedures are easier in $d=2$.

\subsection{Wedge algebras and Borchers triples}
\label{Section:WedgeAlgebras}

Having clarified the geometrical preliminaries, we now turn to studying algebras of observables localized in wedges. This discussion will take place in the setting of a vacuum representation of a quantum field theory on Minkowski space $\Rl^d$, $d\geq2$ (see for example \cite{Haag:1996}). We therefore consider a Hilbert space $\Hil$, carrying a strongly continuous (anti-)unitary positive energy representation $U$ of the proper Poincar\'e group with a $U$-invariant unit vector $\Om$, implementing the vacuum state. The observables of a quantum field theory are represented as operators on $\Hil$: Associated with any localization region $\OO\subset\Rl^d$, we have the $C^*$-algebra $\frA(\OO)\subset\B(\Hil)$ of all observables localized in $\OO$, and the usual assumptions of isotony, locality and covariance under $U$ are assumed to hold for the net $\OO\mapsto\frA(\OO)$. As we are in a 
vacuum representation, we also assume that $\Om$ is cyclic and separating for $\frA(\OO)''$, for each double cone $\OO$ (Reeh-
Schlieder property \cite{Haag:1996}).

In this setting, we introduce the von Neumann algebra $\M$ associated with the right wedge $W_R$ as the smallest von Neumann algebra containing all $\frA(\OO)$, $\OO\subset W_R$,
\begin{align}
	\M
	:=
	\bigvee_{\OO\subset W_R}\frA(\OO)''
	\,.
\end{align}
This algebra has a number of properties which reflect the geometric properties of $W_R$, and which follow directly from the definition of $\M$ and the properties of the net $\OO\mapsto\frA(\OO)$: For any Poincar\'e transformation $g$ with $gW_R\subset W_R$, we have $U(g)\M U(g)^{-1}\subset\M$, and for any Poincar\'e transformation $\gti$ with $\gti W_R\subset W_R'$, we have $U(\gti)\M U(\gti)^{-1}\subset\M'$. Furthermore, the vacuum vector $\Om$ is cyclic for $\M$ (because $W_R$ contains a double cone) and separating  for $\M$ (because $W_R'$ also contains a double cone).

As these properties will be essential in the following, we isolate them in a definition\footnote{Note that we deviate here slightly from the definition in \cite[Def.~4.1]{BuchholzLechnerSummers:2011}, where the term ``causal Borchers triple'' has been used. Also note that in \cite{GuidoLongoWiesbrock:1998}, there is a related but different definition of the term ``Borchers triple''. We will always stick to the definition given here.}. The term ``Borchers triple'', in honor of H.-J.~Borchers who studied such systems (see, e.g., \cite{Borchers:1992}), was suggested in  \cite{BuchholzLechnerSummers:2011}.

\begin{definition}\label{Definition:BorchersTriple}
	A $d$-dimensional {\em Borchers triple} $(\M,U,\Om)$ relative to $W\in\W$ consists of 
	\begin{enumerate}
		\item a strongly continuous (anti-)unitary positive energy representation $U$ of the proper Poincar\'e group $\PG_+$ of $\Rl^d$ on some Hilbert space $\Hil$,
		\item a unit vector $\Om\in\Hil$ that is invariant under $U$, and
		\item\label{item:BorchersTriple-PoincareCondition} a von Neumann algebra $\M\subset\B(\Hil)$ which has $\Om$ as a cyclic and separating vector and which satisfies
		\begin{align}
			U(g)\M U(g)^{-1}&\subset\M\quad\text{\;for all}\;g\in\PG_+ \text{ with }\; gW\subset W\,,
			\label{eq:WedgeIsotony}
			\\
			U(g)\M U(g)^{-1}&\subset\M'\quad\text{for all}\;g\in\PG_+ \text{ with }\; gW\subset W'\,.
			\label{eq:WedgeLocality}
		\end{align}
	\end{enumerate}
\end{definition}

A von Neumann algebra $\M$ in a Borchers triple relative to $W$ describes quantum observables that are localized (in the specified sense) in the wedge~$W$, and will also be referred to as {\em wedge algebra} when $U$ and $\Om$ are fixed and clear from the context. For the sake of concise formulations, we agree to drop the specification ``relative to $W$'' for Borchers triples relative to our standard reference wedge $W=W_R$, or if $W$ is clear from the context.

In comparison to a full quantum field theory, described by an infinite collection of algebras in specific relative positions, the data $(\M,U,\Om)$ of a Borchers triple are much simpler. However --- and this observation is central for all that follows --- one can reconstruct a full net of local algebras from a Borchers triple, essentially by Poincar\'e symmetry (cf. in particular \cite{BisognanoWichmann:1975,BisognanoWichmann:1976,Borchers:1992}, \cite[Sec.~7.3.6]{BaumgrtelWollenberg:1992}), as shall be explained below.
\\\\
Setting ourselves the task to define a local net $\OO\mapsto\A(\OO)$ of von Neumann algebras corresponding to a given Borchers triple $(\M,U,\Om)$, we have to give definitions of the algebras $\A(\OO)$, $\OO\subset\Rl^d$. Assuming for the sake of concreteness that the Borchers triple is relative to the right wedge $W_R$, one first sets
\begin{align}\label{eq:WedgeAlgebrasByCovariance}
	\A(\La W_R+x)
	:=
	U(x,\La)\M\,U(x,\La)^{-1}
	\,,\qquad
	(x,\La)\in\PG_+\,.
\end{align}
This defines for any wedge $W\in\W$ a von Neumann algebra $\A(W)\subset\B(\Hil)$ (Note that \eqref{eq:WedgeAlgebrasByCovariance} is well-defined because of \eqref{eq:WedgeIsotony}). Making use of the properties of the Borchers triple, one checks that \eqref{eq:WedgeAlgebrasByCovariance} yields a map $\W\ni W\mapsto\A(W)\subset\B(\Hil)$ from wedges on $\Rl^d$ to von Neumann algebras in $\B(\Hil)$ that a) is inclusion preserving (isotony), b)~maps spacelike separated wedges to commuting algebras (locality), and c) transforms covariantly under the adjoint action of $U$ by its very definition. Moreover, $\Om$ is cyclic and separating for any $\A(W)$, $W\in\W$.

To proceed from the algebras $\A(W)$ to algebras associated with smaller regions, we first consider double cones. Any double cone $\OO$ is an intersection of wedges, $\OO=\bigcap_i W_i$, where $i$ runs over some index set and $W_i\in\W$. We associate with it the von Neumann algebra
\begin{align}\label{eq:GeneralWedgeIntersection}
	\A(\,\bigcap_i W_i)
	:=
	\bigcap_i \A(W_i)\,.
\end{align}
Making use of the properties listed in Proposition~\ref{Proposition:WedgesInMinkowskiSpace}, one can show that this yields a map from double cones to von Neumann algebras which inherits the isotony, locality, and covariance properties from the wedge net $\W\ni W\mapsto\A(W)\subset\B(\Hil)$ \cite{BisognanoWichmann:1975}. 

Finally, given an arbitrary bounded region $\OO$, we define $\A(\OO)$ as the smallest von Neumann algebra containing all $\A(D)$, where $D\subset\OO$ is a double cone. Also this step preserves isotony, locality, and covariance. We therefore note:

\begin{proposition}\label{Proposition:BasicNetFromBorchersTriple}
	Any $d$-dimensional Borchers triple $(\M,U,\Om)$ defines a local net $\OO\mapsto\A(\OO)$ of von Neumann algebras on $\Rl^d$ such that
	\begin{enumerate}
		\item $\OO\mapsto\A(\OO)$ is isotonous and local, and transforms covariantly under $U$.
		\item $\A(W_R)=\M$ is the algebra associated with the right wedge $W_R$.
		\item For any $\Wti\in\W$, the vacuum vector $\Om$ is cyclic and separating for $\A(\Wti)$.
	\end{enumerate}
\end{proposition}

In view of this observation, the construction of local nets of von Neumann algebras, i.e. models of quantum field theories, is closely related to the construction of Borchers triples $(\M,U,\Om)$. We will see below (p.~\pageref{page:2dConstruction}) that in $d=2$ dimensions, a slight variation of this construction is available.

The representation $U$ (and the vector $\Om$) describe the Poincar\'e symmetry of the model theory constructed from $(\M,U,\Om)$. By decomposition into irreducible components, $U$ yields a list of all species of stable particles in this theory, and can thus be thought of as data implementing kinematic properties. From the point of view of constructing examples, the representation $U$ and vacuum vector $\Om$ pose no problems --- In fact, it follows from Haag-Ruelle scattering theory that in any theory, one may use a standard Fock space construction for realizing these data in terms of a single particle representation $U_1$ of the Poincar\'e group \cite{Araki:1999}, cf. Section~\ref{Section:Modular-Free}.

The dynamics and interaction of the model corresponding to $(\M,U,\Om)$ is encoded in an indirect manner in the choice of von Neumann algebra~$\M$, completing $U,\Om$ to a Borchers triple. In the present general context, where in particular no link to a classical Lagrangian or equation of motion is made, there is currently no general principle known to select $\M$ in such a way that the model constructed along the lines described above exhibits features of a particular type of interaction. As we will see later in Section~\ref{Section:IntegrableModels}, it is however well possible to realize $\M$ in terms of a two-particle S-matrix in the setting of integrable models on two-dimensional Minkowski space, or to modify a given $\M$ to a new one by a deformation procedure (Section~\ref{Section:Warping}).

Independent of concrete construction ideas for $\M$, it should be noted that the conditions in Def.~\ref{Definition:BorchersTriple} impose strong restrictions on $\M$ in general. This is illustrated by a theorem of Longo \cite[Thm.~3]{Longo:1979} (see also Driessler~\cite{Driessler:1975}), which in the context of Borchers triples reads as follows.

\begin{theorem}\label{Theorem:WedgeAlgebrasAreTypeIII1Factors}
	Let $(\M,U,\Om)$ be a Borchers triple. If $\M\neq\Cl$, then $\M$ is a factor of type III$_1$.
\end{theorem}

Since the hyperfinite type III$_1$ factor is known to be unique \cite{Haagerup:1987}, this implies that the internal algebraic structure of $\M$ is almost uniquely fixed by Def.~\ref{Definition:BorchersTriple}. 

A result of a similar nature is a famous theorem of Borchers, stating that there exist also strong restrictions on the modular data of a wedge algebra~\cite{Borchers:1992}. Since $\Om$ is cyclic and separating for the wedge algebra $\M$ of a Borchers triple 
$(\M,U,\Om)$, we can consider the corresponding modular data\footnote{Here and in many places in the following text, we will make use of the Tomita-Takesaki modular theory of von Neumann algebras with cyclic separating vector, see for example \cite{BratteliRobinson:1987} for an introduction.}  $J_W$, $\Delta_W$. In the present setting, Borchers' result can be formulated in the following way. (See \cite{Borchers:1992} for the original work, \cite{Florig:1998} for a simplified proof, and \cite{BuchholzLechnerSummers:2011} for a discussion in the context of Borchers triples.)
\begin{theorem}\label{Theorem:Borchers}
	Let $(\M,U,\Om)$ be a $d$-dimensional Borchers triple. Then the modular conjugation $J_W$ and modular unitaries $\Delta_W^{it}$ act on the translations $U(x):=U(x,1)$ according to, $t\in\Rl$, $x\in\Rl^d$,
	\begin{align}
		\Delta_W^{it} U(x) \Delta_W^{-it} &= U(\La_W(t)x)\,,
		\\
		J_WU(x)J_W&=U(j_Wx)\,,
	\end{align}
	where $j_W$, $\La_W(t)$ denote the Lorentz transformations associated with $W$ \eqref{eq:jWR-LambdaWR}.
\end{theorem}

According to this theorem, the modular data can not be distinguished from the represented reflections $U(j_W)$ and boosts $U(\La_W(t))$ via their action on the translations. This is in line with theorems of Bisognano and Wichmann \cite{BisognanoWichmann:1975,BisognanoWichmann:1976}, who showed that for wedge algebras generated by {\em Wightman fields} \cite{StreaterWightman:1964}, one has\footnote{This statement is stronger than the one of Thm.~\ref{Theorem:Borchers}, which does not yield equality of the modular data $J_W,\Delta_W^{it}$ with the Lorentz transformations $U(j_W)$, $U(\La_W(t))$. However, in the context of a local net satisfying further assumptions, including asymptotic completeness, Mund proved that the Bisognano-Wichmann property {\em does} follow from Borchers' theorem \cite{Mund:2001}.}
\begin{align}\label{eq:BisognanoWichmann}
	\Delta_W^{it}
	&=
	U(\La_W(t))
	\,,\qquad
	J_W=U(j_W)\qquad
	\text{(Bisognano-Wichmann property)}.
\end{align}
In comparison to these strong results on the inner structure of wedge algebras, and on their modular data, very little is known about the double cone algebras $\A(\OO)$, which are given rather indirectly as intersections of wedge algebras \eqref{eq:GeneralWedgeIntersection}. In the general situation described here, it is in particular not known whether the vacuum vector $\Om$ is cyclic for the double cone algebras, and not even if these algebras are non-trivial in the sense that $\A(\OO)\neq\Cl\cdot1$. 
\\\\
\indent Physically interesting models complying with the principle of causality have many local observables. One therefore has to add extra conditions on a Borchers triple, implying a sufficiently rich local structure, and in particular non-triviality of the intersections \eqref{eq:GeneralWedgeIntersection}. This question will be discussed in Section~\ref{Section:RelativeCommutants}.
\\\\\\
\noindent{\bf Constructing local nets from Borchers triples in $\boldsymbol{d=2}$.}\label{page:2dConstruction}
Borchers' theorem also applies in situations where there is no Lorentz symmetry present a priori. Namely, if $(\M,T,\Om)$ is a Borchers triple with only translational symmetry --- that is, $T$ is a positive energy representation of the translation group $\Rl^d$ instead of the Poincar\'e group, eqn.~\eqref{eq:WedgeIsotony} in Def.~\ref{Definition:BorchersTriple}~c) is required to hold only for translations $x$ with $W+x\subset x$, and condition \eqref{eq:WedgeLocality} drops out because there is no translation mapping a wedge $W$ into its causal complement $W'$ --- the conclusion of Thm.~\ref{Theorem:Borchers} still holds. In $d=2$ dimensions, this circumstance can then be used to extend $T$ to a representation $U$ of the proper Poincar\'e group by taking \eqref{eq:BisognanoWichmann} as the {\em definition} of $U(\La_W(
t))$, $U(j_W)$. Since $\Delta^{it}\M\Delta^{-it}=\M$, $t\in\Rl$, and $J\M J=\M'$ by Tomita's Theorem, it then follows that $(\M,U,\Om)$ is a Borchers triple in the usual sense. 

This observation also brings us to the previously mentioned variation of constructing a local net from a Borchers triple in $d=2$. Here it can be advantageous to start from a Borchers triple $(\M,T,\Om)$ with only translational symmetry, as just described, and define the net $\OO\mapsto\A(\OO)$ via $\M$ and the Poincar\'e representation $U$ generated by $T$ and the modular data $\Delta^{it}$, $J$ of $(\M,\Om)$. In that case, one observes that the definition of the wedge algebras \eqref{eq:WedgeAlgebrasByCovariance} is Haag-dual, i.e. satisfies $\A(W')=\A(W)'$ for all $W\in\W$, and the definition of the double cone algebras \eqref{eq:GeneralWedgeIntersection} amounts to
\begin{align}\label{eq:DefinitionDoubleConeAlgebrasInD=2}
	\A(\OO_{x,y})
	=
	\M(x)\cap\M'(y)\,,
\end{align}
where $\M(x)=U(x)\M U(x)^{-1}$, $\M'(y)=U(y)\M'U(y)^{-1}$, $(y-x)\in W_R$. This intersection is the relative commutant of the inclusion $\M(y)\subset\M(x)$, and closely resembles the geometric situation, where the double cone $\OO_{x,y}$ coincides with the relative causal complement of the inclusion $W_R+y\subset W_R+x$ \eqref{eq:DoubleConesAsIntersectionsOfWedges}. In our subsequent analysis, this will be of advantage in comparison to the general construction, where $\A(\OO_{x,y})$ is only a subalgebra of the relative causal complement.

\subsection{Standard pairs and free field theories}
\label{Section:Modular-Free}

The simplest quantum field theories are those describing particles without any interaction (``free'' theories). Such models are very thoroughly studied, and can be presented in many different ways. Whereas the usual approach is to present them as quantized versions of free field theories \cite{Weinberg:1995}, we would like to stress here that free field theories can be constructed without any reference to their classical counterparts, and perfectly fit into the setting of Borchers triples. The construction of such ``free'' Borchers triples will not only give us first examples of Borchers triples and the construction procedure based on them, but it will also serve as the starting point for the construction of interacting theories, considered later.

As one might expect in a free theory, the only required input is a description of the single particle spectrum. Such single particle data can, together with a suitable notion of localization, be described conveniently in terms of a so-called {\em standard pair}. Recall for the following definition that a real-linear subspace $\K_1\subset\Hil_1$ of a complex Hilbert space $\Hil_1$ is called {\em standard} if it is cyclic in the sense that $\K_1+i\,\K_1\subset\Hil_1$ is dense, and separating in the sense that $\K_1\cap i\,\K_1=\{0\}$ \cite{Longo:2008}.

\begin{definition}\label{Definition:StandardPair}{\bf (Standard pairs)}
	A $d$-dimensional standard pair $(\K_1,U_1)$ (with Poin\-car\'e symmetry) relative to a wedge $W\in\W$ consists of a closed real standard subspace $\K_1\subset\Hil_1$ of some complex Hilbert space $\Hil_1$, which carries a unitary strongly continuous positive energy representation $U_1$ of $\PG_+$ such that 
	\begin{align}
		U_1(g)\K_1\subset\K_1\quad&\text{for all}\;g\in\PG_+ \text{ with }\; gW\subset W\,,
		\label{eq:WedgeIsotonyStd}
		\\
		U_1(g)\K_1\subset\K_1'\quad&\text{for all}\;g\in\PG_+ \text{ with }\; gW\subset W'\,,
		\label{eq:WedgeLocalityStd}
	\end{align}
	where $\K_1'=\{\psi\in\Hil_1\,:\,{\rm Im}\langle\psi,\xi\rangle=0\;\text{for all } \xi\in\K_1\}$ is the symplectic complement of $\K_1$ in $\Hil_1$ w.r.t. the symplectic form ${\rm Im}\langle\,\cdot\,,\,\cdot\,\rangle$.
\end{definition}
We added the term ``with Poincar\'e symmetry'' here because standard pairs are often considered with translational symmetry only  \cite{LongoWitten:2010,BischoffTanimoto:2011,LechnerLongo:2014}. In this text, we will however always consider standard pairs with Poincar\'e symmetry, and therefore suppress this term from now on. Just as for Borchers triples, we will also drop the specification ``relative to $W$'' in case the reference region is the right wedge $W_R$, or clear from the context.

The relation between standard pairs and Borchers triples is two-fold. We first consider the step from a Borchers triple to a standard pair.
\begin{lemma}\label{Lemma:ProjectBT}
	Let $(\M,U,\Om)$ be a Borchers triple on a Hilbert space $\Hil$, and
	\begin{align}
		\K:=\{A\Om\,:\,A=A^*\in\M\}^{\|\cdot\|}\,.
	\end{align}
	Let furthermore $Q\in\B(\Hil)$ be an orthogonal projection which commutes with the representation $U$. Then $(Q\K,U|_{Q\Hil})$ is a standard pair on $Q\Hil$.	
\end{lemma}

The proof of this lemma uses the modular theory of standard subspaces, where to any real standard subspace $\K_1$ one associates a Tomita operator $S_{\K_1}:\K_1+i\,\K_1\mapsto \K_1+i\,\K_1$, $S_{\K_1}(k+i\ell):=k-i\ell$, which in turn defines $\K_1$ by $\K_1=\ker(1-S_{\K_1})$. These objects satisfy properties closely analogous to the von Neumann algebra case, see \cite{Longo:2008} for more details and an account of the literature. In the situation of the above lemma, with $Q=1$, the modular data $J,\Delta$ of $(\M,\Om)$ coincide with those of the standard subspace, $J=J_{\K_1}$, $\Delta=\Delta_{\K_1}$. Since $J_{\K_1}$ is known to map $\K_1$ onto its symplectic complement $\K_1'$ by the subspace version of Tomita's Theorem, the conclusion follows in this case. The generalization to $Q\neq1$ is straightforward; one uses that $Q$ commutes with the modular data by virtue of Theorem~\ref{Theorem:Borchers}.

Lemma~\ref{Lemma:ProjectBT} can be applied to extract single particle information from a Borchers triple. To illustrate this, consider the case that in the representation $U$ of the Borchers triple, there exists an isolated eigenvalue $m>0$ of the mass operator, and take the projection $Q:=E_{\{m\}}$ as the corresponding spectral projection. Then $\Hil_1:=Q\Hil$ describes single particle vectors of mass $m$, and $\K_1$ the ``single particle vectors localized in $W$''. In this projection process, a lot of information is lost\footnote{This is even the case for the projection $Q=1$.}, and only single particle data remain. Hence many QFTs give rise to the same standard pairs by projecting their Borchers triples to the single particle level. 

However, for each standard pair one can without further input construct a {\em specific} Borchers triple, representing an interaction-free theory. This brings us to the link in the other direction, namely from a standard pair to a Borchers triple. This step can be carried out by second quantization. In this context, we denote by $\Gamma(\Hil_1)$ the Bose Fock space over a Hilbert space $\Hil_1$, and by $V(\xi)$, $\xi\in\Hil_1$, the Weyl operators on $\Gamma(\Hil_1)$ \cite{BratteliRobinson:1997}, characterized by the Weyl relation $V(\xi)V(\psi)=e^{-\frac{i}{2}{\rm Im}\langle\xi,\psi\rangle}\,V(\xi+\psi)$ and $V(\xi)\Om=e^{-\frac{1}{4}\|\xi\|^2}\,e_\otimes^\xi$, with $e_\otimes^\xi=\bigoplus_{n=0}^\infty\xi\tp{n}/\sqrt{n!}$, and $\Om$ the Fock vacuum.

\begin{proposition}\label{Proposition:BT-SecondQuantization}
	Let $(\K_1,U_1)$ be a standard pair on a Hilbert space $\Hil_1$. On the Fock space $\Gamma(\Hil_1)$, consider the second quantization $\Gamma(U_1)$ of $U_1$, the Fock vacuum $\Om$, and the von Neumann algebra
	\begin{align}\label{eq:WeylAlgebra}
		\M:=\{V(\xi)\,:\,\xi\in\K_1\}''\,.
	\end{align}
	Then $(\M,U,\Om)$ is a Borchers triple, and projecting it with $Q=P_1$ (the projection onto $\Hil_1$) returns the standard pair $(\K_1,U_1)$.
\end{proposition}

This relation between real standard spaces and the algebras of a free field are known from the work of Araki \cite{Araki:1963,Araki:1964}. Their modular data were shown to be of second quantized form by Eckmann and Osterwalder \cite{EckmannOsterwalder:1973}, see also \cite{LeylandsRobertsTestard:1978}. 

Whereas the version presented here is suitable for Bosonic systems with commuting fields at spacelike separation, there is also a version adapted to the Fermionic case, where fields anticommute \cite{Foit:1983,BaumgartelJurkeLledo:2002}. This formulation makes use of so-called ``twisted duality'' \cite{DoplicherHaagRoberts:1969}, and requires only minor modifications. We will not discuss it any further here.
\\
\\
In view of these relations between standard pairs and Borchers triples, all that is required for the construction of a free (second quantization) Borchers triple is a corresponding (single particle) standard pair. This requires in particular a (single particle) representation $U_1$ of the proper Poincar\'e group and a standard subspace $\K_1$, which, as mentioned above, is determined by its modular data according to $\K_1=\ker(1-J_{\K_1}\Delta_{\K_1}^{1/2})$.

For a concrete construction of this space, one can therefore anticipate the Biso\-gna\-no-Wichmann relation \eqref{eq:BisognanoWichmann} between geometric data and modular data, and use it as a {\em definition} for the modular data, and hence the standard subspace. This idea is known as {\em modular localization} \cite{FassarellaSchroer:2002,BrunettiGuidoLongo:2002,MundSchroerYngvason:2006}, see also \cite{BaumgrtelJurkeLled:1995}.
\\
\\
In more detail, Brunetti, Guido and Longo consider a (anti-)unitary strongly continuous positive energy representation $U_1$ of the proper Poincar\'e group on a Hilbert space $\Hil_1$, and the one parameter groups $\La_{W}(t)$ and reflections $j_{W}$, $W\in\W$, in this representation \cite{BrunettiGuidoLongo:2002}. By Stone's theorem, there exists a selfadjoint generator $R_W$ such that $U_1(\La_W(t))=e^{itR_W}$, and one defines for the right wedge $W=W_R$
\begin{align}\label{eq:SingleParticleGeoemtricModularData}
	\Delta_1:=e^{-\pi R_{W_R}}\,,\qquad J_1:=U(j_{W_R})\,,\qquad S_1:=J_1\Delta_1^{1/2}\,.
\end{align}
By comparison with \eqref{eq:BisognanoWichmann}, we see that this assignment mimics the Bisognano-Wichmann relation, and one defines further
\begin{align}\label{eq:SingleParticleStandardSubspace}
	\K_1:=\{\psi\in\dom\Delta_1^{1/2}\,:\,S_1\psi=\psi\}	\,.
\end{align}

\begin{theorem}{\bf \cite{BrunettiGuidoLongo:2002}}
	$(\K_1,U_1)$ is a standard pair, with Tomita operator $S_{\K_1}=S_1$.
\end{theorem}

The main point of this theorem is to demonstrate the inclusion property \eqref{eq:WedgeIsotonyStd}, which is linked to the positive energy condition of $U_1$. For the proof of this, and further results, see \cite{BrunettiGuidoLongo:2002}.

In view of this theorem, we have, for any considered representation $U_1$ of $\PG_+$, an associated standard pair and thus also an associated second quantization Borchers triple. These triples can now be used in the general construction outlined in Section~\ref{Section:WedgeAlgebras} to generate local nets of von Neumann algebras, corresponding to free QFT models.

At the end of Section~\ref{Section:WedgeAlgebras}, we mentioned the problem that the algebras corresponding to smaller regions, defined as intersections of wedge algebras \eqref{eq:DoubleConesAsIntersectionsOfWedges}, are not guaranteed to be non-trivial in the setting of a general Borchers triple. In the present context of free field constructions, this problem can however be resolved. To begin with, Brunetti, Guido and Longo have shown that for any of the representations $U_1$ considered here, the von Neumann algebras corresponding to spacelike cones have the Fock vacuum as a cyclic vector\footnote{In case $U_1$ does not contain the trivial representation, as is adequate for a single particle representation, the algebras corresponding to spacelike cones are also known to be factors of type III$_1$ \cite{BrunettiGuidoLongo:2002,FiglioliniGuido:1994}.} \cite{BrunettiGuidoLongo:2002}. 

Moreover, in the case of ``usual'' representations $U_1$, i.e. direct sums of mass $m\geq0$ finite spin $s$ representations, also cyclicity of the Fock vacuum for algebras associated with double cones is known. In fact, the net resulting from the Borchers triple by application of the procedure in Section~\ref{Section:WedgeAlgebras} (or its twisted version in the Fermionic case) is then a known free field net, which in particular has $\Om$ as a cyclic vector for each double cone algebra. 

As an aside, we mention that there also exist ``continuous spin'' representations of the Poincar\'e group, for which Wightman fields do not exist \cite{Yngvason:1970}. The algebraic construction presented here applies also to such representations, and in fact there exist models of free quantum fields which are localizable only in spacelike cones \cite{MundSchroerYngvason:2006}. The double cone algebras in such models are currently under investigation, and it seems that they might be trivial in such models \cite{Koehler:2015}.

\subsection{Relative commutants of wedge algebras}
\label{Section:RelativeCommutants}

After this excursion to free field models and modular localization, we return to the setting of a general Borchers triple $(\M,U,\Om)$, and the question how to ensure large double cone algebras in the construction in Section~\ref{Section:WedgeAlgebras}. As mentioned earlier, there are no efficient tools available for analyzing intersections of general families of von Neumann algebras, and therefore we focus on the more particular situation of a relative commutant $\M_1'\cap\M_2$ of an inclusion $\M_1\subset\M_2$. Since double cones are relative causal complements of wedges in two dimensions~\eqref{eq:DoubleConesAsIntersectionsOfWedges}, such an analysis directly applies to double cone algebras in $d=2$. In higher dimensions $d>2$, relative commutants of wedge algebras correspond to cylinder like regions which are unbounded in $(d-2)$ perpendicular directions.
\\\\
The best studied type of inclusions of von Neumann algebras are so-called split inclusions \cite{DoplicherLongo:1984}, and, as we shall see, they will also play a prominent role in our present context. We first recall the definition of split inclusions and some of their most important properties before we discuss applications of these concepts to relative commutants of wedge algebras.

\begin{definition}
	Let $\M_1\subset\M_2\subset\B(\Hil)$ be an inclusion of von Neumann algebras on some Hilbert space $\Hil$. 
	\begin{enumerate}
		\item[a)] $\M_1\subset\M_2$ is called split if there exists a type I factor\footnote{That is, a von Neumann algebra isomorphic to $\B(\tilde{\Hil})$ for some Hilbert space $\tilde{\Hil}$.} $\NN$ such that
		\begin{align}
			\M_1\subset\NN\subset\M_2\,.
		\end{align}
		\item[b)] $\M_1\subset\M_2$ is called standard if there exists a vector which is cyclic and separating for $\M_1$, $\M_2$, and the relative commutant $\M_1'\cap\M_2$.
	\end{enumerate}	
\end{definition}

In the standard case, split inclusions can be characterized as follows \cite{DAntoniLongo:1983,DoplicherLongo:1984}.

\begin{lemma}\label{Lemma:Split}
  Let $\M_1\subset\M_2$ be a standard inclusion of von Neumann algebras on the Hilbert space $\Hil$. Then $\M_1\subset\M_2$ is split if and only if there exists a unitary $V:\Hil\to\Hil\otimes\Hil$ such that
  \begin{align}\label{split-tensor}
	VA_1A_2'V^* &= A_1\otimes A_2',\qquad\qquad A_1\in\M_1,\;\; A_2'\in\M_2'\;.
   \end{align}
\end{lemma}

\noindent{\em Remark:} Note that if the assumptions of this lemma are satisfied, the inclusion under consideration has a large relative commutant, namely $\M_1\cap\M_2'\cong\M_1\ot\M_2'$.
\\\\
In view of Lemma~\ref{Lemma:Split}, the split property of an inclusion $\M_1\subset\M_2$ can be understood as a form of statistical independence between the subsystems described by the commuting algebras $\M_1$ and $\M_2'$ of the larger system identified with $\M_1\vee\M_2'$ (see the review \cite{Summers:1990} for a detailed discussion of these matters, and references to the original literature). Namely, it implies that for any pair of normal states $\varphi_1$ on $\M_1$ and $\varphi_2$ on $\M_2'$, there exists a normal state $\varphi$ on $\M_1\vee\M_2'$ such that $\varphi|_{\M_1}=\varphi_1$, $\varphi|_{\M_2'}=\varphi_2$, expressing the fact that states in the subsystems $\M_1$ and $\M_2'$ can be prepared independently of each other. Moreover, $\varphi$ can be chosen in such a way that there are no correlations between ``measurements'' in $\M_1$ and $\M_2'$, i.e. as a product state
\begin{align*}
	\varphi(A_1 A_2')	&=	\varphi_1(A_1)\cdot \varphi_2(A_2')\,,\qquad A_1\in\M_1\,,\;A_2'\in\M_2'\,.
\end{align*}
Taking $\M_1=\A(\OO_1)$ and $\M_2'=\A(\OO_2)$ as the observable algebras of two spacelike separated regions $\OO_1\subset\OO_2'$ in a quantum field theory given by a net $\A$, some form of statistical independence between $\M_1$ and $\M_2'$ can be expected on physical grounds. For the massive free field, the existence of normal product states for such pairs of local algebras was shown by Buchholz \cite{Buchholz:1974-2}. A corresponding analysis for algebras of free Fermi fields, and for the Yukawa$_2+P(\varphi)_2$ model has been carried by Summers \cite{Summers:1982}.

Examples of theories violating the split property can be obtained by considering models with a non-compact global symmetry group, or certain models with infinitely many different species of particles \cite{DoplicherLongo:1984}. Such theories have an immense number of local degrees of freedom, and according to the analysis of Buchholz and Wichmann \cite{BuchholzWichmann:1986}, it is precisely this feature which is responsible for the breakdown of the split property.

One can therefore expect that the split property (for proper inclusions of double cones) holds in theories which do not exhibit pathologically large numbers of local degrees of freedom. Such theories, in turn, can be expected to have a reasonable thermodynamical behavior. In the literature, there exist several ``nuclearity'' conditions \cite{BuchholzWichmann:1986,BuchholzPorrmann:1990,BuchholzDAntoniLongo:1990-1,BuchholzDAntoniLongo:2007}, reminiscent of the trace class condition Tr$(e^{-\beta H})<\infty$ for Gibbs states in quantum mechanics, which are related to the split property and thermodynamical properties.

For applications to relative commutants of wedge algebras, the relevant condition is the so-called ``modular nuclearity condition'' \cite{BuchholzDAntoniLongo:1990-1,BuchholzDAntoniLongo:1990}. Given a Borchers triple $(\M,U,\Om)$, one considers the inclusions
\begin{align}
	\M(x):=U(x,1)\M U(x,1)^{-1}\subset\M\,,\qquad x\in W_R,
\end{align}
and defines the maps
\begin{align}\label{eq:DefinitionXi(x)}
	\Xi(x):\M\to\Hil\,,\qquad \Xi(x)A:=\Delta^{1/4}U(x)A\Om\,.
\end{align}
Here $\Delta$ is the modular operator of $(\M,\Om)$. Using elementary properties of modular theory, it is easy to see that $\Xi(x)$ is a bounded operator between the Banach spaces $\M$ (equipped with the operator norm of $\B(\Hil)$) and $\Hil$. 

If $\Xi(x)$ is even compact, and more particularly, {\em nuclear} (i.e. $\Xi(x)$ can be written as a norm convergent sum of rank one operators), then one has the following result.

\begin{theorem}{\bf \cite{BuchholzDAntoniLongo:1990-1}}
	Let $(\M,U,\Om)$ be a Borchers triple and assume that for some $x\in W_R$, the map $\Xi(x)$ \eqref{eq:DefinitionXi(x)} is nuclear. Then the inclusion $\M(x)\subset\M$ is split. Conversely, if $\M(x)\subset\M$ is split, then $\Xi(x)$ is compact.
\end{theorem}

This theorem provides a sufficient condition for an inclusion to be split. However, it must be noticed that the split property is a very strong condition. It is a reasonable assumption for inclusions of {\em local} algebras in theories which satisfy some rough bound on the number of their local degrees of freedom, but some care is needed when dealing with unbounded regions like wedges, even in such theories. In fact, there is an argument by Araki \cite[p. 292]{Buchholz:1974-2} to the effect that inclusions of wedge algebras cannot be split if the spacetime dimension is larger than two. Araki's argument exploits the translation invariance of wedges along their edges and does not apply in two dimensions, where these edges are zero-dimensional points.

In two dimensions, the split property for wedges is known to hold in the theory of a free, scalar, massive field \cite{Muger:1998, BuchholzLechner:2004}. It is, however, not fulfilled for arbitrary mass spectra. For example, the split property for wedges does not hold in massless theories, and is also violated in 
the model of a generalized free field with continuous mass spectrum \cite{DoplicherLongo:1984}. But for models describing finitely many species of massive particles, there is no a priori reason for the split property for wedges not to hold. We can therefore take it as a tentative assumption (to be verified in concrete models), and now discuss its consequences.

\begin{proposition}\label{Proposition:SplitBorchersTriplesGivesTypeIII1}{\bf \cite{BuchholzLechner:2004}}
	Let $(\M,U,\Om)$ be a two-dimensional Borchers triple, and $x\in W_R$. If the inclusion $\M(x)\subset\M$ is split, then $\M$, $\M(x)$ and the relative commutant $\M(x)'\cap\M$ are all isomorphic to the unique hyperfinite type III$_1$ factor. In particular, the relative commutant has cyclic vectors, and $\M(x)\subset\M$ is standard.
\end{proposition}

In the light of this result, we can view the split property as a sufficient condition for non-trivial relative commutants of inclusions of wedge algebras. Whereas non-triviality of local algebras is a minimal requirement in a local theory, also the Reeh-Schlieder property, i.e. cyclicity of the vacuum vector for algebras of observables localized in arbitrarily small regions, is of importance in quantum field theory.

To arrive at such a statement from the split property, we first recall that on the basis of Lemma~\ref{Lemma:Split} one can easily show that a standard split inclusion is {\em normal}, i.e.  $(\M_1'\cap\M_2)'\cap\M_2=\M_1$ \cite{DoplicherLongo:1984}. By similar arguments in the setting of a Borchers triple $(\M,U,\Om)$ for which $\M(x)\subset\M$ is split for some $x\in W$, it follows that $\M$ is {\em locally generated}, i.e. it coincides with the smallest von Neumann algebra containing all relative commutants $\M\cap\M'(x)$, $x\in W$ \cite{Lechner:2008}. In combination with a result of M\"uger \cite{Muger:1998}, stating that algebras corresponding to double cones of different sizes are closely related, this provides sufficient information for application of the usual Reeh-Schlieder arguments \cite{ReehSchlieder:1961}, making use of positivity of the energy. One arrives at the following statement \cite{Lechner:2008}.

\begin{proposition}\label{Proposition:SplitGivesCyclicity}
	Let $(\M,U,\Om)$ be a Borchers triple, and $x\in W_R$. If the inclusion $\M(x)\subset\M$ is split, then $\Om$ is cyclic for the relative commutant $\M\cap\M'(x)$.
\end{proposition}

Thus the modular nuclearity and split conditions yield nets satisfying all the basic assumptions of algebraic quantum field theory \cite{Haag:1996}. Below we summarize these and additional results in a theorem, which strengthens Proposition~\ref{Proposition:BasicNetFromBorchersTriple} under the assumption of the split property for wedges. In its formulation, we make use of the {\em diameter} of a two-dimensional double cone $\OO_{x,y}=(W_R+x)\cap(W_L+y)$, defined as $d(\OO_{x,y}):=\sqrt{-(x-y)^2}\geq0$.

\begin{theorem}\label{Theorem:SPWNets}
	Let $(\M,U,\Om)$ be a two-dimensional Borchers triple, such that the inclusion $\M(x)\subset\M$ is split for some $x\in W_R$ (this is in particular the case if the map $\Xi(x)$ is nuclear). Let $s:=\sqrt{-x^2}>0$ be the ``splitting distance''. Then the net $\A$ constructed from $(\M,U,\Om)$ has (in addition to what is stated in Proposition~\ref{Proposition:BasicNetFromBorchersTriple}) the following properties: For any double cone $\OO$ with $d(\OO)>s$,
	\begin{enumerate}
		\item $\A(\OO)$ is isomorphic to the hyperfinite type III$_1$ factor.
		\item The vacuum vector $\Om$ is cyclic and separating for $\A(\OO)$. 
		\item Haag duality holds, i.e.  $\A(\OO)'=\A({\OO}')$.
		\item Weak additivity holds, i.e. 
		\begin{align}
		 \bigvee_{x\in\Rl^2}\A(\OO+x)=\B(\Hil)\,.
		\end{align}
		\item The time slice property (in its von Neumann version) holds above the splitting distance: If $t_0,t_1\in\Rl$ with $t_1-t_0>s$, then the algebra $\A(S(t_0,t_1))$ associated to the time slice $S=\{x\in\Rl^2\,:\,t_0<x_0<t_1\}$ is $\A(S(t_0,t_1))=\B(\Hil)$.
	\end{enumerate}
\end{theorem}

In this theorem, a), b) follow from Prop.~\ref{Proposition:SplitBorchersTriplesGivesTypeIII1} and Prop.~\ref{Proposition:SplitGivesCyclicity}. For c) and d), see \cite{Lechner:2008}, and for e), \cite{Muger:1998}. Note that the above theorem gives slightly generalized statements over the ones found in, say, \cite{Muger:1998,Lechner:2008}: In these works, the split property was assumed to hold for arbitrarily small splitting distances $s>0$. However, the corresponding results for finite splitting distance are straightforward to obtain by the same arguments. Also see the work of M\"uger for further implications of the split property for wedges, in particular with regard to superselection theory \cite{Muger:1998}.

Although it is possible to construct models of algebraic quantum field theory in which there exists a minimal length in the sense that $\A(\OO)=\Cl\cdot1$ for double cones below a minimal diameter \cite{LechnerLongo:2014}, this is not an expected feature in typical QFT models. We mostly stated the theorem in the above form because in certain models, discussed in Section~\ref{Section:IntegrableModels}, the modular nuclearity condition can so far only be proven for large enough splitting distance. Of course, Theorem~\ref{Theorem:SPWNets} also applies to the case where the split property holds for {\em all} the inclusions $\M(x)\subset\M$, $x\in W_R$, and then gives the usual unrestricted forms of cyclicity, additivity, duality, and the time slice property.

Theorem~\ref{Theorem:SPWNets} can be seen as the abstract form of a general construction scheme, which we can summarize as ``first construct a Borchers triple, then check modular nuclearity for its wedge inclusions''. It provides physically reasonable properties of the emerging net under assumptions which are both natural (for massive theories, in $d=2$) and manageable in concrete models. Its main drawback is that it does not apply to more than two spacetime dimensions. Finding conditions that suitably weaken the split property for wedges and apply to dimension $d>2$ is currently an open question and subject of ongoing research.

\section{Integrable models and inverse scattering theory}\label{Section:IntegrableModels}

In this section we discuss a concrete implementation of the general construction scheme of QFT models via Borchers triples, presented in Thm.~\ref{Theorem:SPWNets}. As we saw in Section~\ref{Section:Modular-Free}, free theories can be completely described in terms of their (stable) single particle content, formalized as a specific representation $U_1$ of the Poincar\'e group and its associated net $W\mapsto\K_1(W)$ of real standard subspaces of the single particle space $\Hil_1$. Here we will describe how to proceed in a similar manner for certain {\em interacting} theories in two dimensions. 

From a mathematical perspective, we will start from a specific single particle representation $U_1$ of $\PG_+$ as before, and then ``deform'' the second quantization step (cf. Prop.~\ref{Proposition:BT-SecondQuantization}) from the subspace net $W\mapsto\K_1(W)$ to the net of von Neumann algebras $W\mapsto\A(W)$. The ``deformation parameter'' will take the form of a unitary $S$ on $\Hil_1\ot\Hil_1$ with specific properties, which enters into both, the definition 
of the multi particle Hilbert space, and the definition of the wedge algebras. 

Physically speaking, $S$ describes the two-particle S-matrix, and we are considering theories in which this two-particle scattering operator completely fixes the full (multi particle) S-matrix. These are theories in which no particle production occurs in collision processes of arbitrary energy. Examples of such models are well known as {\em integrable quantum field theories}, referring to the existence of an infinite number of conversation laws which constrain the dynamics in such a way that each collision process factorizes into two-particle processes (``factorizing S-matrix'', see \cite{Iagolnitzer:1993}). Specific examples are field theories like the Sinh-Gordon model, the Ising model, the Sine-Gordon model, the Thirring model, and many more \cite{AbdallaAbdallaRothe:1991,Smirnov:1992}.
\\\\
Since we are starting our construction from the (two particle) S-matrix $S$, we consider the {\em inverse scattering problem}, in contrast to the canonical approach, where the interaction is specified in terms of a classical Lagrangian or Hamiltonian density, which is then quantized \cite{Weinberg:1995}. In fact, in the case of integrable models, the S-matrix is typically much simpler than the Lagrangian (which is not of polynomial form), and therefore $S$ suggests itself as a suitable quantity for characterizing the interaction. 

This inverse scattering point of view also lies at the heart of the form factor program, see the monograph \cite{Smirnov:1992} and review article \cite{BabujianKarowski:2004}. In that approach, one characterizes local fields/observables by their expectation values in scattering states (``form factors''), which are severely restricted by factorizability and analyticity of the S-matrix. In many cases, it is possible to obtain explicit expressions for form factors, see \cite{BabujianFoersterKarowski:2012,Smirnov:1994,FringMussardoSimonetti:1993,BalogHauer:1994} for just some sample articles. After the determination of the form factors, the next step in the form factor program is to compute Wightman $n$-point functions, which are given by series of integrals over form factors. Controlling these complicated series, as required for a complete construction of a model, has so far only been possible in very few examples \cite{BabujianFoersterKarowski:2006}.

In comparison, the algebraic approach presented here circumvents the explicit construction of local field operators (see however the end of Section~\ref{Section:Discussion} for results in that direction), and analyzes the local observable content via the modular nuclearity property from Tomita-Takesaki theory.
\\\\
In Section~\ref{Section:S}, we introduce the class of S-matrices we consider, and then use them to construct Borchers triples in Section~\ref{Section:S>BT}. The known results pertaining to the modular nuclearity condition are reviewed in Section~\ref{Section:Nuc}. In Section~\ref{Section:Discussion}, it is then shown that this construction solves the inverse scattering problem and yields asymptotically complete theories. Finally, in Section~\ref{Section:Mass0} we discuss certain massless versions of these models, and compare to related constructions in the literature.

\subsection{Factorizing S-Matrices}
\label{Section:S}

Just as in Section~\ref{Section:Modular-Free}, the construction of the models we are interested in begins with the specification a single particle representation of the proper Poincar\'e group, fixing the single particle spectrum. Recall that in two dimensions, to any mass $m>0$, there exists a unique (up to unitary equivalence) irreducible, unitary, strongly continuous, positive energy representation $U_{1,m}$ of the proper orthochronous Poincar\'e group $\PGpo$. It can be realized on the representation space $\Hil_{1,m}:=L^2(\Rl,d\te)$ as
\begin{align}\label{eq:U1m}
	(U_{1,m}(x,\la)\psi)(\te)
	:=
	e^{ip_m(\te)\cdot x}\cdot \psi(\te-\la)
	\,,
\end{align}
where $(x,\la)\in\PGpo$ denotes the Poincar\'e transformation consisting of a boost with rapidity $\la\in\Rl$ and a subsequent space-time translation by $x\in\Rl^2$. The variable $\te$ is the {\em rapidity}, which is connected to the on-shell momentum and mass via
\begin{align}\label{eq:palpha}
	p_m(\te)
	:=
	m\left(
	\begin{array}{c}
	\cosh\te\\
	\sinh\te
	\end{array}
	\right)
	\,.
\end{align}
We will allow for several species of particles, and therefore consider a direct sum of several representations $U_{1,m}$ with different masses. 

We can also include the single particle charges in our description, identified with equivalence classes $q$ of unitary irreducible representations of a compact Lie group $G$ (the global gauge group) as usual. As charges carried by single particles, we consider a set $\Q$ of finitely many charges, and to account for antiparticles, we assume that with each class $q\in\Q$, also the conjugate class $\overline{q}$ is contained in $\Q$. We are interested in constructing massive stable quantum field theories and must therefore guarantee that in each sector, the masses are positive isolated eigenvalues of the mass operator. This will in particular be the case when to each charge $q$, there corresponds a single mass $m(q)>0$ (with $m(\overline{q})=m(q)$), and for simplicity, we restrict ourselves to this setting.

The single particle Hilbert space then has the form
\begin{align}\label{eq:OneParticleSpace}
	\Hil_1
	=
	\bigoplus_{q\in\Q}\Hil_{1,q}
	=
	\bigoplus_{q\in\Q} L^2(\Rl,d\te)\ot\VV_q
	=
	L^2(\Rl,d\te)\ot\VV
	\,,
\end{align}
where $\VV=\bigoplus_{q\in\Q}\VV_q$ is a finite-dimensional Hilbert space, $D:=\dim\VV$. Picking unitary irreducible representations $V_{1,q}$ of $G$ in the class $q\in\Q$, the representations of the Poincar\'e group $\PGpo$ and the gauge group $G$ on $\Hil_1$ are
\begin{align}\label{eq:U1}
	U_1
	&:=
	\bigoplus_{q\in\Q} \left(U_{1,m(q)}\ot{\rm id}_{\VV_q}\right)
	\,,\qquad
	V_1
	:=
	\bigoplus_{q\in\Q} \left({\rm id}_{L^2(\Rl,d\te)}\ot V_{1,q}\right)
	\,.
\end{align}
In the following, we will always tacitly refer to a fixed particle spectrum given by the data $G,\Q,\{V_{1,q}\}_{q\in\Q},\{m_q\}_{q\in\Q}$ and complying with the above assumptions. It will be convenient to use a particular orthonormal basis for $\VV$ \eqref{eq:OneParticleSpace}: For each subspace $\VV_q$ of fixed charge, we choose an orthonormal basis, and denote their direct sum by $\{e^\alpha\,:\,\alpha=1,...,D\}$. We can thus associate with each index $\alpha$ a definite charge $q_{[\alpha]}$ and mass $m_{[\alpha]}:=m(q_{[\alpha]})$. The corresponding components of vectors $\Psi_1\in\Hil_1$ will be denoted by $\te\mapsto\Psi_1^\alpha(\te)$. Using standard multi index notation, this notation is extended to tensor products: We write $\xi^\balpha$, $\balpha=(\alpha_1,...,\alpha_n)$, for the components of vectors $\xi\in\VV^{\ot n}$, $T^\balpha_\bbeta$, $\balpha=(\alpha_1,...,\alpha_n)$, $\bbeta=(\beta_1,...,\beta_n)$, for the components of tensors $T\in\B(\VV^{\ot n})$, and $\Rl^n\ni\bte\mapsto\Psi_n^\balpha(
\bte)$ for the component functions of $\Psi_n\in\Hil_1\tp{n}$, $n\in\Nl$.

To proceed as in Section~\ref{Section:Modular-Free}, we need a representation of the {\em proper} Poincar\'e group, i.e. we still need to define a single particle TCP operator $J_1$, implementing the spacetime reflection $j(x):=-x$. In view of our above assumption regarding conjugate charges $q,\overline{q}\in\Q$, there exists such an operator on $\Hil_1$ (see for example \cite{DoplicherHaagRoberts:1974, BuchholzFredenhagen:1982, GuidoLongo:1995, Mund:2001}). It is the product of a charge conjugation operator exchanging the representation spaces $\VV_q$ and $\VV_{\overline{q}}$, and a space-time reflection, acting by complex conjugation on $L^2(\Rl,d\te)$. When working in the basis chosen above, this simply means that we have an
involution $\alpha\mapsto\overline{\alpha}$ of $\{1,...,D\}$ (that is, a permutation of $D$ elements with $\overline{\overline{\alpha}}=\alpha$) such that $m_{[\overline{\alpha}]}=m_{[\alpha]}$ and
$q_{[\overline{\alpha}]}=\overline{q_{[\alpha]}}$, and the TCP operator reads
\begin{align}\label{eq:J1}
	(J_1\Psi_1)^\alpha(\te)
	:=
	\overline{\Psi_1^{\overline{\alpha}}(\te)}
	\,.
\end{align}
By straightforward calculation, one checks that $J_1$ is an antiunitary involution which commutes with $V_1$ and extends the representation $U_1$ to the proper Poincar\'e group $\PGp$ via $U_1(j):=J_1$.

As in \eqref{eq:SingleParticleGeoemtricModularData}, we introduce the ``geometric modular group'' $\Delta_1^{it}:=U_1(0,-2\pi t)$ and the real standard subspace \eqref{eq:SingleParticleStandardSubspace}
\begin{align}
  \K_1:=\ker(1-J_1\Delta_1^{1/2})\subset\Hil_1\,.
\end{align}
For later reference, we recall that $\K_1+i\,\K_1$ coincides with the (vector-valued) Hardy space ${\mathbb H}^2(-\pi,0)\ot\VV\subset\Hil_1$, consisting of all $\Psi_1\in\Hil_1$ such that $\te\mapsto\Psi_1^\alpha(\te)$ is the boundary value of a function analytic in $\Strip(-\pi,0):=\{\zeta\in\Cl\,:\,-\pi<{\rm Im}\zeta<0\}$ with  $$\sup_{-\pi<\la<0}\sum_\alpha\int_\Rl d\te\,|\Psi_1^\alpha(\te+i\la)|^2<\infty,$$
and the real standard subspace is
\begin{align}\label{eq:K1Explicit}
	\K_1=\{\Psi_1\in{\mathbb H}^2(-\pi,0)\ot\VV\,:\,\Psi_1^\alpha(\te+i\pi)=\overline{\Psi_1^{\overline{\alpha}}(\te)}\,\text{ for all } \te,\alpha\},
\end{align}
see, for example, \cite{LechnerLongo:2014}.
\\
\\
We now come to specifying the two-particle S-matrix of the models to be constructed. This S-matrix can be formulated as a unitary $\bS$ on  $\Hil_1\ot\Hil_1$ which has to satisfy a number of compatibility conditions with the single particle data $U_1$, $V_1$, and additional properties. This unitary will have the form $$(\bS\Psi_2)^{\alpha_1\alpha_2}(\te_1,\te_2)=S^{\alpha_1\alpha_2}_{\beta_1\beta_2}(\te_1-\te_2)\Psi_2^{\beta_1\beta_2}(\te_1,\te_2)$$ (here and in the following, we use Einstein's summation convention) with some function $S:\Rl\to\U(\VV\ot\VV)$, expressing the Poincar\'e invariance of $\bS$. The properties of $\bS$ can therefore be most explicitly formulated in terms of the function $S$, as we shall do below. For a manifestly basis-independent formulation in terms of $\bS$, see \cite{Bischoff:2012,BischoffTanimoto:2013}. All the listed properties are standard in the context of integrable models, and for example discussed in \cite{AbdallaAbdallaRothe:1991, Mussardo:1992, Dorey:1998}.

\begin{definition}\label{definition:SMatrix}
     A (two-particle) $S$-matrix is a continuous bounded function\footnote{The continuity assumption can be relaxed, cf. \cite{LechnerSchlemmerTanimoto:2013}.} $S:\overline{\Strip(0,\pi)}\to\B(\VV\ot\VV)$ which is analytic in the interior of this strip and satisfies for arbitrary $\te,\te'\in\Rl$,
     \begin{enumerate}
	  \item\label{item:Unitarity} Unitarity:
	  \begin{align}
	       S(\te)^*=S(\te)^{-1}
	  \end{align}
	  \item\label{item:HermitianAnalyticity} Hermitian analyticity:
	  \begin{align}
	       S(\te)^{-1}=S(-\te)
	  \end{align}
	  \item\label{item:YangBaxterEqn} Yang-Baxter equation:
	  \begin{align}
	     (S(\te)\ot1_1)(1_1\ot S(\te+\te'))&(S(\te')\ot 1_1)
		\\
		&=
		(1_1\ot S(\te'))(S(\te+\te')\ot 1_1)(1_1\ot S(\te))
		\nonumber
	  \end{align}
	  \item\label{item:TranslationInvariance} Poincar\'e symmetry: 
	  \begin{align}
	       S^{\alpha\beta}_{\gamma\delta}(\te)&=0\;\text{ if }\; m_{[\alpha]} \neq m_{[\delta]}\;\text{ or }\; m_{[\beta]}\neq m_{[\gamma]}\,,\\
	       S^{\alpha\beta}_{\gamma\delta}(\te)
	       &=	       S^{\overline{\delta}\overline{\gamma}}_{\overline{\beta}\overline{\alpha}}(\te)
	       \,.
	  \end{align}
	\item\label{item:GaugeInvariance} Gauge invariance:
	\begin{align}
		[S(\te),\,V_1(g)\ot V_1(g)]=0
		\,,\qquad
		g\in G\,.
	\end{align}
	\item\label{item:Crossing} Crossing symmetry: For all $\alpha,\beta,\gamma,\delta\in\{1,...,D\}$,
	\begin{align}
		S^{\alpha\beta}_{\gamma\delta}(i\pi-\te)
		=
		S^{\overline{\gamma}\alpha}_{\delta\overline\beta}(\te)
	\end{align}
     \end{enumerate}
     The family of all S-matrices will be denoted $\SF$.
\end{definition}

Many examples of such S-matrices are known. In the {\em scalar case}, pertaining to the irreducible representation $U_1=U_{1,m}$ with $\VV=\Cl$, a complete characterization of $S$ can be given \cite{Lechner:2006}. Some typical examples are listed in the following table.

\begin{table}[H]\label{table:ScalarS}
    \begin{tabular}{p{2cm}p{7.4cm}p{2cm}}
	  \hline\noalign{\smallskip}
	  $S(\te)$ & Name of associated QFT model & References  \\
	  \noalign{\smallskip}\hline\noalign{\smallskip}
	  $+1$ & free field theory  & \\
	  $-1$ & Ising model & \cite{BergKarowskiWeisz:1979,Lechner:2005} \\
	  $\frac{\sinh\te-i\sin b(g)}{\sinh\te+i\sin b(g)}$ & Sinh-Gordon model with coupling constant $g$, and $b(g)=\frac{\pi g^2}{4\pi+g^2}$  & \cite{ArinshteinFateevZamolodchikov:1979,AbdallaAbdallaRothe:1991}\\
	  $e^{i\kappa m^2\sinh\te}$ & S-matrix of non-commutative Minkowski space with noncommutativity parameter $\kappa>0$ & \cite{GrosseLechner:2007}\\
	  \noalign{\smallskip}\hline\noalign{\smallskip}
    \end{tabular}
\end{table}

In the tensor case, where $D=\dim\VV>1$, the general solution to the constraints summarized in Definition~\ref{definition:SMatrix} is not known, mostly because of the complicated Yang-Baxter equation~c). However, special S-matrices {\em are} known, for example the S-matrix of the $O(N)$ Sigma models. In this case, $D=N>2$, $G=O(N)$, and $\Q$ consists of the defining representation of $O(N)$ on $\Cl^N$, with S-matrix
	\begin{align}\label{eq:SSigmaModel}
	S(\te)^{\alpha_1\alpha_2}_{\beta_1\beta_2}
	:=
	\sigma_1(\theta)\delta^{\alpha_1\alpha_2}\delta^{\beta_1\beta_2}
	+
	\sigma_2(\theta)\delta^{\alpha_1\beta_2}\delta^{\alpha_2\beta_1}
	+
	\sigma_3(\theta)\delta^{\alpha_1\beta_1}\delta^{\alpha_2\beta_2}
	\,.
\end{align}
Here $\sigma_1,\sigma_2,\sigma_3$ are specific combinations of rational functions and Gamma functions, see \cite[Ch.~8.3.2]{AbdallaAbdallaRothe:1991} for details.

The assumption that $S$ is analytic in the physical strip is not satisfied for all integrable models. In  general $S$ is only expected to be {\em meromorphic} in that strip, with poles signifying the presence of bound states  \cite{AbdallaAbdallaRothe:1991}. The known results are strongest for S-matrices without such bound state poles, and we will therefore restrict to this case in this review. However, some steps of the construction program to follow have also been accomplished in the meromorphic case by Cadamuro and Tanimoto \cite{CadamuroTanimoto:2015}; we will comment on this point later.

\subsection{Construction of Borchers triples from two-particle S-matrices}
\label{Section:S>BT}

Given a single particle spectrum (with representations $U_1$, $V_1$), and a two-particle S-matrix $S\in\SF$, we now set out to construct a corresponding Borchers triple $(\M_S,U_S,\Om_S)$. We first introduce a convenient Hilbert representation space $\Hil_S$, the {\em $S$-symmetric Fock space}\footnote{Picking this particular Hilbert space is a matter of choice, see \cite{Lechner:2012, Alazzawi:2012} for other possibilities.} over our single particle space $\Hil_1$. In a different context, the construction of $\Hil_S$ was first carried out by Liguori and Mintchev \cite{LiguoriMintchev:1995-1}, then for the $S$-matrix case in \cite{Lechner:2003,LechnerSchutzenhofer:2013}. For a more abstract formulation, emphasizing the functorial properties of the construction, see \cite{BischoffTanimoto:2013}.

Starting from the single particle space \eqref{eq:OneParticleSpace}, we consider the $n$-fold tensor products $\Hil_1^{\ot n}=L^2(\Rl^n,d^n\bte)\ot\VV^{\ot n}$, and the subspace $\Hil_{S,n}\subset\Hil_1\tp{n}$ of {\em $S$-symmetric wave functions}, i.e. those $\Psi_n\in\Hil_1\tp{n}$ which satisfy
\begin{align}\label{eq:Dnk}
     S(\te_{k+1}-\te_k)_{k,k+1}\Psi_n(\te_1,..,\te_{k+1},\te_k,..,\te_n)
     =
     \Psi_n(\te_1,..,\te_k,\te_{k+1},..,\te_n)
     \,,
\end{align}
for all $\te_1,...,\te_n\in\Rl$, $k\in\{1,...,n-1\}$. Here the subscript $k,k+1$ on $S$ signifies that this tensor acts on the tensor factors $k$ and $k+1$ of $\VV\tp{n}$.
    
For the constant ``flip'' $S$-matrix $S(\te)=\pm F$ (with $F(v\ot w)=w\ot v$, $v,w\in\VV$), the space $\Hil_{S,n}$ then coincides with the totally symmetric $(+)$ respectively totally antisymmetric $(-)$ $n$-fold tensor product of $\Hil_1$ with itself. For general $S\in\SF$, one can describe $\Hil_{S,n}$ as the invariant subspace of $\Hil_1\tp{n}$ of an $S$-dependent representation of the permutation group \cite{LiguoriMintchev:1995-1,LechnerSchutzenhofer:2013}.
\\\\
We now define the {\em $S$-symmetric Fock space} as the direct sum of these ``$n$-particle Hilbert spaces'',
\begin{align}\label{eq:Hil}
     \Hil_S:=\bigoplus_{n=0}^\infty \Hil_{S,n}
     \,,
\end{align}
where we understand $\Hil_{S,1}:=\Hil_1$ and $\Hil_{S,0}:=\Cl$. The $n$-particle component of a vector $\Psi\in\Hil_S$ will be denoted $\Psi_n\in\Hil_{S,n}$. Occasionally we will also use the ``particle number operator'' $(N\Psi)_n:=n\Psi_n$, and the dense subspace $\DD_S\subset\Hil_S$ of ``finite particle number''. For the time being, these are just names for certain subspaces, their physical interpretation in terms of particle states will be justified later in scattering theory.

Each $\Hil_S$, $S\in\SF$, is a closed subspace of the ``Boltzmann Fock space'' $\widehat{\Hil}:=\bigoplus_{n=0}^\infty\Hil_1\tp{n}$, and we denote by $P_S:\widehat{\Hil}\to\Hil_S$ the corresponding orthogonal projection. The Fock vacuum $\Om_S\in\Hil_S$, given by $(\Om_S)_0=1$, $(\Om_S)_n=0$ for $n\geq1$, will be the vacuum vector of the Borchers triple to be constructed.

The second ingredient of the Borchers triple, the representation of the Poincar\'e group, is given by a variant of the standard second quantization procedure.

\begin{lemma}\label{Lemma:SymmetryGroups}
	\begin{enumerate}
		\item For $(x,\la)\in\PGpo$, let
		\begin{align}
			U_S(x,\la):=P_S \bigoplus_{n=0}^\infty\left(U_1(x,\la)\tp{n}\ot{\rm id}_{\VV\tp{n}}\right)P_S
			\,.
		\end{align}
		Then $U_S$ is a strongly continuous, unitary, positive energy representation of $\PGpo$ on $\Hil_S$, with (unique) invariant vector $\Om_S$.
		\item\label{item:J} Let $J:\Hil_S\to\Hil_S$ be defined as
		\begin{align}\label{eq:J}
			({J}\Psi)_n^\balpha(\bte)
			:=
			\overline{\Psi_n^{\overline{\alpha_n}...\overline{\alpha_1}}(\te_n,...,\te_1)}
			\,.
		\end{align}
		Then $J$ is an antiunitary involution satisfying $JU_S(x,\la)J=U_S(-x,\la)$.
		\item For $g\in G$, let \begin{align}\label{eq:VSecondQuantized}
			V_S(g)
			:=
			P_S\bigoplus_{n=0}^\infty \left({\rm id}_{L^2(\Rl^n)}\ot V_1(g)^{\ot n}\right)P_S
			\,.
		\end{align}
		Then $V_S$ is a unitary representation of the gauge group $G$ on $\Hil_S$, commuting with $U_S$ and $J$. 
	\end{enumerate}
\end{lemma}

The proof of this lemma essentially amounts to showing that the various operators considered here respect the $S$-symmetry, i.e. restrict from $\widehat{\Hil}$ to the subspace $\Hil_S$. This is the case because of property \ref{item:TranslationInvariance} and \ref{item:GaugeInvariance} of $S$ in Def.~\ref{definition:SMatrix}. Explicitly, we have
\begin{align}\label{eq:U}
   [{U}(x,\la)\Psi]_n^{\balpha}(\bte)
   :=
   \exp (i \sum^{n}_{l=1} \; p_{\alpha_l}(\te_l) \cdot x) \;\Psi_n^\balpha(\te_1-\la,...,\te_n-\la)
   \,,
\end{align}
where $p_{\alpha_l}$ is shorthand for $p_{m_{[\alpha_l]}}$. Following our earlier convention also on the multi particle level, we write  $\Delta^{it}:=U(0,-2\pi t)$ for the rescaled boosts.
\\\\
Having fixed the representation $U_S$ and its vacuum vector $\Om_S$, we now turn to the construction of the most important ingredient, the von Neumann algebra $\M_S$ of wedge-local observables, completing $U_S,\Om_S$ to a Borchers triple. As a prerequisite for this, we first introduce creation and annihilation operators on the $S$-symmetric Fock space $\Hil$. 

On $\widehat{\Hil}$, we have the usual unsymmetrized operators $a(\varphi)$, $\ad(\varphi)$, $\varphi\in\Hil_1$. They are defined by linear and continuous extension from
\begin{align}
   \ad(\varphi)\psi_1\ot...\ot\psi_n
   &:=
   \sqrt{n+1}\,\varphi\ot\psi_1\ot...\ot\psi_n\,,\qquad\psi_1,..,\psi_n\in\Hil_1\,,
   \label{eq:AD}
   \\
   a(\varphi)\psi_1\ot...\ot\psi_n
   &:=
   \sqrt{n}\,\langle\varphi,\psi_1\rangle\,\psi_2\ot...\ot\psi_n
   \,,\qquad\;\;
   a(\varphi)\widehat{\Om}:=0\,,
   \label{eq:A}
\end{align}
to $\Hil_1\tp{n}$, and then to the subspace of finite particle number. We introduce their projections onto $\Hil_S$ as
\begin{align}\label{eq:zzd}
     \zd_S(\varphi):=P_S\,\ad(\varphi)P_S\,,\qquad z_S(\varphi):=P_S\,a(\varphi)P_S\,,\qquad \varphi\in\Hil_1,
\end{align}
and the distributional kernels $z^\#_{S,\alpha}(\te)$ of these operators by
\begin{align}\label{eq:ZZDKernels}
     \zd_S(\varphi)
     =
     \sum_\alpha\int d\te\,\varphi_\alpha(\te)\zd_{S,\alpha}(\te)
     \,,\qquad
     z_S(\varphi)
     =\sum_\alpha
     \int d\te\,\overline{\varphi_\alpha(\te)}z_{S,\alpha}(\te)
     \,.
\end{align}

Their essential properties are listed next.
\newpage

\begin{proposition}\label{proposition:ZZD}
     Let $\varphi\in\Hil_1$ and $\Psi\in\DD_S$ be arbitrary.
     \begin{enumerate}
	  \item\label{item:ZZDexplicit} The operators \eqref{eq:zzd} are in general unbounded, but well-defined on $\DD_S$.
	  \item\label{item:ZZDStar} We have
	  \begin{align}\label{eq:ZZDStar}
	       z_S(\varphi)^*\supset\zd_S(\varphi)\,.
	  \end{align}
	  \item\label{item:ZZDCovariance} For $(x,\la)\in\PGpo$ and $g\in G$, we have
	  \begin{align*}
	     U_S(x,\la)z_S^\#(\varphi)U_S(x,\la)^{-1}
	     &=
	     z_S^\#(U_1(x,\la)\varphi)
	     \,,\\
	     V_S(g)z_S^\#(\varphi)V_S(g)^{-1}
	     &=
	     z_S^\#(V_1(g)\varphi)
	     \,,
	  \end{align*}
	   where $z_S^\#$ stands for either $z_S$ or $\zd_S$.
	  \item\label{item:ZZDNBounds} With respect to the particle number operator $N$, there hold the bounds
	  \begin{align}\label{eq:NBoundsOnZZD}
	       \|z_S(\varphi)\Psi\| \leq \|\varphi\| \|N^{1/2}\Psi\|, \qquad \|z_S^{\dagger}(\varphi)\Psi\| \leq \|\varphi\| \|(N+1)^{1/2}\Psi\|\,.
	  \end{align}
	  \item\label{item:ZFAlgebra} $z_S,\zd_S$ form a representation of the Zamolodchikov-Faddeev algebra with $S$-matrix~$S$: The distributional kernels $z^\#_{S,\alpha}(\te)$ satisfy
	  \begin{align}
	       z_{S,\alpha}(\te) z_{S,\beta}(\te') - S^{\beta\alpha}_{\delta\gamma}(\te-\te') z_{S,\gamma}(\te') z_{S,\delta}(\te)  &=0,
		\label{eq:ZZ-Kernel-Commutator}
	       \\
	       \zd_{S,\alpha}(\te) \zd_{S,\beta}(\te') - S^{\gamma\delta}_{\alpha\beta}(\te-\te') \zd_{S,\gamma}(\te')\zd_{S,\delta}(\te)  &=0,
	       \\
	       z_{S,\alpha}(\te) \zd_{S,\beta}(\te') - S^{\alpha\gamma}_{\beta\delta}(\te'-\te) \zd_{S,\gamma}(\te') z_{S,\delta}(\te)
	       &=
	       \delta^{\alpha\beta}\delta(\te-\te')\cdot 1\,.
	  \end{align}
     \end{enumerate}
\end{proposition}

The algebraic relations in e) are known as the {\em Zamolodchikov--Faddeev} algebra \cite{ZamolodchikovZamolodchikov:1979, Faddeev:1984}, and are frequently used in the context of integrable quantum field theories (see for example \cite{BabujianFoersterKarowski:2006, Smirnov:1992}, and references cited therein). Note in particular that for the constant $S$-matrices $S^{\alpha\beta}_{\gamma\delta}(\te)=\pm\delta^{\alpha}_\delta\delta^\beta_\gamma$, they coincide with the familiar CCR/CAR relations.

The exchange relations in Prop.~\ref{proposition:ZZD}~e) were proposed by the Zamolodchikov brothers \cite{ZamolodchikovZamolodchikov:1979}. Their heuristic basis is that a product $\zd_{S,\alpha_1}(\te_1)\cdots\zd_{S,\alpha_n}(\te_n)$ of $n$ creation operators (acting on a vacuum vector) represents a configuration of $n$ particles with rapidities $\te_j$ and inner degrees of freedom $\alpha_j$, such that the order of factors in the product corresponds to the ordering of the particles on the line (from left to right). An exchange of two particles should correspond to a two-particle scattering process and thus produce an $S$ factor. Faddeev completed this structure \cite{Faddeev:1984} by adding a corresponding annihilation operator and the exchange relation between $z_S$ and $\zd_S$.

In our subsequent analysis, we will not rely on this motivation. Rather, we take the operators $z_S^\#$ as a tool for defining (generators of) a wedge algebra, and will then {\em derive} the intuitive picture drawn by the Zamolodchikov's from Haag-Ruelle scattering theory (Section~\ref{Section:Discussion}).
\\
\\
In the CCR situation $S=F$ (tensor flip), we can form the usual free field as the (selfadjoint closure of) the sum of a creation and an annihilation operator, and use the unitaries $\exp(i(\zd_F(\xi)+z_F(\xi))$, $\xi\in\K_1$, to generate the wedge algebra $\M_F$ as in~\eqref{eq:WeylAlgebra}. In this case, the same field operator can be used to generate the observables in left and right wedges, because the covariance statement in Proposition~\ref{proposition:ZZD}~c) also holds for the TCP operator $J$.

For other S-matrices $S$, however, this is not the case: Proposition~\ref{proposition:ZZD}~c) does {\em not} hold if $U(x,\la)$ is replaced by $J$, and different generators are needed for left and right wedges. We therefore introduce
\begin{align}
     \zd_S(\varphi)':=J\zd_S(J\varphi)J
     \,,\qquad
     z_S(\varphi)':=Jz_S(J\varphi)J
     \,.
\end{align}
Taking into account Lemma~\ref{Lemma:SymmetryGroups}~\ref{item:J}, it becomes apparent that items \ref{item:ZZDStar}--\ref{item:ZZDNBounds} of Proposition \ref{proposition:ZZD} apply to the $z_S^\#(\varphi)'$ without any changes.

The TCP-reflected creation/annihilation operators satisfy commutation relations analogous to the ones listed in Proposition \ref{proposition:ZZD} \ref{item:ZFAlgebra} with the only difference that $S^{\alpha\beta}_{\gamma\delta}(\te)$ has to be replaced by $S^{\beta\alpha}_{\delta\gamma}(-\te)$. For controlling commutators between observables in left and right wedges, it will be important to know the  {\em relative} commutation relations of the $z_S^\#(\varphi)$ and $z_S^\#(\varphi)'$. They are stated next, see also \cite{Niedermaier:1998} for a related analysis.

\begin{proposition}{\bf\cite{LechnerSchutzenhofer:2013}} \label{proposition:ZZ'CommutationRelations}
    Let $\varphi,\psi\in\Hil_1$. Then, on the domain $\DD_S$,
		\begin{align}
		[z_S(\varphi)',\, z_S(\psi)]
		&=
		0,\label{eq:ZZ'Commutator}
		\\
		[\zd_S(\varphi)',\, \zd_S(\psi)]
		&=0,
		\label{eq:ZDZD'Commutator}
		\\
		[z_S(\varphi)',\, \zd_S(\psi)]
		&=
		K_S(\varphi,\psi),
		\label{eq:Z'ZDCommutator}
		\\
		[\zd_S(\varphi)',\, z_S(\psi)]
		&=
		L_S(\varphi,\psi),
		\label{eq:ZD'ZCommutator}
		\end{align}
		where $K_S(\varphi,\psi)$ and $L_S(\varphi,\psi)$ are bounded operators on $\Hil_S$ which commute with~$N$.
    If $\varphi\in\K_1$ and $\psi'\in \K_1'$, then 
		\begin{align}\label{eq:KernelsEqual}
			L_S(\varphi,\psi')
			=
			-K_S(\varphi,\psi')\,.
		\end{align}
\end{proposition}

The proof of the first part of this proposition follows by explicit calculation of the commutators, which also gives the explicit form of the operators $K_S(\varphi,\psi)$ and $L_S(\varphi,\psi)$ as multiplication operators with certain tensor-valued bounded functions $K_{S,n}(\varphi,\psi)$, $L_{S,n}(\varphi,\psi)$ on each $n$-particle space $\Hil_{S,n}$. These functions are given by integrating the components of $\varphi$ and $\psi$ against a kernel which consists of a product of $S$-factors. To establish \eqref{eq:KernelsEqual}, one relies on the analyticity properties of both, $\varphi,J\psi'\in\K_1$ \eqref{eq:K1Explicit} and $S$, as well as their boundary conditions $J\Delta^{1/2}\varphi=\varphi$, $J\Delta^{-1/2}\psi'=\psi'$ and the crossing symmetry in Def.~\ref{definition:SMatrix}~\ref{item:Crossing}. These features allow for a contour shift in the integrals defining $K_S,L_S$, and lead to \eqref{eq:KernelsEqual}. 

When $S$ has first order poles in the physical strip, \eqref{eq:KernelsEqual} fails, which is the reason why we restrict ourselves to analytic $S$. However, for certain S-matrices with poles, Cadamuro and Tanimoto found a way to modify the generators $\Phi_S$ (see below) to cancel these residue contributions. This modification, and the emerging subtle domain questions, are explained in \cite{CadamuroTanimoto:2015}.
\\
\\
Sticking to the case of analytic $S$ and following Schroer \cite{Schroer:1999}, we now define the field operators we are interested in as
\begin{align}
	\Phi_S(\xi)
	&:=
	\zd_S(\xi)'+z_S(J\Delta^{1/2}\xi)'
	\,,\\
	\Phi'_S(\xi)
	&:=
	\zd_S(\xi)+z_S(J\Delta^{-1/2}\xi)
	\,.
\end{align}
The counter intuitive placement of the primes is chosen here to have $\Phi_S$ generate $\M_S$ (instead of $\M_S'$), but also be consistent with the literature.

The above defined operators depend complex-linearly on their arguments, and it is clear that $\Phi_S(\xi)$ is a well-defined operator on $\DD_S$ for $\xi\in\dom\Delta_1^{1/2}$, whereas $\Phi_S'(\xi)$ is well-defined for $\xi\in\dom\Delta_1^{-1/2}$. It also follows quickly from the definitions that
\begin{align}
	J\Phi_S(\xi)J=\Phi_S'(J\xi)\,,\qquad \xi\in\dom\Delta^{1/2}\,.
\end{align}
\begin{proposition}\label{proposition:PhiPhi'}
	Let $\xi\in\dom\Delta^{1/2}$.
	\begin{enumerate}
		\item\label{item:PhiAdjoint} $\Phi_S(\xi)^*\supset\Phi_S(J\Delta^{1/2}\xi)$.
		\item\label{item:PhiAnalyticVectors} All vectors in $\DD_S$ are entire analytic for $\Phi_S(\xi)$. For $\xi\in\K_1$, the field operator $\Phi_S(\xi)$ is essentially selfadjoint.
		\item\label{item:PhiCovariance} $\Phi_S$ transforms covariantly under the proper orthochronous Poincar\'e group~$\PGpo$ and the gauge group $G$, i.e. on $\DD_S$, there holds
		\begin{align}\label{eq:PhiCovariance}
			U_S(x,\la)\Phi_S(\xi)U_S(x,\la)^{-1}
			&=
			\Phi_S(U_1(x,\la)\xi)
			\,,\qquad
			(x,\la)\in\PGpo\,,\\
			V_S(g)\Phi_S(\xi)V_S(g)^{-1}
			&=
			\Phi_S(V_1(g)\xi)\,,\qquad
			g\in G\,.
		\end{align}
		\item\label{item:PhiReehSchlieder} The vacuum vector $\Om_S$ is cyclic for $\Phi_S$, i.e.
		the subspace
		\begin{equation*}
			\Cl{\rm \text{-}span}\{\Phi_S(\xi_1)\cdots\Phi_S(\xi_n)\Omega_S\;:\; \xi_1,...,\xi_n\in\K_1,\,n\in\Nl_0\}
		\end{equation*}
		is dense in $\Hil_S$.
	\end{enumerate}
	All these statements also hold if $\Phi_S$ is exchanged with $\Phi'_S$, and $\K_1$ with $\K_1'$.
\end{proposition}

All these properties follow from straightforward calculations, except part \ref{item:PhiAnalyticVectors} which uses the bound from Proposition~\ref{proposition:ZZD}~\ref{item:ZZDNBounds} and an application of Nelson's Theorem.

Part~\ref{item:PhiAdjoint} of this proposition explains the factors $J\Delta^{\pm1/2}$ appearing in the definition of $\Phi_S$ and $\Phi_S'$, anticipating the Bisognano-Wichmann property of a wedge algebra $\M_S$ generated by $\Phi_S(\xi)$. In fact, we have, $\xi\in\dom\Delta^{1/2}$,
\begin{align*}
	J\Delta^{1/2}\Phi_S(\xi)\Om_S
	=
	J\Delta^{1/2}\xi
	=
	\Phi_S(J\Delta^{1/2}\xi)\Om_S
	=
	\Phi_S(\xi)^*\Om_S\,,
\end{align*}
and $J\Delta^{-1/2}\Phi_S'(\xi')\Om_S=\Phi_S'(\xi')^*\Om_S$ for $\xi'\in\dom\Delta^{-1/2}$. This is consistent with $\Phi_S(\xi)$, $\xi\in\dom\Delta^{1/2}$, being affiliated to a von Neumann algebra which has $\Om_S$ as a cyclic separating vector, and modular data $J,\Delta$, as well as $\Phi_S'(\xi')$, $\xi'\in\dom\Delta^{-1/2}$, being affiliated with the commutant of that algebra.

We therefore define the von Neumann algebra
\begin{align}
	\M_S
	:=
	\{e^{i\Phi_S(\xi)}\,:\,\xi\in\K_1\}''\,,
\end{align}
and want to convince ourselves that $(\M_S,U_S,\Om_S)$ is a Borchers triple. To this end, we first note that in view of $U_1(x,\la)\K_1\subset\K_1$ for $x\in\overline{W_R}$, $\la\in\Rl$, and Proposition~\ref{proposition:PhiPhi'}~\ref{item:PhiCovariance}, we have $U(x,\la)\M_S U(x,\la)^{-1}\subset\M_S$ for $x\in\overline{W_R}$, $\la\in\Rl$. Next, by using Proposition~\ref{proposition:PhiPhi'}~\ref{item:PhiReehSchlieder} and standard techniques, one can show that $\Om_S$ is cyclic for $\M_S$. 

The crucial point is to show that $\Om_S$ is also separating, which amounts to showing that $\M_S$ has a large commutant in $\B(\Hil_S)$. At this point, the second field $\Phi_S'$ comes into play: For $\varphi\in\K_1$, $\psi'\in\K_1'$, we find the following equality on $\DD_S$:
\begin{align*}
 [\Phi_S(\varphi),\Phi_S'(\psi')]
 &=
 [\zd_S(\varphi)'+z_S(\varphi)',\,\zd_S(\psi')+z_S(\psi')]
 \\
 &=
 L_S(\varphi,\psi')+K_S(\varphi,\psi')
 \\
 &=0\,.
\end{align*}
Here we have first used that $J\Delta^{1/2}\varphi=\varphi$, $J\Delta^{-1/2}\psi'=\psi'$, and then the commutation relations of Proposition~\ref{proposition:ZZ'CommutationRelations}. By linearity, it then follows that $\Phi_S(\varphi)$ and $\Phi_S'(\psi')$ commute on $\DD_S$ for any $\varphi\in\dom\Delta_1^{1/2}$, $\psi'\in\dom\Delta_1^{-1/2}$, so that bounded functions of $\Phi_S'(\psi')$ are good candidates for operators commuting with $\M_S$. Indeed, by a calculation on analytic vectors \cite{Alazzawi:2014, Lechner:2012}, one finds
\begin{align}
	\left[e^{i\Phi_S(\varphi)},\,e^{i\Phi_S'(\psi')}\right]=0
	\,,\qquad
	\varphi\in\K_1,\,\psi'\in\K_1'\,.
\end{align}
This implies that the von Neumann algebra
\begin{align}
	\widetilde{\M_S}
		:=
		\{e^{i\Phi'_S(\xi')}\,:\,\xi'\in\K_1'\}''\,,
\end{align}
commutes with $\M_S$. As $\Om_S$ is cyclic for $\widetilde{\M_S}$ by the same arguments as for $\M_S$, we arrive at the following result.

\begin{theorem}\label{Theorem:SGivesBorchersTriple}
    Let $S\in\SF$. Then the triple $(\M_S,U_S,\Om_S)$ is a Borchers triple. 
\end{theorem}

This result was proven in \cite{Lechner:2003} for the scalar case, and in \cite{LechnerSchutzenhofer:2013} for the general tensor case. It completes the first step in the construction program presented in Section~\ref{Section:Abstract}, and we therefore obtain a local covariant net $\OO\mapsto\A_S(\OO)$  of von Neumann algebras on $\Hil_S$ for each S-matrix $S\in\SF$.

As discussed in Section~\ref{Section:RelativeCommutants}, the next step is to analyze the relative commutants of the inclusions $\M_S(x)\subset\M_S$, $x\in W_R$, by means of the split property and modular nuclearity. In this context, it is important to know the modular data of $(\M_S,\Om_S)$.

\begin{theorem}
    Let $S\in\SF$. 
 	\begin{enumerate}
		\item The Bisognano-Wichmann property holds, i.e. the modular data $\widetilde{J}$, $\widetilde{\Delta}$ of $(\M_S,\Om_S)$ are given by 
		\begin{align}
			\widetilde{\Delta}=\Delta
			\,,\qquad
			\widetilde{J}=J\,.
		\end{align}
	\item The commutant of $\M_S$ in $\B(\Hil_S)$ is ${\M_S}'=\widetilde{\M_S}$.
	\end{enumerate}
\end{theorem}

This theorem was proven in the scalar case in \cite{BuchholzLechner:2004}, and in the tensor case in \cite{Alazzawi:2014}.

\subsection{Regular S-matrices and the modular nuclearity condition}
\label{Section:Nuc}

As discussed in the abstract setting in Section~\ref{Section:Abstract}, the operator-algebraic construction program of a QFT model proceeds in two main steps. The first step is to find a suitable Borchers triple, and has been accomplished in Theorem~\ref{Theorem:SGivesBorchersTriple} for any S-matrix $S\in\SF$. The second step consists of analyzing the local observable content which is (indirectly) defined by the Borchers triple. In this section, we will follow the general strategy explained in Section~\ref{Section:RelativeCommutants}, and summarize the known results on modular nuclearity and the split property in the model theories with Borchers triple $(\M_S,U_S,\Om_S)$, $S\in\SF$. 

We first recall that for split distance $s>0$, the map in question is \eqref{eq:DefinitionXi(x)}
\begin{align}
	\Xi_S(s):\M_S\to\Hil_S\,,\qquad \Xi_S(s)A:=\Delta^{1/4}U_S(s)A\Om_S
	\,,
\end{align}
where $U_S(s)$ is shorthand for the purely spatial translation by $(0,s)\in W_R$. Making use of the explicit form \eqref{eq:U} of the representation $U_S$, one finds
\begin{align}\label{eq:XiConcrete}
	(\Xi_S(s)A)_n^\balpha(\bte)
	=
	\prod_{k=1}^n e^{-m_\alpha s \cosh\te_k}\cdot (A\Om)^\balpha_n(\bte+i\bla_0)
	,\quad
	\bla_0:=\left(-\tfrac{\pi}{2},...,-\tfrac{\pi}{2}\right)\,.
\end{align}
Here the complex argument $\bte+i\bla_0$ has to be understood in terms of analytic continuation, and suggests that for understanding nuclearity properties of the map $\Xi_S(s)$, complex analysis will be essential.

For general $S$, one can not expect $\Xi_S(s)$ to be nuclear. We therefore make the following definition of ``regular'' $S$ \cite{Lechner:2008}, demanding stronger analyticity of $S$ than can be expected from first principles \cite{Martin:1969}. 

\begin{definition}\label{Definition:RegularS}
    An $S$-matrix $S\in\SF$ is called regular if there exists $\eps>0$ such that $\te\mapsto S(\te)$ extends to a bounded analytic function on the wider strip $\Strip(-\eps,\pi+\eps)$. The family of all regular S-matrices will be denoted $\SF_0\subset\SF$.
\end{definition}

This condition demands in particular that all singularities of $S$ lie a minimal distance $\eps>0$ off the physical strip. Poles in $\Strip(-\pi,0)$ are usually interpreted as evidence for unstable particles with a finite lifetime \cite{Weinberg:1995}, with lifetime of such a resonance becoming arbitrarily long if the corresponding pole lies sufficiently close to the real axis. The regularity condition rules out S-matrices with infinitely many resonances with arbitrarily long lifetimes and ``masses'' such that the thermodynamical partition function diverges, a situation in which we cannot expect modular nuclearity to hold \cite{BuchholzDAntoniLongo:1990,BuchholzJunglas:1989}. The additional condition of a {\em bounded} analytic extension of $S$ to $\Strip(-\eps,\pi+\eps)$, $\eps>0$, is a condition on the phase shift that can also be found in other approaches \cite{KarowskiThunTruongWeisz:1977}. In both the scalar and tensor case, there exist many regular S-matrices.

The detailed analysis of the nuclearity properties of the map \eqref{eq:XiConcrete} requires the discussion of many technical points and will not be presented here. Rather, we will give an outline of the strategy which was so far used for proving that $\Xi_S(s)$ is nuclear for regular $S$. It consists of the following three steps.
\\\\
\noindent {\bf Step 1) }{\em Show that for $A\in\M_S$, the functions $(A\Om)_n:\Rl^n\to\VV\tp{n}$ have an analytic continuation to a tube $\Tu_n=\Rl^n+i\B_n\subset\Cl^n$ which contains the point $\bla_0$ in the interior of its base~$\B_n\subset\Rl^n$.}

For a heuristic argument why such analyticity can be expected, take $A$ to be a polynomial in the generators $\Phi_S(\xi_1),...,\Phi_S(\xi_N)$, $\xi_1,...,\xi_N\in\K_1$. (This unbounded operator is not an element of $\M_S$, but affiliated with this algebra.) Considering only the term with only creation operators for concreteness, $\widetilde{A}=\zd_S(\xi_1)\cdots\zd_S(\xi_n)$, we find the $n$-particle wave functions
\begin{align}
  (\widetilde{A}\Om)_n(\bte)
  =
  \frac{1}{n!}\sum_{\pi\in\frS_n}S^\pi(\bte)
  \left(\xi_1(\te_{\pi(1)})\ot...\ot\xi_n(\te_{\pi(n)})\right)\,,
\end{align}
where the $S^\pi(\bte)\in\B(\VV\tp{n})$ are tensors consisting of several S-factors, depending on differences $\te_k-\te_l$ of rapidities. As each $\xi_k$ is analytic in $\Strip(-\pi,0)$ (cf. \eqref{eq:K1Explicit}), we see that $(\widetilde{A}\Om)_n$ is analytic in a tube containing an open neighborhood of $\bla_0$ in its base, provided $S$ is regular.

For general observables $A\in\M_S$, such analyticity properties were shown to hold for regular $S$ in \cite{Lechner:2008} in the scalar case, and by S.~Alazzawi in the general tensor case \cite{Alazzawi:2014}. A generalization of the detailed analytic structure of these wave functions has later also proven to be important in the work of Bostelmann and Cadamuro on characterizations of local observables \cite{BostelmannCadamuro:2014}.
\\\\
\noindent {\bf Step 2) }{\em Show that there exists $0<C_n(2s)<\infty$ such that for all $A\in\M$}
\begin{align}
	\sup_{\la\in\B_n}\left(\int_{\Rl^n}d^n\bte\,\|(U_S(s)A\Om)_n(\bte+i\bla)\|^2\right)^{1/2}
	\leq
	C_n(2s)\cdot\|A\|\,.
\end{align}
Such Hardy type bounds were established in \cite{Lechner:2008, Alazzawi:2014}.

It then follows that the linear map $$\Sigma_n(s) :\M_S\to{\mathbb H}^2(\Tu_n)\,,\qquad\Sigma_n(s)A:=(U(\tfrac{s}{2})A\Om)_n,$$ from the wedge algebra $\M_S$ into the (vector-valued) Hardy space ${\mathbb H}^2(\Tu_n)$ on the tube $\Tu_n$, is bounded with $\|\Sigma_n(s)\|\leq C_n(s)<\infty$.

To formulate the last step, we first note that by employing tools from complex analysis, one can show that the map $E_n(s):{\mathbb H}^2(\Tu_n)\to L^2(\Rl^n,d^n\bte)$ given by $E_n(s)(F)(\bte):=\prod_{k=1}^n e^{ism\sinh\te_k}F(\bte+i\bla_0)$ is nuclear\footnote{It is in fact $p$-nuclear for any $p>0$, i.e. its singular values decay faster than any inverse power.}. As $\Xi_S(s)=\sum_n E_n(s)\circ\Sigma_S(s)$, this gives the nuclearity bound
\begin{align}
    \|\Xi_S(s)\|_1
    \leq
    \sum_{n=0}^\infty \|E_n(s)\Sigma_n(s)\|_1
    \leq
    \sum_{n=0}^\infty C_n(s)\|E_n(s)\|_1
    \,.
\end{align}
To prove nuclearity of $\Xi_S(s)$, one has to estimate the norm bound $C_n(s)$ and nuclear norm $\|E_n(s)\|_1$ sharp enough to have this series converging. This is only possible if one properly takes into account the statistics ($S$-symmetry). Therefore the last step is:
\\\\
\noindent {\bf Step 3) } {\em Exploit the $S$-symmetry to obtain bounds $C_n(s)$ and $\|E_n(s)\|_1$ sharp enough such that $\sum_{n=0}^\infty C_n(s)\|E_n(s)\|_1<\infty$, at least for sufficiently large split distance $s>0$.}
\\\\
In the scalar case, Step~3 has been accomplished in case $S(0)=-1$, where an effective Pauli principle becomes available\footnote{Note that the stronger results claimed in \cite{Lechner:2008} contain an incorrect estimate \cite{Lechner:Erratum2015}, spotted by S.~Alazzawi. At the time of writing, the result stated here is the strongest one with a complete proof.} \cite{Lechner:2008}. 

\begin{theorem}\label{Theorem:Nuclearity-For-Regular-S}
    Let $S\in\SF_0$ be a scalar S-matrix with $S(0)=-1$. Then there exists a splitting distance $s_0>0$ such that $\Xi_S(s)$ is nuclear at least for $s>s_0$.  
\end{theorem}

In the tensor case, a similar situation occurs when demanding $S$ to be regular and $S(0)=-F$ (with $F$ the flip on $\VV\ot\VV$ as before), these assumptions are in particular met by the S-matrices of the $O(N)$ sigma models. There is good evidence that this will give rise to a proof of modular nuclearity, as in the scalar case \cite{Alazzawi:2014}.

The additional assumption $S(0)=-1$ (only the two possibilities $S(0)=\pm1$ exist for regular scalar $S$) amounts to a kind of ``hard core'' condition, and is satisfied in most of the interacting models known from Lagrangian formulations.

\subsection{Discussion of the constructed models}
\label{Section:Discussion}

Having constructed a large class of QFT models, one would next like to analyze their properties, a minimum requirement being that these actually do describe non-trivial interaction\footnote{That this is a non-trivial issue can be seen (in the setting of Wightman QFT) at the example of a family of complicated Wightman functions \cite{Read:1996} which were only later realized to be equivalent to generalized free fields \cite{Rehren:1996}.}. 

In the case at hand, where an S-matrix is the input into the construction, it is most natural to consider scattering theory. Whenever there exist non-trivial observables interpolating between the vacuum and single particle states and localized in some double cone (of arbitrary size), one is in the position to apply Haag-Ruelle scattering theory (see, for example, \cite{Araki:1999}). In the models at hand, such observables exist, but are given only indirectly as elements of intersections of wedge algebras. However, it turns out that explicit knowledge of local operators is not necessary, one can express all collision states in terms of the explicit wedge-local fields by making use of the adaptation of Haag-Ruelle theory to the case of wedge-local observables \cite{BorchersBuchholzSchroer:2001}. The following result holds.

\begin{theorem}
    Let $S\in\SF$ be a two-particle S-matrix such that the vacuum vector is cyclic for some double cone (This is in particular the case for $S$ satisfying the assumptions of Thm.~\ref{Theorem:Nuclearity-For-Regular-S}). Then the collision operator of the QFT model generated from the Borchers triple $(\M_S,U_S,\Om_S)$ is the factorizing S-matrix with two-particle S-matrix $S$. Furthermore, this theory is then asymptotically complete. 
\end{theorem}

This result shows that the presented construction provides a solution to the inverse scattering problem for two-particle S-matrix $S\in\SF$. Under the assumption that $\Om_S$ is cyclic for some double cone (of arbitrary size), one can also explicitly compute incoming and outgoing $n$-particle scattering states. Restricting to the scalar case for simplicity, one finds the (improper) asymptotic collision states of $n$ particles of rapidities $\te_1,...,\te_n\in\Rl$, $n\in\Nl_0$,
\begin{align}
	|\te_1,...,\te_n\rangle_{S, \rm out}
	&=
	\zd_S(\te_1)\cdots\zd_S(\te_n)\Om_S\,,\qquad \text{if }\; \te_1<...<\te_n
	\,,\\
	|\te_1,...,\te_n\rangle_{S, \rm in}
	&=
	\zd_S(\te_1)\cdots\zd_S(\te_n)\Om_S\,,\qquad \text{if }\; \te_1>...>\te_n\,.
\end{align}
These identifications can be proven with the tools of Haag-Hepp-Ruelle scattering theory \cite{Haag:1958,Ruelle:1962,Hepp:1965}. They are in perfect agreement with the intuition behind the Zamolodchikov-Faddeev algebra, identifying products of $n$ creation operators in order of increasing (decreasing) rapidities as creating outgoing (incoming) collision states, and rearrangements of these operators as two-particle scattering processes, producing $S$-factors \cite{ZamolodchikovZamolodchikov:1979,Faddeev:1984}.
\\
\\
We conclude this section with a comparison with other approaches to solving the inverse scattering problem for integrable QFTs, notably the form factor program \cite{BabujianFoersterKarowski:2006,Smirnov:1992}. In that approach, one aims at constructing a {\em Wightman} QFT \cite{StreaterWightman:1964,Jost:1965} by specifying the $n$-point functions of local field operators. These are {\em not} the field operators $\Phi_S$ appearing in the approach presented here, which can also be formulated as operator-valued distributions over Minkowski space, but are localized only in wedges $W_{R/L}+x$ rather than points $\{x\}$ \cite{LechnerSchutzenhofer:2013}. The role of the wedge-local fields $\Phi_S$ is to serve as generators of Borchers triples. It might well happen that the emerging net of von Neumann algebras can equivalently be generated by certain point-local Wightman fields, see \cite{Bostelmann:2005,Bostelmann:2005-2} for the right tools for analyzing the local observable content. However, there is no 
straightforward connection between the wedge-local fields and point like fields. 

Hence the approach presented here is complementary to the form factor program, and also to conventional constructive QFT \cite{GlimmJaffe:1987}: It does give a rigorous solution of the inverse scattering problem by operator-algebraic methods, but does not provide explicit formulae for strictly local quantities like Wightman $n$-point functions. Comparison with the conventional approach to constructive QFT is therefore indirect: Only if a Lagrangian model is conjectured to have a certain S-matrix (such as the Sinh-Gordon model, the $O(N)$-Sigma models, etc) then the approach starting from the Lagrangian can be compared to the one starting from the S-matrix. For most of the S-matrices $S\in\SF$, however, no corresponding Lagrangian is known, and the classical counterparts of these theories are therefore unknown.

The problem of characterizing the local observables $A\in\A_S(\OO)$ more explicitly has been taken up by Bostelmann and Cadamuro. Generalizing the well-known fact that on the totally symmetric Bose Fock space, any bounded operator can be expanded into a series of normal ordered creation and annihilation (CCR) operators~\cite{Araki:1963}, they showed that for any quadratic form $A$ on $\Hil_S$, where $S\in\SF$ is a scalar S-matrix and $A$ is subject to certain regularity assumptions, one has \cite{BostelmannCadamuro:2012}
\begin{align}\label{eq:FF-expansion}
    A
    =
    \sum_{m,n=0}^\infty\int\frac{d\bte\,d\eta}{m!n!}\,f^{[A]}_{m,n}(\bte,\eta)\,\zd_S(\te_1)\cdots\zd_S(\te_m)\,z_S(\eta_1)\cdots z_S(\eta_n)
    \,.
\end{align}
The expansion coefficients $f^{[A]}_{m,n}$ are sums of contracted matrix elements of $A$. It is interesting to note that such distributions also appear in the context of proving the modular nuclearity condition \cite{Lechner:2008} (for wedge-local bounded $A$). The expansion \eqref{eq:FF-expansion} is essentially the form factor expansion. We refer to \cite{BostelmannCadamuro:2012} for details, also regarding the convergence properties of this series.

The expansion \eqref{eq:FF-expansion} holds for any (sufficiently regular) $A$, independent of its localization properties. If the expanded operator $A$ is an element of a local algebra, one has more information on the expansion coefficients\footnote{However, boundedness of $A$ is not directly reflected in its coefficients $f^{[A]}_{n,m}$ because the expansion \eqref{eq:FF-expansion} involves the unbounded operators $z_S^\#$.} -- for example, compact localization in spacetime leads to strong analyticity properties of the $f^{[A]}_{m,n}$. In \cite{BostelmannCadamuro:2014}, Bostelmann and Cadamuro give complete, necessary and sufficient, conditions on the coefficients $f^{[A]}_{n,m}$ for $A$ to be localized in a double cone $\OO_r$ of radius $r>0$ around the origin. These conditions are rather involved, and solutions are 
currently known only for the constant scalar $S$-matrices $S=1$ (the free field) and $S=-1$ (the Ising model) \cite{Cadamuro:2012}. The Ising model S-matrix $S=-1$, although looking almost trivial as an S-matrix, already gives rise to rather complicated local operators, see also \cite{BergKarowskiWeisz:1979} for a related analysis in the form factor program.

This state of affairs is typical for the comparison of the algebraic approach and the form factor program: Whereas the operator-algebraic tools are more efficient for questions like proving existence of local observables, the form factor methods (and also the expansion \eqref{eq:FF-expansion}) give more explicit information about these local quantities, see for example \cite{BostelmannCadamuroFewster:2013,BostelmannCadamuro:2015_2} for applications to quantum energy inequalities and energy densities in the models considered here.

\subsection{Massless models}
\label{Section:Mass0}

Up to this point we have considered theories with purely massive particle spectrum. While many integrable models are of this type, and scattering theory is best understood in the massive case\footnote{However, see \cite{Buchholz:1975-1,Buchholz:1977,Dybalski:2005} for results on scattering involving massless particles.} , there are also good reasons to consider massless theories. On the one hand, for an analysis of phenomena like confinement or asymptotic freedom, one has to consider short distance (scaling) limits, in which the masses go to zero (see \cite{BuchholzVerch:1998,DAntoniMorsellaVerch:2004,BostelmannDAntoniMorsella:2009} for discussions within the algebraic framework). Since certain integrable QFTs are believed to be asymptotically free (see, for example, \cite{AbdallaAbdallaRothe:1991}), this point of view is also relevant here. On the other hand, massless integrable models often exhibit conformal symmetry, providing a link to conformal QFT \cite{Rehren:2015} with all its specialized tools, 
which can give additional insight into their structure. In this section, we therefore consider various massless versions of the constructions presented so far. Due to page constraints, we will be rather brief, and often refer to the literature for details.

As for massive models, we first consider the interaction-free case, and recall the Borchers triple of the free massless current. Since the definition of the rapidity \eqref{eq:palpha} depends on the mass, we work here instead in the momentum picture, and consider the Hilbert space $\Hil_1=\Hil^+_1\oplus\Hil^-_1$, with $\Hil^\pm_1:=L^2(\Rl_\pm,dp/|p|)$ and the mass zero Poincar\'e representation $U_1=U_1^+\oplus U_1^-$, given by $(U^\pm_1(x,\la)\psi)(p)=e^{i(\pm x_0-x_1)p}\,\psi(e^{\mp\la}p)$. The massless free current decomposes into a direct sum of chiral operators $j_\pm$, each depending only on one of the light cone coordinates $x_0\pm x_1$. Since the right wedge $W_R$ is in these coordinates the set $x_0+x_1>0$, $x_0-x_1<0$, one considers the von Neumann algebras $\M_\pm=\{\exp(ij_\pm(f))\,:\,f\in C_{c,\Rl}^\infty(\Rl_\pm)\}''$ on the Bose Fock spaces $\Hil^\pm$ over $\Hil^\pm_1$. Denoting the respective Fock vacua by $\Om^\pm\in\Hil^\pm$, one obtains on $\Hil:=\Hil^+\ot\Hil^-$ the Borchers triple
\begin{align}\label{eq:BT-chiral}
  (\M_{0}^+\ot\M_{0}^-,U^+\ot U^-,\Om^+\ot\Om^-)\,.
\end{align}
This construction is equivalent to the one presented in the preceding sections, with the trivial scalar S-matrix $S=1$, and the modification that instead of the field $\Phi_S$, one has to consider its  directional derivatives (current) in order to avoid the infrared singularity of the measure $dp/|p|$.

Since all data decompose into light cone coordinates, one extends in this setting the definition of wedges, Borchers triples and standard pairs to the case of dimension $d=1$. The space $\Rl^1$ is thought of as a light ray, the affine group of $\Rl$ (``$ax+b$ group'') plays the role of the Poincar\'e group, and ``positive energy representation'' means that the generator of the translation subgroup is positive. Replacing the right wedge by the right half line $\Rl_+$, and double cones by intervals $I\subset\Rl$, the framework presented in Sections~\ref{Section:Wedges} and \ref{Section:WedgeAlgebras} also applies to the one-dimensional case.

To motivate the following construction, we first recall a result by Longo and Witten on endomorphisms of standard pairs \cite{LongoWitten:2010}. In the setting of a one-dimensional standard pair $(\K,U)$ on a Hilbert space $\Hil$, they defined an endomorphism to be a unitary $V\in\B(\Hil)$ such that $V\K\subset \K$ and $[V,U(x)]=0$ for all translations $x\in\Rl$. In their result, they make use of symmetric inner functions on the upper half plane, that is, analytic bounded functions $\varphi$ on the upper half plane such that
\begin{align}\label{eq:sifu}
	\overline{\varphi(p)}=\varphi(p)^{-1}=\varphi(-p)\quad\text{for almost all } p\in\Rl\,.
\end{align}

\begin{theorem}{\bf \cite{LongoWitten:2010}}
	Let $(\K,U)$ be a one-dimensional standard pair with $U$ non-degenerate and irreducible. Then a unitary $V$ is an endomorphism of $(\K,U)$ if and only if $V=\varphi(P)$, where $P>0$ is the generator of the translations, and $\varphi$ is a symmetric inner function on the upper half plane.
\end{theorem}

For generalizations of this theorem, see \cite{LechnerLongo:2014}. 

The equations \eqref{eq:sifu} appearing in the definition of a symmetric inner function are reminiscent of the constraints imposed on a (scalar) S-matrix. In fact, when changing variables from $p$ to $\te=\log p$, the upper half plane is transformed to the strip $\Strip(0,\pi)$ appearing in Definition~\ref{definition:SMatrix}, and \eqref{eq:sifu} translates to $\overline{\psi(\te)}=\psi(\te)^{-1}=\psi(\te+i\pi)$, $\te\in\Rl$. In this form, the similarities to scalar S-matrices are most striking, and reveal a remarkable match between the properties of endomorphisms of standard pairs, and the completely independently defined S-matrices. At the time of writing, no deep reason for this match was known, and we refer to \cite{LechnerLongo:2014} for further discussions of this point.
\\\\
In the context of deformations of Borchers triples, the result by Longo and Witten provides the background to a construction by Tanimoto \cite{Tanimoto:2011-1}, which amounts to twisting one tensor factor in the chiral Borchers triple \eqref{eq:BT-chiral}. He defines
\begin{align}\label{eq:MS0}
	\M_{0,\varphi}
	:=
	(\M_{0}^+\ot 1)\vee S_\varphi(1\ot\M_{0}^-)S_\varphi^*\,,
\end{align}
where $S_\varphi\in\B(\Hil^+\ot\Hil^-)$ is a suitably chosen unitary depending on a symmetric inner function $\varphi$, namely
\begin{align}\label{eq:Sfi}
	[S_\varphi \Psi]_{n,n'}(p_1,..,p_n;q_1,..,q_{n'})
	&:=
	\prod_{{l=1,..,n}\atop{l'=1,..,n'}}{\varphi}(-p_lq_{l'})\cdot\Psi_{n,n'}(p_1,..,p_n;q_1,..,q_{n'})
	\,.
\end{align}
Here the indices $n,n'$ refer to the Fock structure of $\Psi\in\Hil^+\ot\Hil^-$, and the momenta to the spectra of the second quantized generators of translations in the light like directions, see \cite{Tanimoto:2011-1}.

\begin{theorem}{\bf\cite{Tanimoto:2011-1}}
	Let $\varphi$ be a symmetric inner function on the upper half plane, and $S_\varphi$ defined as \eqref{eq:Sfi}. Then the triple $(\M_{0,\varphi},U^+\ot U^-,\Om^+\ot\Om^-)$ is a two-dimensional Borchers triple.
\end{theorem}

The physical significance of the operator $S_\varphi$ appearing here is that of a ``wave S-matrix'' \cite{Buchholz:1975}, as has been shown by Dybalski and Tanimoto \cite{DybalskiTanimoto:2010}. In the sense of scattering of waves, one even has asymptotic completeness in this situation, and can recover the Borchers triple from its S-matrix and an asymptotic algebra \cite{DybalskiTanimoto:2010,Tanimoto:2011-1}.

Tanimoto's deformation formula \eqref{eq:MS0} is a result of the chiral (tensor product) structure present in the massless case, and has no direct analogue in the massive situation. Nonetheless, the twisted chiral triple is equivalent to a massless version of the inverse scattering construction discussed in the preceding sections. We refrain from giving the details of the massless version, which can be found in \cite{BostelmannLechnerMorsella:2011}. The equivalence proof is contained in \cite{LechnerSchlemmerTanimoto:2013}.
\\\\
The twist \eqref{eq:MS0} is not the only possibility to obtain deformations by symmetric inner functions in the massless case. In fact, since the mass shell decomposes into two half light rays which are both invariant under Lorentz boosts, one can use {\em three} such functions $\varphi,\varphi_+,\varphi_-$ \cite{BischoffTanimoto:2013,LechnerSchlemmerTanimoto:2013}. Here $\varphi$ corresponds to a twist ``between the two light rays'' which mixes the two tensor factors as in \eqref{eq:MS0}, with the interpretation of wave scattering. The functions $\varphi_\pm$, on the other hand, are used to perform a construction of a ``deformed field theory'' on a line in close analogy to the (scalar) construction presented in Section~\ref{Section:S>BT} \cite{BostelmannLechnerMorsella:2011}. This amounts to deformations of the individual half line algebras $\M_0^\pm\leadsto (\M_0^\pm)_{\varphi_\pm}$, leaving the tensor product structure between them unchanged.

A combination of these effects appears in the short distance scaling limits of the massive two-dimensional integrable models. As argued in \cite{BostelmannLechnerMorsella:2011}, if for a regular, scalar S-matrix $S$ the limits $\lim_{\te\to\pm\infty}S(\te)$ exist, then the integrable model with mass $m>0$ and S-matrix $S$ has a well-defined short distance scaling limit (performed at the level of the generating fields $\Phi_S$, $\Phi_S'$). The resulting limit theory has zero mass and decomposes into a (possibly twisted) chiral tensor product, which differs from the free current triple \eqref{eq:BT-chiral} by the deformations $\M_0^\pm\leadsto(\M_0^\pm)_{{\varphi_\pm}}$, with $\varphi_\pm=S$, and a constant twist $\varphi=\pm1$ (this sign depends on the limit of $S$). At the time of writing, these limit theories were completely analyzed only in the simple cases $S=\pm1$ \cite{BostelmannLechnerMorsella:2011}. The main challenge is to analyze the relative commutants of the half line inclusions $(\M_0^\pm)_{\
varphi_\pm}(x_\pm)\subset (\M_0^\pm)_{\varphi_\pm}$, $x_\pm>0$, which, in contrast to the massive two-dimensional case, are not split.
\\\\
A common feature of all the constructions discussed in this chapter is the fact that they are based on data which respect the particle number: Factorizing S-matrices correspond to scattering processes without particle production, and the deformations based on Longo-Witten endomorphisms of standard pairs have a structure akin to second quantization. From this point of view, it is not surprising that the constructed models have the typical features of integrable models, and their higher-dimensional generalizations \cite{Lechner:2012} exhibit non-local features (see also the following section for such effects).

Extensions of this construction program to interactions with particle productions, as they are typical for relativistic quantum field theories in more than two dimensions \cite{Aks:1965}, require new ideas. In this context, it is promising to note that in \cite{BischoffTanimoto:2011}, Bischoff and Tanimoto already found a wave S-matrix which does not preserve the Fock structure of its representation space. Although this can not be seen as particle creation due to the considered theory being massless, it is an indication that substantial generalizations of the program presented here might well be possible.

\section{Deformations of quantum field theories by warped convolution}
\label{Section:Warping}

This section is devoted to another class of examples of operator-algebraic constructions of QFT models. As before, we follow the general approach of Section~\ref{Section:Abstract} and proceed by identifying certain Borchers triples. However, in contrast to the inverse scattering approach in Section~\ref{Section:IntegrableModels}, the dimension $d\geq2$ of Minkowski space is arbitrary here, and the input into the construction does not consist of an $S$-matrix, but rather of a representation of the translation group.

The point of view taken here is that of {\em deformation theory}, i.e. we will start with a given (arbitrary) Borchers triple $(\M,U,\Om)$, and then construct new triples by ``deforming'' the algebra\footnote{As we saw in Thm.~\ref{Theorem:WedgeAlgebrasAreTypeIII1Factors}, wedge algebras are always type III$_1$ factors, and are in fact expected to be isomorphic to the unique hyperfinite type III$_1$ factor in generic cases. Thus ``deforming $\M$'' does not mean deforming the (fixed) internal algebraic structure of $\M$, as in other algebraic deformation procedures \cite{Gerstenhaber:1964}. Rather, ``deforming $\M$'' means deforming/changing the position of $\M$ inside $\B(\Hil)$, i.e. deforming the subfactor $\M\subset\B(\Hil)$.} $\M$. In the situations we consider, $U$ and $\Om$ will be held fixed\footnote{However, it is entirely possible to ``deform'' Borchers triples by keeping the wedge algebra fixed while changing $U$ and $\Om$. On two-dimensional de Sitter space, this is the strategy followed by Barata,
 J\"akel, and Mund \cite{BarataJakelMund:2013}.}. Recently, several related deformation methods have been investigated (see, for example, \cite{Lechner:2012, Alazzawi:2012,Plaschke:2012}), of which the approach considered here is a particularly general and representative example.

The mathematical method to be used for this deformation, the warped convolution, has its roots in non-commutative geometry, and will be reviewed in Section~\ref{Section:Warping+Rieffel}. The application to Borchers triples is then presented in Section~\ref{Section:Warping-and-Borchers-Triples}.

\subsection{Warped convolutions and Rieffel deformations}
\label{Section:Warping+Rieffel}

In this section we concentrate on the mathematical backbone of the deformation procedure, and begin by recalling Rieffel's deformations of $C^*$-algebras (see \cite{Rieffel:1992} for Rieffel's original work, and \cite{Kasprzak:2009,LechnerWaldmann:2011} for generalizations in several different directions). 

The setting is a $C^*$-algebra $\frA$ with a strongly continuous automorphic $\Rl^d$-action~$\alpha$. One chooses a non-degenerate inner product on $\Rl^d$ which is normalized to determinant $\pm1$ and will be denoted $px$ for $p,x\in\Rl^d$. As a deformation parameter, we pick a real matrix $Q\in\Rl^{d\times d}_-$ which is skew-symmetric w.r.t. this inner product. In this setting, one wants to deform $\frA$ by introducing a new product, while keeping the linear structure and ${}^*$-involution unchanged. Motivated by the desire to formulate an abstract, $C^*$-algebraic version of quantization, one considers the integral expression
\begin{align}\label{eq:RieffelProduct}
	A\times_Q B
	=
	(2\pi)^{-d}\int_{\Rl^d}\int_{\Rl^d}dp\,dx\,e^{-ipx}\,\alpha_{Qp}(A)\alpha_x(B)\,,
\end{align}
which is strongly reminiscent of the Weyl-Moyal star product of deformation quantization\footnote{The Weyl-Moyal product (see, for example \cite{Gracia-BondiaVarilly:1988}) features prominently in deformation quantization \cite{Waldmann:2007}, where it describes the transition from classical mechanics to quantum mechanics. In that setting, one considers suitable number-valued functions $f,g$ on the simple classical phase space $\Rl^d$ ($d$ even) and equips them with the non-commutative product $$(f\star g)(y)=(2\pi)^{-d}\int_{\Rl^d}dp\int_{\Rl^d}dx\,e^{-ipx}f(y+\hbar\te p)g(y+x),$$where the antisymmetric matrix $\te$ is given by the Poisson bracket.}.
For $A,B$ in the dense subalgebra $\frA^\infty\subset\frA$ of all elements for which $x\mapsto\alpha_x(A)$ is smooth, the above integral exists in a precise oscillatory sense, defined as a limit of removing a smooth cutoff. This limit exists in the (Fr\'echet) topology of $\frA^\infty$, i.e. in particular $A\times_Q B\in\frA^\infty$. Furthermore, the new product $\times_Q$ is compatible with the star involution and  identity element (if it exists) of $\frA$, and there exists a $C^*$-norm $\|\cdot\|_Q$ on $(\frA^\infty,\times_Q)$, completing in which yields the Rieffel-deformed $C^*$-algebra $(\frA_Q,\times_Q)$. In this procedure, $Q$ plays the role of a deformation parameter: Setting $Q=0$ results in the old ``undeformed'' product $A\times_0 B=AB$, and the $C^*$-algebras $\frA_Q$ depend on $Q$ in a continuous manner (see \cite{Rieffel:1992} for details).
\\\\
The {\em warped convolution} was introduced in \cite{BuchholzSummers:2008} as a generalization of a deformation procedure in \cite{GrosseLechner:2007}. It was then thoroughly studied in \cite{BuchholzLechnerSummers:2011}, and we refer to this article for proofs of all the claims made in this section. Because of its origin in quantum field theory on non-commutative, ``quantized'' Minkowski spacetime \cite{GrosseLechner:2007}, it is closely related to the abstract quantization procedure of Rieffel\footnote{See also \cite{Neshveyev:2013} for another recent closely related approach, drawing from \cite{Kasprzak:2009}.}.

For defining the warped convolution, one uses a concrete setting instead of an abstract $C^*$-algebra, namely a Hilbert space $\Hil$ with a unitary strongly continuous representation $U$ of the translation group $\Rl^d$, $d\geq2$. As in the Rieffel setting, one fixes a non-degenerate inner product on $\Rl^d$ (in the applications to QFT, typically the Minkowski inner product), and a skew-symmetric real matrix $Q\in\Rl^{d\times d}_-$. 

It is then the aim to deform operators $A$ on $\Hil$ to new operators $A_Q$ on the same Hilbert space. That is, the action of $A,\Psi\to A\Psi$ on vectors $\Psi\in\Hil$ is changed to $A,\Psi\to A_Q\Psi$, but the product of operators is unchanged. As in the Rieffel setting, this is accomplished by an integral formula, namely
\begin{align}\label{eq:WarpingFormula}
	A_Q\Psi
	:=
	(2\pi)^{-d}\int_{\Rl^d}dp\int_{\Rl^d}dx\,e^{-ipx}\,U(Qp)AU(Qp)^{-1}U(x)\Psi
	\,.
\end{align}
If $A$ is smooth in the sense that $x\mapsto U(x)AU(x)^{-1}$ is smooth in norm (that is, $A\in\C^\infty$, where $\C\subset\B(\Hil)$ denotes the $C^*$-algebra on which the adjoint action of $U$ acts strongly continuously), and $\Psi$ is smooth in the sense that $x\mapsto U(x)\Psi$ is smooth ($\Psi\in\Hil^\infty$), then this integral exists in an oscillatory sense, and defines a smooth vector $A_Q\Psi\in\Hil^\infty$. Thus \eqref{eq:WarpingFormula} yields a densely defined (smooth) operator $A_Q$. It is easy to see that setting $Q=0$ results in the old operator, $A_0=A$, so that one can think of the mapping $A\mapsto A_Q$ as a deformation of operators. 

The mathematical status of these operators is settled in the following theorem.

\begin{theorem}
	Let $A\in\C^\infty$. Then the map $\Hil^\infty\ni\Psi\mapsto A_Q\Psi$ \eqref{eq:WarpingFormula} extends to bounded (smooth) operator $A_Q\in\C^\infty$. More precisely, the map $A\mapsto A_Q$ is a linear bijection of $\C^\infty$ onto itself, such that, $A\in\C^\infty$,
	\begin{align}
		\|A_Q\|=\|A\|_Q<\infty\,,
	\end{align}
	where $\|A\|_Q$ denotes the norm of the Rieffel-deformed $C^*$-algebra $\C_Q$, corresponding to the undeformed $C^*$-algebra $\C$ and action $\alpha_x(A)=U(x)AU(x)^{-1}$.
\end{theorem}

\begin{definition}
	Let $A\in\C^\infty$ and $Q\in\Rl^{d\times d}_-$. The warped convolution of $A$ is the operator $A_Q\in\C^\infty$ defined by extending \eqref{eq:WarpingFormula} to all of $\Hil$.
\end{definition}

Having defined the warped convolution (or ``warping'', for short), we now summarize its properties. We begin with the algebraic aspects.

\begin{theorem}\label{Theorem:Warping-Representation-Properties}
	The warped convolution extends to a representation of the Rieffel-deformed $C^*$-algebra $\C_Q$. In particular, 
	\begin{enumerate}
		\item\label{item:Warping-And-Rieffel-Product} Warping carries the operator product into the Rieffel product,
		\begin{align}
			A_QB_Q=(A\times_Q B)_Q\,,\qquad A,B\in\C^\infty\,,
		\end{align}
		\item\label{item:Warping-Adjoints} Warping commutes with taking adjoints, 
		\begin{align}
			{A_Q}^*={A^*}_Q\,,\qquad A\in\C^\infty\,.
		\end{align}
		\item Furthermore, any $U$-invariant operator $A=U(x)AU(x)^{-1}$ is invariant under warping, $A_Q=A$. In particular,
		\begin{align}
			1_Q=1\,.
		\end{align}
	\end{enumerate}
\end{theorem}

In the light of these results, the warped convolution seems very similar to Rieffel deformations, and in fact, it can be viewed as a module version of the Rieffel deformation \cite{LechnerWaldmann:2011}. But despite these similarities, warping has some advantages over the abstract Rieffel setting in application to QFT, in particular when it comes to comparing different deformation parameters $Q_1,Q_2$, or when spectral data of the translation representation are needed.

The following proposition, listing covariance properties of the warped convolution, shows that in typical QFT situations (in contrast to the situation in deformation quantization), several different deformation parameters must appear.

\begin{proposition}\label{Proposition:Warping-And-Covariance}
	Let $A\in\C^\infty$ be smooth and $Q\in\Rl^{d\times d}_-$.
	\begin{enumerate}
		\item The warping procedure is $U$-covariant,
		\begin{align}
			U(x)A_QU(x)^{-1}=\left(U(x)AU(x)^{-1}\right)_Q\,.
		\end{align}
		\item\label{item:Warping-And-The-Vacuum} If $\Om\in\Hil$ is a $U$-invariant vector, then
		\begin{align}
			A_Q\Om=A\Om\,.
		\end{align}
		\item\label{item:Warping-and-Lorentz-transformations} If $V$ is a unitary or antiunitary operator on $\Hil$ such that $VU(x)V^{-1}=U(Mx)$ for some invertible matrix $M\in{\rm GL}(d,\Rl)$, then 
		\begin{align}
			VA_QV^{-1}=\left(VAV^{-1}\right)_{\pm MQM^T}\,,
		\end{align}
		where $M^T$ is the transpose of $M$ w.r.t. the inner product used in the oscillatory factor in \eqref{eq:WarpingFormula}, and the sign is ``$+$'' for unitary $V$, and ``$-$'' for antiunitary $V$.
	\end{enumerate}
\end{proposition}

The last statement of this proposition applies in particular to the situation where we have a representation of the Poincar\'e group, and carry out the warped convolution with the representation of the translation subgroup. In that case, one sees that by introducing a warping w.r.t. some matrix $Q$, and insisting on Lorentz symmetry, one automatically ends up with all Lorentz transformed matrices $\La Q \La^T$, $\La\in\LGpo$ as well.

It will therefore be important to also consider situations where (at least) two different deformation parameters $Q_1,Q_2\in\Rl^{d\times d}_-$ enter. We first mention the following simple lemma.

\begin{lemma}
	Let $A\in\C^\infty$ and $Q_1,Q_2\in\Rl^{d\times d}_-$. Then 
	\begin{align}
		(A_{Q_1})_{Q_2}=A_{Q_1+Q_2}\,.
	\end{align}
	In particular, the inverse of the warping map $A\mapsto A_Q$ is given by the negative parameter, i.e. $A\mapsto A_{-Q}$.
\end{lemma}

In the context of a Borchers triple $(\M,U,\Om)$, we will consider operators $A\in\M$ and generate a new wedge algebra $\M_Q$ by the warped convolutions $A_Q$. Thinking of locality and commutators between observables in spacelike separated wedges (see Def.~\ref{Definition:BorchersTriple}~\ref{item:BorchersTriple-PoincareCondition}), it is clear that also commutators $[A_{Q_1},B_{Q_2}]$ between operators with different warping parameters will be relevant.

In general, commuting operators $[A,B]=0$ do not give rise to commuting warped convolutions, i.e. in general $[A_{Q_1},B_{Q_2}]\neq0$. This holds in particular for $Q_1=Q_2$, as is clear from the origin of warping in quantization, where a commutative algebra is deformed into a non-commutative one. However, there does exist an interesting commutation theorem for warped convolutions with {\em opposite} deformation parameters $Q$ and $-Q$. In this theorem, due to Buchholz and Summers \cite{BuchholzSummers:2008}, a spectral condition enters. To understand the relevance of this spectral information, let us first  mention that the warped convolution can also be expressed as
\begin{align}\label{eq:Warping-Spectral}
	A_Q
	=
	\int dE(x) \, \alpha_{Qx}(A)
	=
	\int \alpha_{Qx}(A) dE(x)
	\,,
\end{align}
where $E$ is the joint spectral measure of the generators $P_0,...,P_{d-1}$ of the translations $U(x)=\exp iP_\mu x^\mu$ (with support the joint spectrum, denoted ${\rm Sp}\,U$), and $\alpha_x(A)=U(x)AU(x)^{-1}$ as before. Both the above integrals can be defined as strong limits of suitably cut off expressions, and then coincide with each other and the warped convolution $A_Q$.

Whereas in most calculations, the oscillatory form \eqref{eq:WarpingFormula} is more convenient to work with, the spectral integral form \eqref{eq:Warping-Spectral} is better suited for establishing the following compatibility result on the warped convolution $A\mapsto A_Q$ and its inverse $A\mapsto A_{-Q}$. 

\begin{proposition}\label{Proposition:Warping-Commutation-Theorem}
	Let $A,B\in\C^\infty$ be operators such that $[\alpha_{Qp}(A), \alpha_{-Qq}(B)] = 0$ for all $p,q\in {\rm Sp} \, U$. Then
	\begin{align}
		[ A_Q, B_{-Q}] = 0 \,.
	\end{align}
\end{proposition}

This result completes the list of properties of the warped convolution that we will need in the next subsection, where application to deformations of Borchers triples are discussed. Besides that application, warped convolutions have by now also been used in a variety of other situations, such as Wightman QFT \cite{GrosseLechner:2008}, QFT on curved spacetimes \cite{Morfa-Morales:2012}, Wick rotation on non-commutative spacetime \cite{GrosseLechnerLudwigVerch:2013}, deformations of quantum mechanical Hamiltonians \cite{Much:2013_2}, and quantum measurement theory \cite{Andersson:2013}.

\subsection{Borchers triples and warped convolutions}
\label{Section:Warping-and-Borchers-Triples}

We now apply the warped convolution to deformations of Borchers triples. We therefore start with a fixed but arbitrary $d$-dimensional Borchers triple $(\M,U,\Om)$, $d\geq2$, according to Def.~\ref{Definition:BorchersTriple}. The warped convolution will always be carried out w.r.t. the translations $U(x,1)$, $x\in\Rl^d$, from the representation $U$ of the triple, the inner product in the oscillatory integrals will be the Minkowski inner product, and the deformation parameter $Q$ skew-symmetric w.r.t. this inner product.
\\\\
The basic idea for deforming $(\M,U,\Om)$ is to keep $U$ and $\Om$ unchanged (as in Section~\ref{Section:IntegrableModels}), and replace $\M$ by an algebra containing ``all $A_Q$, $A\in\M$''. Some remarks are in order here: First, one checks that as a consequence of  its half-sided translational invariance, the wedge algebra $\M$ contains a strongly dense subalgebra $\M^\infty=\M\cap\C^\infty$ of smooth elements. Thus the warped convolutions $A_Q$ are well-defined for sufficiently many $A\in\M$. However, products of such operators are typically not of the form $C_Q$ for some $C\in\M$. This is so because $A_QB_Q=(A\times_Q B)_Q$ (Theorem~\ref{Theorem:Warping-Representation-Properties}~\ref{item:Warping-And-Rieffel-Product}), and $A\times_Q B$ involves integration over translations in all directions \eqref{eq:RieffelProduct}. Thus the set $\{A_Q\,:\,A\in\M^\infty\}$ is typically not an algebra. One therefore passes to the algebra generated, and writes (with a slight abuse of notation)
\begin{align}\label{eq:Deformed-Wedge-Algebra}
	\M_Q:=\{A_Q\,:\,A\in\M^\infty\}''\,.
\end{align}
Note that in view of Thm.~\ref{Theorem:Warping-Representation-Properties}~\ref{item:Warping-Adjoints}, $\M_Q$ is a von Neumann algebra, and using Proposition~\ref{Proposition:Warping-And-Covariance}~\ref{item:Warping-And-The-Vacuum}, one can show that it has $\Om$ as a cyclic vector. 

In order for the so defined triple $(\M_Q,U,\Om)$ to be a Borchers triple, the deformation parameter $Q$ has to be chosen in a suitable manner, essentially by adapting it to the geometry of the right wedge $W_R$. Looking at Def.~\ref{Definition:BorchersTriple}, it is clear that we only have to check the conditions in part~\ref{item:BorchersTriple-PoincareCondition}. Since (orthochronous) Poincar\'e transformations $U(x,\La)$ act on warped convolutions according to (cf.~Proposition~\ref{Proposition:Warping-And-Covariance}~\ref{item:Warping-and-Lorentz-transformations})
\begin{align}
	U(x,\La)A_QU(x,\La)^{-1}
	=
	\left(U(x,\La)AU(x,\La)^{-1}\right)_{\La Q\La^{-1}}
	\,,
\end{align}
and this is required to be an element of $\M_Q$ for $\La W_R+x\subset W_R$, we should choose $Q$ in such a way that $\La Q\La^{-1}=Q$ for any Lorentz transformation $\La$ with $\La W_R=W_R$. Furthermore, since the only commutativity result we have for warped convolutions is Proposition~\ref{Proposition:Warping-Commutation-Theorem}, we ought to choose $Q$ in such a way that also $\La Q\La^{-1}=-Q$ for any Lorentz transformation $\La$ mapping $W_R$ onto its causal complement, $\La W_R=-W_R$. Finally, we have to take into account the energy momentum spectrum, lying in the forward light cone, which appears in Proposition~\ref{Proposition:Warping-Commutation-Theorem}. These considerations suggest to consider only {\em admissible} matrices $Q$, defined by the following three conditions:
\begin{enumerate}
	\item[i)] $\La Q\La^{-1}=Q$ for any $\La\in\LGpo$ such that $\La W_R=W_R$,
	\item[ii)] $\La Q\La^{-1}=-Q$ for any $\La\in\LGpo$ such that $\La W_R=-W_R$, and $jQj=Q$ (with $j$ the reflection at the edge of $W_R$),
	\item[iii)] $Q\overline{V^+}\subset\overline{W_R}$.
\end{enumerate}
\vspace*{2mm}
These conditions drastically constrain the form of $Q$. In fact, $Q$ is admissible if and only if  \cite{GrosseLechner:2007}

\begin{equation} \label{q1}
Q
=
\left( \begin{array}{ccccc}  0 & \kappa & 0 & \cdots & {0} \\ 
                       \kappa &    0   & 0 & \cdots & {0} \\
                            0 &    0   & 0 & \cdots & {0} \\    
                       \vdots & \vdots & \vdots & \ddots &  \vdots \\
                       {0} &    {0}   & {0} & \cdots & {0} \end{array} \right)
\end{equation}
for some $\kappa\geq 0$, in case the spacetime dimension is $d\neq4$. In the physically most 
interesting case $d=4$, there is a little more freedom\footnote{This is related to the fact that $Q$ is skew-symmetric and for $d>4$, the edge of $W_R$ is fixed by the subgroup SO$(d-2)$ of rotations in the edge.} for choosing $Q$. In this case, $Q$ is admissible if and only if
\begin{equation*}  \label{q2}
Q
=
\left( \begin{array}{cccc}  0 & \kappa & 0 & 0 \\ 
                          \kappa   &  0 & 0 & 0 \\
                            0 &    0   & 0 & \kappa' \\
                            0 &    0   & -\kappa' & 0 \end{array} \right)\,,
\end{equation*}
with parameters $\kappa\geq0$, $\kappa'\in\Rl$.
\vspace*{4mm}

The condition (iii) on admissible $Q$'s implies that for $A\in\M$ (localized in $W_R$) and $B\in\M'$ (localized in $-W_R$), one does not only have $[A,B]=0$, but even $[\alpha_{Qp}(A),\alpha_{-Qq}(B)]=0$ for all $p,q\in\overline{V^+}$. Since the energy-momentum spectrum is contained in $\overline{V^+}$, we are in the situation to apply Proposition~\ref{Proposition:Warping-Commutation-Theorem}, which is the key to the following result.
\begin{theorem}
	Let $(\M,U,\Omega)$ be a $d$-dimensional Borchers triple, $d\geq2$, and let $Q$ be admissible (see above).
	\begin{enumerate}
		\item The warped triple $(\M_Q,U,\Omega)$ is again a $d$-dimensional Borchers triple.
		\item The modular data $\Delta_Q,J_Q$ of $(\M_Q,\Om)$ coincide with those of the original triple, i.e.
		\begin{align}
			\Delta_Q = \Delta\,,\qquad J_Q = J\,. 
		\end{align}
		\item The commutant of the deformed wedge algebra is the inverse deformation of the original commutant,
		\begin{align*}
			{\M_Q}'={\M'}_{-Q}\,.
		\end{align*}
	\end{enumerate}
\end{theorem}

In view of this theorem, warping yields a one-parameter family (two-parameter family in dimension $d=4$) of ``new'' Borchers triples, which can again be taken to generate QFT models. In particular, when applying this method to a Borchers triple which is explicitly known (such as that of a free field theory, or of any other completely constructed QFT), it yields new QFT models in which $Q$ plays the role of a coupling constant.
\\\\
In comparison to the construction in Section~\ref{Section:IntegrableModels}, where we started from an S-matrix, the input into the construction by warped convolution is of a more abstract nature. It is therefore necessary to investigate the physical properties of these models. To begin with, there is the question how much the theory given by the deformed triple $(\M_Q,U,\Om)$ depends on $Q$. Since the algebras $\M_Q$ are generically expected to be isomorphic to each other (generically being isomorphic to the unique hyperfinite type III$_1$ factor), any argument for showing that the Borchers triples $(\M,U,\Om)$ and $(\M_Q,U,\Om)$, $Q\neq0$, are not equivalent, must also take into account the representation $U$. Indirect arguments to this effect exist. 

On the one hand, one can consider massive theories, in which scattering states can be constructed. Then, using methods developed in \cite{BorchersBuchholzSchroer:2001}, one can show that the S-matrix elements of collision processes with two incoming and two outgoing particles depend on $Q$ via a factor $e^{ipQq}$ (with $p,q$ the momenta of the incoming particles) \cite{GrosseLechner:2007, BuchholzSummers:2008}. Hence at least in this case, one does have a true dependence on $Q$, and inequivalence of the models with different values of $Q$. In particular, one sees in the scattering data that the warping procedure changes the interaction of the model under consideration\footnote{There is also a sharp difference between the undeformed and deformed case in the thermal equilibrium states. In the deformed $(Q\neq0$) situation, the thermal representation leads to a decoupling of the noncommutativity parameters $\La Q\La^{-1}$ related to different wedges \cite{LechnerSchlemmer:2015}.}.

For comparison with the approach taken in Section~\ref{Section:IntegrableModels}, it is also instructive to consider the two-dimensional case. In this case, the deformation parameter must be of the form
\begin{equation*}  \label{q2}
Q
=
\left( \begin{array}{cc}  0 & \kappa \\ 
                          \kappa   &  0
                          \end{array} \right)\,,\qquad \kappa\geq0\,.
\end{equation*}
In the case of the representation $U=U_m$ given by the second quantization of the irreducible $U_{1,m}$ with fixed mass $m>0$ \eqref{eq:U1m}, and using the rapidity parametrization \eqref{eq:palpha}, one notes that 
\begin{align}\label{eq:NC-S}
	e^{ip_m(\te')Qp_m(\te)}=e^{i\kappa m^2\sinh(\te-\te')}
	=:S(\te-\te')
\end{align}
is a (scalar) S-matrix in the sense of Definition~\ref{definition:SMatrix}. In fact, one can show that for the Borchers triple $(\M,U,\Om)$ of the massive scalar field in two dimensions, its deformation $(\M_Q,U,\Om)$ gives rise to a QFT which is unitarily equivalent to the integrable model constructed in Section~\ref{Section:IntegrableModels}, with the S-matrix \eqref{eq:NC-S} \cite{GrosseLechner:2007}.

Note, however, that this S-matrix is {\em not} regular in the sense of Definition~\ref{Definition:RegularS}, because it is unbounded on any strip $\Strip(-\eps,\pi+\eps)$, $\eps>0$. Hence the modular nuclearity results of Section~\ref{Section:Nuc} do not apply, so that we cannot conclude existence of local observables for this model.

The models obtained by warped convolution have certain non-local features in general, which can also be seen by other arguments: In application to Wightman QFT, one observes that warping modifies the $n$-point functions in a non-local manner \cite{Soloviev:2008,GrosseLechner:2008}. In the general case, one can show (in dimension $d\geq3$, and under a mild assumption on the energy-momentum spectrum, see \cite{BuchholzLechnerSummers:2011}) that if one starts from a QFT which has the vacuum $\Om$ as a cyclic vector for its double cone algebras, then $\Om$ will {\em not} be cyclic for the double cone algebras of the QFT generated by the deformed Borchers triple $(\M_Q,U,\Om)$, $Q\neq0$.

These non-local aspects of the warped models show that the deformation procedure employed here is still too simple to yield realistic quantum field theory models in physical spacetime. To find deformation methods which are also compatible with strict localization in four dimensions is the subject of ongoing research.

In conclusion, we also mention how the inherent non-locality can be understood by considering the origin of the warped convolution in quantum field theory on noncommutative Minkowski space~\cite{GrosseLechner:2007}: The initial motivation to consider the warped convolution was to define quantum field operators on a space in which the coordinates $X_0,...,X_{d-1}$ do not commute, but rather satisfy a relation of the form
\begin{align}
    [X_\mu,X_\nu]=i\,Q_{\mu\nu}\,,\qquad [Q_{\mu\nu},X_\kappa]=0\,,
\end{align}
where $Q$ is either a (non-zero) matrix times an identity element in the algebra generated by the $X_\mu$, or an operator with spectrum consisting of a Lorentz orbit of such matrices \cite{DoplicherFredenhagenRoberts:1995,BahnsDoplicherMorsellaPiacitelli:2015}. On a space with non-commuting coordinates, localization in bounded regions is impossible. This fact is reflected in the weaker than usual localization properties of the algebras constructed by warped convolution, and shows that more refined deformation methods are needed to obtain strictly local quantum field theories in higher dimensions.

\vspace*{5mm}
\noindent{\bf Acknowledgements.} {The publications reviewed in this article include joint work with my colleagues S.~Alazzawi, H.~Bostelmann, D.~Buchholz, C.~Dappiaggi, H.~Grosse, R.~Longo, T.~Ludwig, E.~Morfa-Morales, G.~Morsella, J.~Schlemmer, C.~Sch\"utzenhofer, S.~J.~Summers, Y.~Tanimoto, R.~Verch, and S.~Waldmann. I wish to thank them all for fruitful and enjoyable collaborations.}

\footnotesize
\addcontentsline{toc}{section}{References}

\newcommand{\etalchar}[1]{$^{#1}$}


\begin{thebibliography}{KTTW77}
\providecommand{\href}[1]{\texttt{#1}{link}}

\bibitem[AAR91]{AbdallaAbdallaRothe:1991}
E.~Abdalla, C.~Abdalla, and K.~Rothe.
\newblock {\em {Non-perturbative methods in 2-dimensional quantum field
  theory}}.
\newblock World Scientific, 1991.

\bibitem[AFZ79]{ArinshteinFateevZamolodchikov:1979}
A.~E. Arinshtein, V.~A. Fateev, and A.~B. Zamolodchikov.
\newblock {Quantum S-Matrix of the (1+1)-Dimensional Toda Chain}.
\newblock \href{http://dx.doi.org/10.1016/0370-2693(79)90561-6}{{\em Phys.
  Lett.} \textbf{B87} (1979)  389--392}.

\bibitem[{\AA}ks65]{Aks:1965}
S.~{\AA}ks.
\newblock {Proof that scattering implies production in quantum field theory}.
\newblock \href{http://dx.doi.org/10.1063/1.1704305}{{\em J. Math. Phys.}
  \textbf{6} (1965)  516--532}.

\bibitem[Ala13]{Alazzawi:2012}
S.~Alazzawi.
\newblock {Deformations of Fermionic Quantum Field Theories and Integrable
  Models}.
\newblock \href{http://dx.doi.org/10.1007/s11005-012-0576-3}{{\em Lett. Math.
  Phys.} \textbf{103} (2013) no.~1, 37--58}.
\newblock \href{http://arxiv.org/abs/1203.2058v1}{\texttt{[open access]}}.

\bibitem[Ala14]{Alazzawi:2014}
S.~Alazzawi.
\newblock {\em {Deformations of Quantum Field Theories and the Construction of
  Interacting Models}}.
\newblock PhD thesis, University of Vienna, 2014.
\newblock \href{http://www.arxiv.org/abs/1503.00897}{\texttt{[open access]}},
  arXiv:1503.00897.

\bibitem[And13]{Andersson:2013}
A.~Andersson.
\newblock {Operator Deformations in Quantum Measurement Theory}.
\newblock {\em Lett. Math. Phys.} \textbf{104} (2013) no.~4, 415--430.
\newblock \href{http://arxiv.org/abs/1304.2806v2}{\texttt{[open access]}}.

\bibitem[Ara63]{Araki:1963}
H.~Araki.
\newblock {A Lattice of von {N}eumann algebras Associated with the Quantum
  Theory of a Free {B}ose Field}.
\newblock {\em J. Math. Phys.} \textbf{4} (1963) no.~11, 1343--1362.

\bibitem[Ara64]{Araki:1964}
H.~Araki.
\newblock {Von {N}eumann Algebras of Local Observables for Free Scalar Field}.
\newblock {\em J. Math. Phys.} \textbf{5} (1964) no.~1, 1--13.

\bibitem[Ara99]{Araki:1999}
H.~Araki.
\newblock {\em {Mathematical Theory of Quantum Fields}}.
\newblock {Int. Series of Monographs on Physics}. Oxford University Press,
  Oxford, 1999.

\bibitem[BB99]{BorchersBuchholz:1999}
H.~Borchers and D.~Buchholz.
\newblock {Global properties of vacuum states in de Sitter space}.
\newblock {\em Annales Poincare Phys. Theor.} \textbf{A70} (1999)  23--40.
\newblock \href{http://arxiv.org/abs/gr-qc/9803036}{\texttt{[open access]}}.

\bibitem[BBS01]{BorchersBuchholzSchroer:2001}
H.~Borchers, D.~Buchholz, and B.~Schroer.
\newblock {Polarization-free generators and the {S}-matrix}.
\newblock \href{http://dx.doi.org/10.1007/s002200100411}{{\em Comm. Math.
  Phys.} \textbf{219} (2001)  125--140}.
\newblock \href{http://arxiv.org/abs/hep-th/0003243}{\texttt{[open access]}}.

\bibitem[BC12]{BostelmannCadamuro:2012}
H.~Bostelmann and D.~Cadamuro.
\newblock {An operator expansion for integrable quantum field theories}.
\newblock \href{http://dx.doi.org/10.1088/1751-8113/46/9/095401}{{\em J. Phys.
  A: Math. Theor.} \textbf{46} (2012)  095401}.

\bibitem[BC14]{BostelmannCadamuro:2014}
H.~Bostelmann and D.~Cadamuro.
\newblock {Characterization of local observables in integrable quantum field
  theories}.
\newblock {\em Preprint} (2014)  .
\newblock arXiv:1402.6127v1,
  \href{http://arxiv.org/abs/1402.6127v1}{\texttt{[open access]}}.

\bibitem[BC15]{BostelmannCadamuro:2015_2}
H.~Bostelmann and D.~Cadamuro.
\newblock {Negative energy densities in integrable quantum field theories at
  one-particle level}.
\newblock {\em Preprint arXiv:1502.01714} (2015)  .
\newblock \href{http://www.arxiv.org/abs/1502.01714}{\texttt{[open access]}}.

\bibitem[BCF13]{BostelmannCadamuroFewster:2013}
H.~Bostelmann, D.~Cadamuro, and C.~J. Fewster.
\newblock {Quantum energy inequality for the massive Ising model}.
\newblock {\em Physical Review D} \textbf{88} (2013) no.~2, 025019.
\newblock
  \href{http://journals.aps.org/prd/abstract/10.1103/PhysRevD.88.025019;
  http://arxiv.org/pdf/1304.7682}{\texttt{[open access]}}.

\bibitem[BDF{\etalchar{+}}09]{BrunettiDuetschFredenhagen:2009}
R.~Brunetti, M.~D\"{u}tsch, K.~Fredenhagen, et~al.
\newblock {Perturbative algebraic quantum field theory and the renormalization
  groups}.
\newblock {\em Advances in Theoretical and Mathematical Physics} \textbf{13}
  (2009) no.~5, 1541--1599.
\newblock \href{http://arxiv.org/abs/0901.2038}{\texttt{[open access]}}.

\bibitem[BDL90a]{BuchholzDAntoniLongo:1990}
D.~Buchholz, C.~D'Antoni, and R.~Longo.
\newblock {Nuclear Maps and Modular Structures 2: Applications to Quantum Field
  Theory}.
\newblock {\em Comm. Math. Phys.} \textbf{129} (1990)  115.

\bibitem[BDL90b]{BuchholzDAntoniLongo:1990-1}
D.~Buchholz, C.~D'Antoni, and R.~Longo.
\newblock {Nuclear maps and modular structures. I. General properties.}
\newblock {\em J. Funct. Anal.} \textbf{88} (1990)  233--250.

\bibitem[BDL07]{BuchholzDAntoniLongo:2007}
D.~Buchholz, C.~D'Antoni, and R.~Longo.
\newblock {Nuclearity and Thermal States in Conformal Field Theory}.
\newblock {\em Comm. Math. Phys.} \textbf{270} (2007)  267--293.
\newblock \href{http://arxiv.org/abs/math-ph/0603083}{\texttt{[open access]}}.

\bibitem[BDM09]{BostelmannDAntoniMorsella:2009}
H.~Bostelmann, C.~D'Antoni, and G.~Morsella.
\newblock {Scaling algebras and pointlike fields: A nonperturbative approach to
  renormalization}.
\newblock {\em Comm. Math. Phys.} \textbf{285} (2009)  763--798.
\newblock \href{http://arxiv.org/abs/0711.4237}{\texttt{[open access]}}.

\bibitem[BDMP15]{BahnsDoplicherMorsellaPiacitelli:2015}
D.~Bahns, S.~Doplicher, G.~Morsella, and G.~Piacitelli.
\newblock {Quantum Spacetime and Algebraic Quantum Field Theory}.
\newblock {\em arXiv:1501.03298} (2015)  .
\newblock \href{http://arxiv.org/abs/1501.03298}{\texttt{[open access]}}.

\bibitem[BF82]{BuchholzFredenhagen:1982}
D.~Buchholz and K.~Fredenhagen.
\newblock {Locality and the Structure of Particle States}.
\newblock {\em Comm. Math. Phys.} \textbf{84} (1982)  1.
\newblock \href{http://projecteuclid.org/euclid.cmp/1103921044}{\texttt{[open
  access]}}.

\bibitem[BFK06a]{BabujianFoersterKarowski:2012}
H.~M. Babujian, A.~Foerster, and M.~Karowski.
\newblock {SU(N) and O(N) off-shell nested Bethe ansatz and form factors}.
\newblock {\em arXiv:hep-th/0611012} (2006)  .
\newblock \href{http://www.arxiv.org/abs/hep-th/0611012}{\texttt{[open
  access]}}.

\bibitem[BFK06b]{BabujianFoersterKarowski:2006}
H.~M. Babujian, A.~Foerster, and M.~Karowski.
\newblock {The Form Factor Program: a Review and New Results - the Nested SU(N)
  Off-Shell Bethe Ansatz}.
\newblock {\em SIGMA} \textbf{2} (2006)  082.
\newblock \href{http://arxiv.org/abs/hep-th/0609130}{\texttt{[open access]}}.

\bibitem[BGL02]{BrunettiGuidoLongo:2002}
R.~Brunetti, D.~Guido, and R.~Longo.
\newblock {Modular localization and Wigner particles}.
\newblock \href{http://dx.doi.org/10.1142/S0129055X02001387}{{\em Rev. Math.
  Phys.} \textbf{14} (2002)  759--786}.
\newblock \href{http://arxiv.org/abs/math-ph/0203021}{\texttt{[open access]}}.

\bibitem[BH94]{BalogHauer:1994}
J.~Balog and T.~Hauer.
\newblock {Polynomial form-factors in the O(3) nonlinear sigma model}.
\newblock {\em Phys. Lett.} \textbf{B337} (1994)  115--121.

\bibitem[Bis12]{Bischoff:2012}
M.~Bischoff.
\newblock {\em {Construction Of Models In Low-Dimensional Quantum Field Theory
  Using Operator Algebraic Methods}}.
\newblock PhD thesis, Rome, 2012.

\bibitem[BJ89]{BuchholzJunglas:1989}
D.~Buchholz and P.~Junglas.
\newblock {On the Existence of Equilibrium States in Local Quantum Field
  Theory}.
\newblock {\em Comm. Math. Phys.} \textbf{121} (1989)  255--270.

\bibitem[BJL95]{BaumgrtelJurkeLled:1995}
H.~Baumg\"{a}rtel, M.~Jurke, and F.~Lled\'{o}.
\newblock {On Free Nets Over Minkowski Space}.
\newblock \href{http://dx.doi.org/10.1016/0034-4877(96)83512-9}{{\em Rept.
  Math. Phys.} \textbf{35} (1995) no.~1, 101--127}.
\newblock
  \href{http://www.sciencedirect.com/science?_ob=ArticleURL&_udi=B6VN0-3YMWKX2-J&_user=162904&_rdoc=1&_fmt=&_orig=search&_sort=d&view=c&_acct=C000013078&_version=1&_urlVersion=0&_userid=162904&md5=eda241c67a6639fb4bd486d8ad51ecdc}{\texttt{[open
  access]}}.

\bibitem[BJL02]{BaumgartelJurkeLledo:2002}
H.~Baumg\"{a}rtel, M.~Jurke, and F.~Lled\'{o}.
\newblock {Twisted duality of the CAR algebra}.
\newblock \href{http://dx.doi.org/10.1063/1.1483376}{{\em J. Math. Phys.}
  \textbf{43} (2002)  4158--4179}.

\bibitem[BJM13]{BarataJakelMund:2013}
J.~C.~A. Barata, C.~D. J\"{a}kel, and J.~Mund.
\newblock {The ${\mathscr P}(\varphi)_2$ Model on the de Sitter Space}.
\newblock {\em arXiv:1311.2905v1} (2013)  .
\newblock \href{http://arxiv.org/abs/1311.2905v1}{\texttt{[open access]}}.

\bibitem[BK04]{BabujianKarowski:2004}
H.~M. Babujian and M.~Karowski.
\newblock {Towards the construction of Wightman functions of integrable quantum
  field theories}.
\newblock {\em Int. J. Mod. Phys.} \textbf{A19S2} (2004)  34--49.
\newblock \href{http://arxiv.org/abs/hep-th/0301088}{\texttt{[open access]}}.

\bibitem[BKW79]{BergKarowskiWeisz:1979}
B.~Berg, M.~Karowski, and P.~Weisz.
\newblock {Construction of Green's Functions from an Exact S-Matrix}.
\newblock {\em Phys. Rev. D} \textbf{19} (1979) no.~8, 2477--2479.

\bibitem[BL04]{BuchholzLechner:2004}
D.~Buchholz and G.~Lechner.
\newblock {Modular nuclearity and localization}.
\newblock \href{http://dx.doi.org/10.1007/s00023-004-0190-8}{{\em Annales Henri
  Poincar\'{e}} \textbf{5} (2004)  1065--1080}.
\newblock \href{http://arxiv.org/abs/math-ph/0402072}{\texttt{[open access]}}.

\bibitem[BLM11]{BostelmannLechnerMorsella:2011}
H.~Bostelmann, G.~Lechner, and G.~Morsella.
\newblock {Scaling limits of integrable quantum field theories}.
\newblock \href{http://dx.doi.org/10.1142/S0129055X11004539}{{\em Rev. Math.
  Phys.} \textbf{23} (2011) no.~10, 1115--1156}.
\newblock \href{http://arxiv.org/abs/1105.2781}{\texttt{[open access]}}.

\bibitem[BLS11]{BuchholzLechnerSummers:2011}
D.~Buchholz, G.~Lechner, and S.~J. Summers.
\newblock {Warped Convolutions, Rieffel Deformations and the Construction of
  Quantum Field Theories}.
\newblock \href{http://dx.doi.org/10.1007/s00220-010-1137-1}{{\em Comm. Math.
  Phys.} \textbf{304} (2011)  95--123}.
\newblock \href{http://arxiv.org/abs/1005.2656}{\texttt{[open access]}}.

\bibitem[BMS01]{BuchholzMundSummers:2001}
D.~Buchholz, J.~Mund, and S.~J. Summers.
\newblock {Transplantation of Local Nets and Geometric Modular Action on
  Robertson-Walker Space-Times}.
\newblock {\em Fields Inst. Commun.} \textbf{30} (2001)  65--81.
\newblock \href{http://arxiv.org/abs/hep-th/0011237}{\texttt{[open access]}}.

\bibitem[Bor92]{Borchers:1992}
H.~Borchers.
\newblock {The CPT theorem in two-dimensional theories of local observables}.
\newblock \href{http://dx.doi.org/10.1007/BF02099011}{{\em Comm. Math. Phys.}
  \textbf{143} (1992)  315--332}.
\newblock \href{http://projecteuclid.org/euclid.cmp/1104248958}{\texttt{[open
  access]}}.

\bibitem[Bor00]{Borchers:2000}
H.~Borchers.
\newblock {On revolutionizing quantum field theory with Tomita's modular
  theory}.
\newblock {\em J. Math. Phys.} \textbf{41} (2000)  3604--3673.

\bibitem[Bos05a]{Bostelmann:2005-2}
H.~Bostelmann.
\newblock {Operator product expansions as a consequence of phase space
  properties}.
\newblock \href{http://dx.doi.org/10.1063/1.2007567}{{\em J.Math.Phys.}
  \textbf{46} (2005)  082304}.
\newblock \href{http://arxiv.org/abs/math-ph/0502004v3}{\texttt{[open
  access]}}.

\bibitem[Bos05b]{Bostelmann:2005}
H.~Bostelmann.
\newblock {Phase space properties and the short distance structure in quantum
  field theory}.
\newblock {\em J. Math. Phys.} \textbf{46} (2005)  052301.
\newblock \href{http://arxiv.org/abs/math-ph/0409070}{\texttt{[open access]}}.

\bibitem[BP90]{BuchholzPorrmann:1990}
D.~Buchholz and M.~Porrmann.
\newblock {How Small Is the Phase Space in Quantum Field Theory?}
\newblock {\em Annales Poincare Phys.Theor.} \textbf{52} (1990)  237.

\bibitem[BR87]{BratteliRobinson:1987}
O.~Bratteli and D.~W. Robinson.
\newblock {\em {Operator Algebras and Quantum Statistical Mechanics I}}.
\newblock Springer, 1987.

\bibitem[BR97]{BratteliRobinson:1997}
O.~Bratteli and D.~W. Robinson.
\newblock {\em {Operator Algebras and Quantum Statistical Mechanics II}}.
\newblock Springer, 1997.

\bibitem[BS04]{BuchholzSummers:2004-2}
D.~Buchholz and S.~J. Summers.
\newblock {Stable quantum systems in anti-de Sitter space: Causality,
  independence and spectral properties}.
\newblock \href{http://dx.doi.org/10.1063/1.1804230}{{\em J. Math. Phys.}
  \textbf{45} (2004)  4810--4831}.
\newblock \href{http://www.arxiv.org/abs/math-ph/0407011}{\texttt{[open
  access]}}.

\bibitem[BS08]{BuchholzSummers:2008}
D.~Buchholz and S.~J. Summers.
\newblock {Warped Convolutions: A Novel Tool in the Construction of Quantum
  Field Theories}.
\newblock In E.~Seiler and K.~Sibold, editors, {\em {Quantum Field Theory and
  Beyond: Essays in Honor of Wolfhart Zimmermann}}, pages 107--121. World
  Scientific, 2008.
\newblock \href{http://arxiv.org/abs/0806.0349}{\texttt{[open access]}}.

\bibitem[BT13]{BischoffTanimoto:2011}
M.~Bischoff and Y.~Tanimoto.
\newblock {Construction of wedge-local nets of observables through Longo-Witten
  endomorphisms. II}.
\newblock \href{http://dx.doi.org/10.1007/s00220-012-1593-x}{{\em Comm. Math.
  Phys.} \textbf{317} (2013) no.~3, 667--695}.
\newblock \href{http://arxiv.org/abs/1111.1671v1}{\texttt{[open access]}}.

\bibitem[BT15]{BischoffTanimoto:2013}
M.~Bischoff and Y.~Tanimoto.
\newblock {Integrable QFT and Longo-Witten endomorphisms}.
\newblock \href{http://dx.doi.org/10.1007/s00023-014-0337-1}{{\em Annales Henri
  Poincar\'{e}} \textbf{16} (2015) no.~2, 569--608}.
\newblock \href{http://arxiv.org/abs/1305.2171}{\texttt{[open access]}}.

\bibitem[Buc74]{Buchholz:1974-2}
D.~Buchholz.
\newblock {Product States for Local Algebras}.
\newblock {\em Comm. Math. Phys.} \textbf{36} (1974)  287--304.

\bibitem[Buc75a]{Buchholz:1975-1}
D.~Buchholz.
\newblock {Collision Theory for Massless Fermions}.
\newblock {\em Comm. Math. Phys.} \textbf{42} (1975)  269.

\bibitem[Buc75b]{Buchholz:1975}
D.~Buchholz.
\newblock {Collision Theory for Waves in Two Dimensions and a Characterization
  of Models with Trivial S-Matrix}.
\newblock {\em Comm. Math. Phys.} \textbf{45} (1975)  1--8.

\bibitem[Buc77]{Buchholz:1977}
D.~Buchholz.
\newblock {Collision Theory for Massless Bosons}.
\newblock {\em Comm. Math. Phys.} \textbf{52} (1977)  147.
\newblock \href{http://projecteuclid.org/euclid.cmp/1103900494}{\texttt{[open
  access]}}.

\bibitem[BV98]{BuchholzVerch:1998}
D.~Buchholz and R.~Verch.
\newblock {Scaling algebras and renormalization group in algebraic quantum
  field theory. II: Instructive examples}.
\newblock {\em Rev. Math. Phys.} \textbf{10} (1998)  775--800.
\newblock \href{http://arxiv.org/abs/hep-th/9708095}{\texttt{[open access]}}.

\bibitem[BW75]{BisognanoWichmann:1975}
J.~J. Bisognano and E.~H. Wichmann.
\newblock {On the Duality Condition for a Hermitian Scalar Field}.
\newblock {\em J. Math. Phys.} \textbf{16} (1975)  985--1007.

\bibitem[BW76]{BisognanoWichmann:1976}
J.~J. Bisognano and E.~H. Wichmann.
\newblock {On the Duality Condition for Quantum Fields}.
\newblock {\em J. Math. Phys.} \textbf{17} (1976)  303--321.

\bibitem[BW86]{BuchholzWichmann:1986}
D.~Buchholz and E.~H. Wichmann.
\newblock {Causal Independence and the Energy Level Density of States in Local
  Quantum Field Theory}.
\newblock {\em Comm. Math. Phys.} \textbf{106} (1986)  321.
\newblock \href{http://projecteuclid.org/euclid.cmp/1104115703}{\texttt{[open
  access]}}.

\bibitem[BW92]{BaumgrtelWollenberg:1992}
H.~Baumg\"{a}rtel and M.~Wollenberg.
\newblock {\em {Causal Nets of Operator Algebras}}.
\newblock Akademie Verlag, 1992.

\bibitem[Cad12]{Cadamuro:2012}
D.~Cadamuro.
\newblock {\em {A Characterization Theorem for Local Operators in Factorizing
  Scattering Models}}.
\newblock PhD thesis, G\"{o}ttingen University, 2012.
\newblock \href{http://arxiv.org/abs/1211.3583v1;
  http://arxiv.org/pdf/1211.3583v1}{\texttt{[open access]}}.

\bibitem[CT15]{CadamuroTanimoto:2015}
D.~Cadamuro and Y.~Tanimoto.
\newblock {Wedge-local fields in integrable models with bound states}.
\newblock {\em arXiv:1502.01313} (2015)  .
\newblock \href{http://arxiv.org/abs/1502.01313v1}{\texttt{[open access]}}.

\bibitem[DF01]{DtschFredenhagen:2001-1}
M.~D\"{u}tsch and K.~Fredenhagen.
\newblock {Algebraic quantum field theory, perturbation theory, and the loop
  expansion}.
\newblock {\em Comm. Math. Phys.} \textbf{219} (2001)  5--30.
\newblock \href{http://arxiv.org/abs/hep-th/0001129}{\texttt{[open access]}}.

\bibitem[DFR95]{DoplicherFredenhagenRoberts:1995}
S.~Doplicher, K.~Fredenhagen, and J.~E. Roberts.
\newblock {The Quantum structure of space-time at the {P}lanck scale and
  quantum fields}.
\newblock {\em Comm. Math. Phys.} \textbf{172} (1995)  187--220.
\newblock \href{http://arxiv.org/abs/hep-th/0303037}{\texttt{[open access]}}.

\bibitem[DHR69]{DoplicherHaagRoberts:1969}
S.~Doplicher, R.~Haag, and J.~E. Roberts.
\newblock {Fields, observables and gauge transformations. I}.
\newblock {\em Comm. Math. Phys.} \textbf{13} (1969)  1--23.
\newblock \href{http://projecteuclid.org/euclid.cmp/1103841481}{\texttt{[open
  access]}}.

\bibitem[DHR74]{DoplicherHaagRoberts:1974}
S.~Doplicher, R.~Haag, and J.~E. Roberts.
\newblock {Local observables and particle statistics. {II}}.
\newblock {\em Comm. Math. Phys.} \textbf{35} (1974)  49--85.
\newblock \href{http://projecteuclid.org/euclid.cmp/1103859518}{\texttt{[open
  access]}}.

\bibitem[DL83]{DAntoniLongo:1983}
C.~D'Antoni and R.~Longo.
\newblock {Interpolation by type I factors and the flip automorphism}.
\newblock \href{http://dx.doi.org/10.1016/0022-1236(83)90018-6}{{\em
  J.Funct.Anal.} \textbf{51} (1983)  361}.

\bibitem[DL84]{DoplicherLongo:1984}
S.~Doplicher and R.~Longo.
\newblock {Standard and split inclusions of von Neumann algebras}.
\newblock {\em Invent. Math.} \textbf{75} (1984)  493--536.

\bibitem[DLMM11]{DappiaggiLechnerMorfaMorales:2010}
C.~Dappiaggi, G.~Lechner, and E.~Morfa-Morales.
\newblock {Deformations of quantum field theories on spacetimes with Killing
  fields}.
\newblock \href{http://dx.doi.org/10.1007/s00220-011-1210-4}{{\em Comm. Math.
  Phys.} \textbf{305} (2011) no.~1, 99--130}.
\newblock \href{http://arxiv.org/abs/1006.3548}{\texttt{[open access]}}.

\bibitem[DMV04]{DAntoniMorsellaVerch:2004}
C.~D'Antoni, G.~Morsella, and R.~Verch.
\newblock {Scaling algebras for charged fields and short-distance analysis for
  localizable and topological charges}.
\newblock \href{http://dx.doi.org/10.1007/s00023-004-0183-7}{{\em Annales Henri
  Poincare} \textbf{5} (2004)  809--870}.
\newblock \href{http://arxiv.org/abs/math-ph/0307048v1}{\texttt{[open
  access]}}.

\bibitem[Dor97]{Dorey:1998}
P.~Dorey.
\newblock {Exact S-matrices}.
\newblock In Z.~H. {L. Palla}, editor, {\em {Conformal Field Theories and
  Integrable Models, Lecture Notes in Physics Volume 498}}, pages 85--125.
  Springer, 1997.
\newblock \href{http://arxiv.org/abs/hep-th/9810026}{\texttt{[open access]}}.

\bibitem[Dri75]{Driessler:1975}
W.~Driessler.
\newblock {Comments on Lightlike Translations and Applications in Relativistic
  Quantum Field Theory}.
\newblock {\em Comm. Math. Phys.} \textbf{44} (1975)  133--141.

\bibitem[DT11]{DybalskiTanimoto:2010}
W.~Dybalski and Y.~Tanimoto.
\newblock {Asymptotic completeness in a class of massless relativistic quantum
  field theories }.
\newblock {\em Comm. Math. Phys.} \textbf{305} (2011)  427--440.
\newblock \href{http://arxiv.org/abs/1006.5430}{\texttt{[open access]}}.

\bibitem[Dyb05]{Dybalski:2005}
W.~Dybalski.
\newblock {Haag-Ruelle scattering theory in presence of massless particles}.
\newblock {\em Lett. Math. Phys.} \textbf{72} (2005)  27--38.
\newblock \href{http://arxiv.org/abs/hep-th/0412226}{\texttt{[open access]}}.

\bibitem[EO73]{EckmannOsterwalder:1973}
J.~Eckmann and K.~Osterwalder.
\newblock {An application of Tomita's theory of modular Hilbert algebras:
  Duality for free Bose fields}.
\newblock \href{http://dx.doi.org/10.1016/0022-1236(73)90062-1}{{\em J. Funct.
  Anal.} \textbf{13} (1973) no.~1, 1--12}.

\bibitem[Fad84]{Faddeev:1984}
L.~D. Faddeev.
\newblock {\em {Quantum completely integrable models in field theory}},
  volume~1 of {\em {Mathematical Physics Reviews}}, pages 107--155.
\newblock 1984.
\newblock In Novikov, S.p. ( Ed.): Mathematical Physics Reviews, Vol. 1,
  107-155.

\bibitem[FG94]{FiglioliniGuido:1994}
F.~Figliolini and D.~Guido.
\newblock {On the type of second quantization factors}.
\newblock {\em Journal of Operator Theory} \textbf{31} (1994) no.~2, 229--252.
\newblock
  \href{http://www.theta.ro/jot/archive/1994-031-002/1994-031-002-003.pdf;
  /scholar?q=info:5bZCfhqH5pcJ:scholar.google.com/&output=instlink&hl=en&as_sdt=0,5&as_ylo=1994&as_yhi=1994&as_vis=1&scillfp=14387801428009440218&oi=lle}{\texttt{[open
  access]}}.

\bibitem[Flo98]{Florig:1998}
M.~Florig.
\newblock {On Borchers' theorem}.
\newblock \href{http://dx.doi.org/10.1023/A:1007546507392}{{\em Lett. Math.
  Phys.} \textbf{46} (1998)  289--293}.

\bibitem[FMS93]{FringMussardoSimonetti:1993}
A.~Fring, G.~Mussardo, and P.~Simonetti.
\newblock {Form-factors of the elementary field in the Bullough-Dodd model}.
\newblock {\em Phys. Lett.} \textbf{B307} (1993)  83--90.
\newblock \href{http://arxiv.org/abs/hep-th/9303108}{\texttt{[open access]}}.

\bibitem[Foi83]{Foit:1983}
J.~J. Foit.
\newblock {Abstract Twisted Duality for Quantum Free Fermi Fields}.
\newblock {\em Publ. Res. Inst. Math. Sci. Kyoto} \textbf{19} (1983)  729--741.
\newblock
  \href{http://www.ems-ph.org/journals/show_abstract.php?issn=0034-5318&vol=19&iss=2&rank=11&srch=searchterm%7CAbstract+Twisted+Duality+for+Quantum+Free+Fermi+Fields}{\texttt{[open
  access]}}.

\bibitem[FR15]{FredenhagenRejzner:2015}
K.~Fredenhagen and K.~Rejzner.
\newblock {Perturbative algebraic quantum field theory}.
\newblock In {\em {Mathematical Aspects of Quantum Field Theories}}, pages
  17--55. Springer, 2015.

\bibitem[FS02]{FassarellaSchroer:2002}
L.~Fassarella and B.~Schroer.
\newblock {Wigner particle theory and local quantum physics}.
\newblock {\em J. Phys.} \textbf{A35} (2002)  9123--9164.
\newblock \href{http://arxiv.org/abs/hep-th/0112168}{\texttt{[open access]}}.

\bibitem[GBV88]{Gracia-BondiaVarilly:1988}
J.~M. Gracia-Bondia and J.~C. Varilly.
\newblock {Algebras of distributions suitable for phase space quantum
  mechanics. I}.
\newblock \href{http://dx.doi.org/10.1063/1.528200}{{\em J. Math. Phys.}
  \textbf{29} (1988)  869--879}.

\bibitem[Ger64]{Gerstenhaber:1964}
M.~Gerstenhaber.
\newblock {On the deformation of rings and algebras}.
\newblock {\em Ann. Math.} \textbf{79} (1964)  59--103.
\newblock \href{http://www.jstor.org/stable/1970484}{\texttt{[open access]}}.

\bibitem[GJ87]{GlimmJaffe:1987}
J.~Glimm and A.~Jaffe.
\newblock {\em {Quantum Physics. A Functional Integral Point Of View}}.
\newblock Springer, 2 edition, 1987.

\bibitem[GL95]{GuidoLongo:1995}
D.~Guido and R.~Longo.
\newblock {An algebraic spin and statistics theorem}.
\newblock \href{http://dx.doi.org/10.1007/BF02101806}{{\em Comm. Math. Phys.}
  \textbf{172} (1995) no.~3, 517}.
\newblock \href{http://projecteuclid.org/euclid.cmp/1104274313}{\texttt{[open
  access]}}.

\bibitem[GL07]{GrosseLechner:2007}
H.~Grosse and G.~Lechner.
\newblock {Wedge-Local Quantum Fields and Noncommutative {M}inkowski Space}.
\newblock \href{http://dx.doi.org/10.1088/1126-6708/2007/11/012}{{\em JHEP}
  \textbf{11} (2007)  012}.
\newblock \href{http://arxiv.org/abs/0706.3992}{\texttt{[open access]}}.

\bibitem[GL08]{GrosseLechner:2008}
H.~Grosse and G.~Lechner.
\newblock {Noncommutative Deformations of Wightman Quantum Field Theories}.
\newblock \href{http://dx.doi.org/10.1088/1126-6708}{{\em JHEP} \textbf{09}
  (2008)  131}.
\newblock \href{http://arxiv.org/abs/0808.3459}{\texttt{[open access]}}.

\bibitem[GLLV13]{GrosseLechnerLudwigVerch:2013}
H.~Grosse, G.~Lechner, T.~Ludwig, and R.~Verch.
\newblock {Wick rotation for quantum field theories on degenerate Moyal space
  (-time)}.
\newblock \href{http://dx.doi.org/10.1063/1.4790886}{{\em J. Math. Phys.}
  \textbf{54} (2013)  022307}.

\bibitem[GLW98]{GuidoLongoWiesbrock:1998}
D.~Guido, R.~Longo, and H.~Wiesbrock.
\newblock {Extensions of conformal nets and superselection structures}.
\newblock \href{http://dx.doi.org/10.1007/s002200050297}{{\em Comm. Math.
  Phys.} \textbf{192} (1998)  217--244}.
\newblock \href{http://arxiv.org/abs/hep-th/9703129}{\texttt{[open access]}}.

\bibitem[Haa58]{Haag:1958}
R.~Haag.
\newblock {Quantum field theories with composite particles and asymptotic
  conditions}.
\newblock {\em Phys. Rev.} \textbf{112} (1958)  669--73.

\bibitem[Haa87]{Haagerup:1987}
U.~Haagerup.
\newblock {Conne{\rq}s bicentralizer problem and uniqueness of the injective
  factor of type III1}.
\newblock \href{http://dx.doi.org/10.1007/BF02392257}{{\em {Acta Mathematica}}
  \textbf{158} (1987) no.~1, 95}.

\bibitem[Haa96]{Haag:1996}
R.~Haag.
\newblock {\em {Local Quantum Physics - Fields, Particles, Algebras}}.
\newblock Springer, second edition, 1996.

\bibitem[Hep65]{Hepp:1965}
K.~Hepp.
\newblock {On the connection between {W}ightman and {LSZ} quantum field
  theory}.
\newblock \href{http://dx.doi.org/10.1007/BF01646494}{{\em Comm. Math. Phys.}
  \textbf{1} (1965)  95--111}.
\newblock \href{http://projecteuclid.org/euclid.cmp/1103758732}{\texttt{[open
  access]}}.

\bibitem[Hol08]{Hollands:2008}
S.~Hollands.
\newblock {Renormalized Quantum Yang-Mills Fields in Curved Spacetime}.
\newblock {\em Rev. Math. Phys.} \textbf{20} (2008)  1033--1172.
\newblock \href{http://arxiv.org/abs/0705.3340}{\texttt{[open access]}}.

\bibitem[Iag93]{Iagolnitzer:1993}
D.~Iagolnitzer.
\newblock {\em {Scattering in Quantum Field Theories}}.
\newblock Princeton University Press, Princeton, 1993.

\bibitem[Jos65]{Jost:1965}
R.~Jost.
\newblock {\em {The General Theory of Quantized Fields}}.
\newblock American Mathematical Society, 1965.

\bibitem[K{\"o}15]{Koehler:2015}
C.~K\"{o}hler.
\newblock PhD thesis, University of Vienna, to appear 2015.

\bibitem[Kas09]{Kasprzak:2009}
P.~Kasprzak.
\newblock {Rieffel Deformation via crossed products}.
\newblock \href{http://dx.doi.org/10.1016/j.jfa.2009.05.013}{{\em J. Funct.
  Anal.} \textbf{257} (2009) no.~5, 1288--1332}.
\newblock \href{http://arxiv.org/abs/math/0606333}{\texttt{[open access]}}.

\bibitem[KL04]{KawahigashiLongo:2004}
Y.~Kawahigashi and R.~Longo.
\newblock {Classification of local conformal nets: Case c $<$ 1}.
\newblock {\em Ann. Math.} \textbf{160} (2004)  493--522.
\newblock \href{http://arxiv.org/abs/math-ph/0201015}{\texttt{[open access]}}.

\bibitem[KTTW77]{KarowskiThunTruongWeisz:1977}
M.~Karowski, H.~Thun, T.~Truong, and P.~Weisz.
\newblock {On the uniqueness of a purely elastic S-matrix in (1+1) dimensions}.
\newblock \href{http://dx.doi.org/10.1016/0370-2693(77)90382-3}{{\em Physics
  Letters B} \textbf{67} (1977) no.~3, 321--322}.

\bibitem[Lec03]{Lechner:2003}
G.~Lechner.
\newblock {Polarization-free quantum fields and interaction}.
\newblock \href{http://dx.doi.org/10.1023/A:1025772304804}{{\em Lett. Math.
  Phys.} \textbf{64} (2003)  137--154}.
\newblock \href{http://arxiv.org/abs/hep-th/0303062}{\texttt{[open access]}}.

\bibitem[Lec05]{Lechner:2005}
G.~Lechner.
\newblock {On the existence of local observables in theories with a factorizing
  S-matrix}.
\newblock \href{http://dx.doi.org/10.1088/0305-4470}{{\em J. Phys.}
  \textbf{A38} (2005)  3045--3056}.
\newblock \href{http://arxiv.org/abs/math-ph/0405062}{\texttt{[open access]}}.

\bibitem[Lec06]{Lechner:2006}
G.~Lechner.
\newblock {\em {On the construction of quantum field theories with factorizing
  S-matrices}}.
\newblock PhD thesis, University of G\"{o}ttingen, 2006.
\newblock \href{http://arxiv.org/abs/math-ph/0611050}{\texttt{[open access]}}.

\bibitem[Lec08]{Lechner:2008}
G.~Lechner.
\newblock {Construction of Quantum Field Theories with Factorizing
  {S}-Matrices}.
\newblock \href{http://dx.doi.org/10.1007/s00220-007-0381-5}{{\em Comm. Math.
  Phys.} \textbf{277} (2008)  821--860}.
\newblock \href{http://arxiv.org/abs/math-ph/0601022}{\texttt{[open access]}}.

\bibitem[Lec12]{Lechner:2012}
G.~Lechner.
\newblock {Deformations of quantum field theories and integrable models}.
\newblock \href{http://dx.doi.org/10.1007/s00220-011-1390-y}{{\em Comm. Math.
  Phys.} \textbf{312} (2012) no.~1, 265--302}.
\newblock \href{http://arxiv.org/abs/1104.1948}{\texttt{[open access]}}.

\bibitem[Lec15]{Lechner:Erratum2015}
G.~Lechner.
\newblock {Erratum}.
\newblock {\em to appear} (2015)  .

\bibitem[LL14]{LechnerLongo:2014}
G.~Lechner and R.~Longo.
\newblock {Localization in Nets of Standard Spaces}.
\newblock \href{http://dx.doi.org/10.1007/s00220-014-2199-2}{{\em to appear in
  Comm. Math. Phys.} (2014)  }.
\newblock \href{http://arxiv.org/abs/1403.1226}{\texttt{[open access]}}.

\bibitem[LM95]{LiguoriMintchev:1995-1}
A.~Liguori and M.~Mintchev.
\newblock {Fock representations of quantum fields with generalized statistics}.
\newblock {\em Comm. Math. Phys.} \textbf{169} (1995)  635--652.
\newblock \href{http://arxiv.org/abs/hep-th/9403039}{\texttt{[open access]}}.

\bibitem[Lon79]{Longo:1979}
R.~Longo.
\newblock {Notes on algebraic invariants for noncommutative dynamical systems}.
\newblock {\em Comm. Math. Phys.} \textbf{69} (1979)  195--207.

\bibitem[Lon08]{Longo:2008}
R.~Longo.
\newblock {Lectures on Conformal Nets - Part 1}.
\newblock In {\em {Von Neumann algebras in Sibiu}}, pages 33--91. Theta, 2008.
\newblock
  \href{http://www.mat.uniroma2.it/~longo/Lecture_Notes_files/LN-Part1.pdf}{\texttt{[open
  access]}}.

\bibitem[LR07]{LauridsenRibeiro:2007}
P.~Lauridsen-Ribeiro.
\newblock {\em {Structural and Dynamical Aspects of the AdS/CFT Correspondence:
  a Rigorous Approach}}.
\newblock PhD thesis, Sao Paulo, 2007.
\newblock \href{http://arxiv.org/abs/0712.0401}{\texttt{[open access]}}.

\bibitem[LRT78]{LeylandsRobertsTestard:1978}
P.~Leylands, J.~E. Roberts, and D.~Testard.
\newblock {Duality for Quantum Free Fields}.
\newblock {\em Preprint} (1978)  .

\bibitem[LS14]{LechnerSchutzenhofer:2013}
G.~Lechner and C.~Sch\"{u}tzenhofer.
\newblock {Towards an operator-algebraic construction of integrable global
  gauge theories}.
\newblock \href{http://dx.doi.org/10.1007/s00023-013-0260-x}{{\em Annales Henri
  Poincar\'{e}} \textbf{15} (2014) no.~4, 645--678}.
\newblock \href{http://arxiv.org/abs/1208.2366v1}{\texttt{[open access]}}.

\bibitem[LS15]{LechnerSchlemmer:2015}
G.~Lechner and J.~Schlemmer.
\newblock {Thermal equilibrium states for quantum fields on non-commutative
  spacetimes}.
\newblock {\em arXiv:1503.01639} (2015)  .
\newblock \href{http://arxiv.org/abs/1503.01639}{\texttt{[open access]}}.

\bibitem[LST13]{LechnerSchlemmerTanimoto:2013}
G.~Lechner, J.~Schlemmer, and Y.~Tanimoto.
\newblock {On the equivalence of two deformation schemes in quantum field
  theory}.
\newblock {\em Lett. Math. Phys.} \textbf{103} (2013) no.~4, 421--437.
\newblock \href{On the equivalence of two deformation schemes in quantum field
  theory}{\texttt{[open access]}}.

\bibitem[LW11a]{LechnerWaldmann:2011}
G.~Lechner and S.~Waldmann.
\newblock {Strict deformation quantization of locally convex algebras and
  modules}.
\newblock {\em arXiv:1109.5950} (2011)  .
\newblock \href{http://arxiv.org/abs/1109.5950}{\texttt{[open access]}}.

\bibitem[LW11b]{LongoWitten:2010}
R.~Longo and E.~Witten.
\newblock {An Algebraic Construction of Boundary Quantum Field Theory }.
\newblock \href{http://dx.doi.org/10.1007/s00220-010-1133-5}{{\em Comm. Math.
  Phys.} \textbf{303} (2011) no.~1, 213--232}.
\newblock \href{http://arxiv.org/abs/1004.0616}{\texttt{[open access]}}.

\bibitem[M{\"u}98]{Muger:1998}
M.~M\"{u}ger.
\newblock {Superselection structure of massive quantum field theories in 1+1
  dimensions}.
\newblock {\em Rev. Math. Phys.} \textbf{10} (1998)  1147--1170.
\newblock \href{http://arxiv.org/abs/hep-th/9705019}{\texttt{[open access]}}.

\bibitem[Mar69]{Martin:1969}
A.~Martin.
\newblock {\em {Can one continue the scattering amplitude through the elastic
  cut?}}, page 113.
\newblock Nauka, Moscow, 1969.

\bibitem[MM12]{Morfa-Morales:2012}
E.~Morfa-Morales.
\newblock {\em {Deformations of Quantum Field Theories on Curved Spacetimes}}.
\newblock PhD thesis, University of Vienna, 2012.

\bibitem[MSY06]{MundSchroerYngvason:2006}
J.~Mund, B.~Schroer, and J.~Yngvason.
\newblock {String-localized quantum fields and modular localization}.
\newblock \href{http://dx.doi.org/10.1007/s00220-006-0067-4}{{\em Comm. Math.
  Phys.} \textbf{268} (2006)  621--672}.
\newblock \href{http://arxiv.org/abs/math-ph/0511042}{\texttt{[open access]}}.

\bibitem[Muc14]{Much:2013_2}
A.~Much.
\newblock {Quantum Mechanical Effects from Deformation Theory}.
\newblock {\em J. Math. Phys.} \textbf{55} (2014)  022302.
\newblock \href{http://arxiv.org/abs/1307.2609v2}{\texttt{[open access]}}.

\bibitem[Mun01]{Mund:2001}
J.~Mund.
\newblock {The Bisognano-Wichmann theorem for massive theories}.
\newblock {\em Annales Henri Poincare} \textbf{2} (2001)  907--926.
\newblock \href{http://arxiv.org/abs/hep-th/0101227}{\texttt{[open access]}}.

\bibitem[Mun12]{Mund:2010}
J.~Mund.
\newblock {An Algebraic Jost-Schroer Theorem for Massive Theories }.
\newblock \href{http://dx.doi.org/10.1007/s00220-012-1546-4}{{\em Comm. Math.
  Phys.} \textbf{315} (2012)  445--464}.
\newblock \href{http://arxiv.org/abs/1012.1454}{\texttt{[open access]}}.

\bibitem[Mus92]{Mussardo:1992}
G.~Mussardo.
\newblock {Off critical statistical models: Factorized scattering theories and
  bootstrap program}.
\newblock \href{http://dx.doi.org/10.1016/0370-1573(92)90047-4}{{\em Phys.
  Rept.} \textbf{218} (1992)  215--379}.

\bibitem[Nes13]{Neshveyev:2013}
S.~Neshveyev.
\newblock
\newblock {Smooth crossed products of Rieffel's deformations}{\em to appear in
  Lett. Math. Phys.} (July, 2013)  .
\newblock \href{http://arxiv.org/abs/1307.2016v1}{\texttt{[open access]}}.

\bibitem[Nie98]{Niedermaier:1998}
M.~R. Niedermaier.
\newblock {A derivation of the cyclic form factor equation}.
\newblock {\em Comm. Math. Phys.} \textbf{196} (1998)  411--428.
\newblock \href{http://arxiv.org/abs/hep-th/9706172}{\texttt{[open access]}}.

\bibitem[Pla13]{Plaschke:2012}
M.~Plaschke.
\newblock {Wedge Local Deformations of Charged Fields leading to Anyonic
  Commutation Relations}.
\newblock \href{http://dx.doi.org/10.1007/s11005-013-0607-8}{{\em Lett. Math.
  Phys.} \textbf{103} (2013) no.~5, 507--532}.
\newblock \href{http://arxiv.org/abs/1208.6141v1}{\texttt{[open access]}}.

\bibitem[Rea96]{Read:1996}
C.~J. Read.
\newblock {Quantum field theories in all dimensions}.
\newblock \href{http://dx.doi.org/10.1007/BF02099541}{{\em Comm. Math. Phys.}
  \textbf{177} (1996) no.~3, 631}.
\newblock \href{http://dx.doi.org/10.1007/BF02099541}{\texttt{[open access]}}.

\bibitem[Reh96]{Rehren:1996}
K.~Rehren.
\newblock {Comments on a recent solution to Wightman's axioms}.
\newblock {\em Comm. Math. Phys.} \textbf{178} (1996)  453--466.

\bibitem[Reh15]{Rehren:2015}
K.-H. Rehren.
\newblock {Algebraic conformal quantum field theory in perspective}.
\newblock {\em arXiv:1501.03313} (2015)  .
\newblock \href{http://arxiv.org/abs/1501.03313}{\texttt{[open access]}}.

\bibitem[Rie92]{Rieffel:1992}
M.~A. Rieffel.
\newblock {\em {Deformation Quantization for Actions of $R^d$}}, volume 106 of
  {\em {Memoirs of the Amerian Mathematical Society}}.
\newblock American Mathematical Society, Providence, Rhode Island, 1992.

\bibitem[RS61]{ReehSchlieder:1961}
H.~Reeh and S.~Schlieder.
\newblock {Bemerkungen zur Unit\"{a}r\"{a}quivalenz von lorentzinvarianten
  Feldern}.
\newblock {\em Il Nuovo Cimento} \textbf{22} (1961) no.~5, 1051--1068.

\bibitem[Rue62]{Ruelle:1962}
D.~Ruelle.
\newblock {On the asymptotic condition in quantum field theory}.
\newblock {\em Helvetica Physica Acta. Physica Theoretica. Societatis Physicae
  Helveticae Commentaria Publica} \textbf{35} (1962)  147--163.

\bibitem[Sch97]{Schroer:1997-1}
B.~Schroer.
\newblock {Modular localization and the bootstrap-formfactor program}.
\newblock {\em Nucl. Phys.} \textbf{B499} (1997)  547--568.
\newblock \href{http://arxiv.org/abs/hep-th/9702145v1}{\texttt{[open access]}}.

\bibitem[Sch99]{Schroer:1999}
B.~Schroer.
\newblock {Modular Wedge Localization and the d=1+1 Formfactor Program}.
\newblock {\em Annals Phys.} \textbf{275} (1999)  190--223.
\newblock \href{http://arxiv.org/abs/hep-th/9712124}{\texttt{[open access]}}.

\bibitem[Smi92]{Smirnov:1992}
F.~A. Smirnov.
\newblock {\em {Form Factors in Completely Integrable Models of Quantum Field
  Theory}}.
\newblock World Scientific, Singapore, 1992.

\bibitem[Smi94]{Smirnov:1994}
F.~A. Smirnov.
\newblock {A New set of exact form-factors}.
\newblock {\em Int. J. Mod. Phys.} \textbf{A9} (1994)  5121--5144.
\newblock \href{http://arxiv.org/abs/hep-th/9312039}{\texttt{[open access]}}.

\bibitem[Sol08]{Soloviev:2008}
M.~A. Soloviev.
\newblock {On the failure of microcausality in noncommutative field theories}.
\newblock \href{http://dx.doi.org/10.1103/PhysRevD.77.125013}{{\em Phys. Rev.}
  \textbf{D77} (2008)  125013}.
\newblock \href{http://arxiv.org/abs/0802.0997}{\texttt{[open access]}}.

\bibitem[Sum82]{Summers:1982}
S.~J. Summers.
\newblock {Normal product states for fermions and twisted duality for CCR- and
  CAR-type algebras with application to the Yukawa2 quantum field model}.
\newblock \href{http://dx.doi.org/10.1007/BF01205664}{{\em Comm. Math. Phys.}
  \textbf{86} (1982)  111--141}.

\bibitem[Sum90]{Summers:1990}
S.~J. Summers.
\newblock {On the independence of local algebras in quantum field theory}.
\newblock \href{http://dx.doi.org/10.1142/S0129055X90000090}{{\em Rev. Math.
  Phys.} \textbf{2} (1990)  201--247}.

\bibitem[Sum12]{Summers:2011}
S.~J. Summers.
\newblock {A Perspective on Constructive Quantum Field Theory}.
\newblock {\em Preprint, arXiv:1203.3991} (2012)  .
\newblock \href{http://arxiv.org/abs/1203.3991}{\texttt{[open access]}}.

\bibitem[SW64]{StreaterWightman:1964}
R.~F. Streater and A.~Wightman.
\newblock {\em {PCT, Spin and Statistics, and All That}}.
\newblock Benjamin-Cummings, Reading, MA, 1964.

\bibitem[Tan12]{Tanimoto:2011-1}
Y.~Tanimoto.
\newblock {Construction of wedge-local nets of observables through Longo-Witten
  endomorphisms}.
\newblock \href{http://dx.doi.org/10.1007/s00220-012-1462-7}{{\em Comm. Math.
  Phys.} \textbf{314} (2012) no.~2, 443--469}.
\newblock \href{http://arxiv.org/abs/1107.2629}{\texttt{[open access]}}.

\bibitem[TW97]{ThomasWichmann:1997}
L.~J. Thomas and E.~H. Wichmann.
\newblock {On the causal structure of {M}inkowski space-time}.
\newblock \href{http://dx.doi.org/10.1063/1.531954}{{\em J. Math. Phys.}
  \textbf{38} (1997)  5044--5086}.

\bibitem[Wal07]{Waldmann:2007}
S.~Waldmann.
\newblock {\em {Poisson-Geometrie und Deformationsquantisierung}}.
\newblock Springer, 2007.

\bibitem[Wei95]{Weinberg:1995}
S.~Weinberg.
\newblock {\em {The Quantum Theory of Fields I - Foundations}}.
\newblock Cambridge University Press, 1995.

\bibitem[Yng70]{Yngvason:1970}
J.~Yngvason.
\newblock {Zero-mass infinite spin representations of the poincare group and
  quantum field theory}.
\newblock {\em Comm. Math. Phys.} \textbf{18} (1970)  195--203.
\newblock \href{http://projecteuclid.org/euclid.cmp/1103842535}{\texttt{[open
  access]}}.

\bibitem[ZZ79]{ZamolodchikovZamolodchikov:1979}
A.~B. Zamolodchikov and A.~B. Zamolodchikov.
\newblock {Factorized {S}-matrices in two dimensions as the exact solutions of
  certain relativistic quantum field models}.
\newblock \href{http://dx.doi.org/10.1016/0003-4916(79)90391-9}{{\em Annals
  Phys.} \textbf{120} (1979)  253--291}.

\end{thebibliography}
\end{document}